\documentclass{aastex}

\usepackage{emulateapj5}
\usepackage{graphicx}
\usepackage{times}

\newcommand{\lsim}{\mbox{\hspace{.2em}\raisebox{.5ex}{$<$}\hspace{-.8em}\raisebox{-.5ex}{$\sim$}\hspace{.2em}}}
\newcommand{\gsim}{\mbox{\hspace{.2em}\raisebox{.5ex}{$>$}\hspace{-.8em}\raisebox{-.5ex}{$\sim$}\hspace{.2em}}}

\def\asca       {{\em ASCA}\/}
\def\chandra    {{\em Chandra}\/}

\def\einstein   {{\em Einstein}\/}
\def\xmm        {XMM-{\em Newton}\/}
\def\rosat      {{\em ROSAT}\/}

\def\hst        {{\em HST}\/}

\def\Ls         {{\em L$_{\ast}$}}

\begin{document}

 
\title{X-ray thermal coronae of galaxies in hot clusters 
--- ubiquity of embedded mini cooling cores}

\author{
M.\ Sun,$^{\!}$\altaffilmark{1}
C.\ Jones,$^{\!}$\altaffilmark{2}
W.\ Forman,$^{\!}$\altaffilmark{2}
A.\ Vikhlinin,$^{\!}$\altaffilmark{2, 3}
M.\ Donahue,$^{\!}$\altaffilmark{1}
M.\ Voit$^{\!}$\altaffilmark{1}
}

\altaffiltext{1}{Department of Physics and Astronomy, MSU, East Lansing, MI 48824; sunm@pa.msu.edu}
\altaffiltext{2}{Harvard-Smithsonian Center for Astrophysics,
60 Garden St., Cambridge, MA 02138}
\altaffiltext{3}{Space Research Institute, Moscow, Russia}

\shorttitle{X-ray coronae in hot clusters}
\shortauthors{SUN ET AL.}

\begin{abstract}

We present a systematic investigation of X-ray thermal coronae in 157
early-type galaxies and 22 late-type galaxies from a survey of 25 hot
($kT >$ 3 keV), nearby ($z <$ 0.05) clusters, based on \chandra\ archival
data. Cool galactic coronae ($kT$ = 0.5 - 1.1 keV generally) have been
found to be very common, $>$ 60\% in NIR selected galaxies that are
more luminous than 2 \Ls, and $>$ 40\% in \Ls\ $< L_{\rm Ks} <$ 2 \Ls\
galaxies. These embedded coronae in hot clusters are generally smaller
(1.5-4 kpc radii), less luminous ($\lsim 10^{41}$ erg s$^{-1}$), and
less massive (10$^{6.5}$-10$^{8}$ M$_{\odot}$) than coronae in poor
environments, demonstrating the negative effects of hot cluster
environments on galactic coronae. Nevertheless, these coronae still
manage to survive ICM stripping, evaporation, rapid cooling, and
powerful AGN outflows, making them a rich source of information about
gas stripping, microscopic transport, and feedback processes in the
cluster environment. Heat conduction across the boundary of the coronae
has to be suppressed by a factor of $\gsim$ 100, which implies the X-ray
gas in early-type galaxies is magnetized and the magnetic field plays
an important role in energy transfer. Stripping through transport
processes (viscosity or turbulence) also needs to be suppressed by
at least a factor of ten at the coronal boundary. The stellar mass
loss inside the corona is key to maintaining the gas balance in coronae.
The luminous, embedded coronae, with high central density (0.1 - 0.4
cm$^{-3}$), are mini-versions of group and cluster cooling cores.
As the prevalence of coronae of massive galaxies implies a long lifetime
($\gsim$ several Gyr), there must be a heat source inside coronae to offset
cooling. While we argue that AGN heating may not generally be the heat source,
we conclude that SN heating can be enough as long as the kinetic energy of
SNe can be efficiently dissipated. We have also observed a connection
between radiative cooling and the SMBH activity of their host galaxies
as many coronae are associated with powerful radio galaxies. Cooling of
the coronal gas may provide fuel for the central SMBH and nuclear star
formation in environments where the amount of galactic cold gas is otherwise
at a minimum.
Diffuse thermal coronae have also been detected in at least 8 of 22 late-type
(Sb or later) galaxies in our sample. Evidence for enhanced star formation
triggered by the ICM pressure has been found in four late-type galaxies.
The fraction of luminous X-ray AGN ($> 10^{41}$ ergs s$^{-1}$) is not small
($\sim$ 5\%) in our sample.

\end{abstract}


\section{Introduction}

One of the most important discoveries by the \einstein\ observatory was
the ubiquity of X-ray coronae of early-type galaxies in the field and
poor environments (Forman, Jones \& Tucker 1985). This discovery changed
our view of early-type galaxies: they are gas-rich instead of gas-poor,
but the gas is in the X-ray phase with a temperature of $\sim 10^{7}$ K. 
The consensus is that the X-ray gas of early-type galaxies originates as
the stellar mass lost from evolved stars and planetary nebulae, accumulating in
the galaxy without much escaping via galactic winds (e.g., Mathews 1990). 
Their origin is different from that of the cluster ICM, which should be
mostly primordial. The gas ejected from evolved stars collides
and passes through shocks. The gas temperature has been raised to the stellar
kinetic temperature determined by the stellar velocity dispersion, and
may further be raised by heating from supernovae (SNe). However, it has
been known that the kinetic energy of SNe is not efficiently dissipated
into the hot ISM gas, especially for less massive galaxies with shallow 
potentials, so galactic winds can form in less massive galaxies and make
them X-ray faint (Mathews \& Baker 1971; David, Forman \& Jones 1991;
Brown \& Bregman 1998). Galactic cooling cores have generally formed
in massive galaxies with deep potentials. Cooling of the X-ray coronal gas
indeed happens in at least some galaxies (e.g., the O VI detections by
{\em FUSE}, Bregman et al. 2005).

It has been long known that the X-ray luminosities ($L_{\rm X}$) of coronae
correlates with the optical $B$ band luminosities ($L_{\rm B}$) of their
host galaxies, but the dispersion is large (Forman et al. 1985; Canizares,
Fabbiano \& Trinchieri 1987; Brown \& Bregman 1998; O'Sullivan et al.
2001). The large dispersion has been
a puzzle for almost two decades. The narrow color-magnitude relation and
the narrow fundamental plane of early-type galaxies imply that local
early-type galaxies are a homogeneous group. Nevertheless, early-type galaxies
are not a one-parameter class, and the X-ray luminosity of coronae is also
affected by properties besides the optical luminosity of the galaxy.
Various factors have been proposed to explain the dispersion of the $L_{\rm X}$ -
$L_{\rm B}$ relation for the X-ray coronae of early-type galaxies, including
SN rate, galactic rotation, metallicity, dark matter halo and environments
(see a review by Mathews \& Brighenti 2003 and the references therein).	
However, no extra parameter has been found to significantly tighten the
correlation, and no single scenario seems to be able to explain the large
dispersion. It is likely that both internal effects (e.g., SN rate \& dark
matter halo) and external effects (environment) contribute to the dispersion.

Limited by its angular resolution (30$''$), the X-ray coronae detected by
\rosat\ are almost all in the field, or poor environments, or cool clusters (e.g.,
Virgo). There were some
attempts to study X-ray coronae of galaxies in hot clusters with the \einstein\
and \rosat\ data (e.g., Canizares \& Blizzard 1991 for Coma; Dow \& White 1995
for Coma; Sakelliou \& Merrifield 1998 for A2634), but the results were generally
non-detections or inconclusive. Furthermore, the \rosat\ results on galaxy
coronae suffer greatly from the contamination by AGN, X-ray binaries and
ICM emission. Prior to the launch of \chandra, no 10$^{7}$ K galaxy coronae (not
including cluster cooling cores) were firmly detected in hot ($T \sim$ 10$^{8}$ K)
clusters, nor were they expected, since evaporation and ram-pressure stripping
by the hot, dense ICM should very efficiently remove gas from galaxies. The
first direct evidence for the survival of galaxy coronae in hot clusters
came from the \chandra\ observations of the Coma cluster. Vikhlinin et al.
(2001, V01 hereafter) found small, but extended ($\sim$ 2 kpc in radius), X-ray coronae
($kT \sim$ 1-2 keV) in the cores of the two dominant Coma galaxies, NGC 4874
and NGC 4889, where the surrounding ICM has a temperature of 8 - 9 keV.
More and more similar embedded coronae have been found since then.
At present, eight more coronae, small but spatially resolved, were unambiguously
revealed by \chandra\ in hot clusters ($>$ 3 keV) and investigated in detail:
two lie in the 3.2 keV cluster A1060 (Yamasaki et al. 2002), four in a 5-6 keV
region of A1367 (Sun et al. 2005, hereafter S05), one in a 6-7 keV region of Perseus
(Sun, Jerius \& Jones 2005, hereafter SJJ05), and one associated with the cD
galaxy of the 4 keV cluster A2670 (Fujita, Sarazin \& Sivakoff 2006). Thermal
emission from cluster galaxies was also detected in several other cases (e.g.,
Smith 2003; Hardcastle, Sakelliou \& Worrall 2005), although detailed analysis
was not done because the X-ray sources are either unresolved or faint. Among all these
detections, the corona associated with NGC~1265, the prototype of narrow-angle
tail (NAT) radio galaxies in Perseus, is the most interesting one, as its corona
survives gas stripping despite motion with a Mach number of $\sim$ 3, ICM
evaporation ($T_{\rm ICM}/T_{\rm ISM} \sim$ 10), fast cooling
($t_{\rm cooling} \sim$ 30 Myr at the center) and powerful AGN outbursts.
These galaxy coronae ($kT \sim$ 1 keV) in rich environments (high ambient
pressure) are pressure confined and small (1.5 - 4.5 kpc, or 3 - 10$''$ in radius
at the distance of the Coma cluster) and require the superior \chandra\ angular
resolution to resolve them.
These small coronae should form before their host galaxies began to evolve
in hot clusters. If they were once destroyed in hot clusters, the hot ICM
should have filled the interstellar space. It is then difficult for galactic cooling
cores to re-form against strong stripping and evaporation.

These embedded galaxy coronae are perfect targets to study the gas physics
and microscopic transport processes involved in the ICM-corona interaction.
Gas stripping and evaporation by the surrounding hot dense ICM should be very
efficient to destroy coronae. However, V01 pointed out that the
survival of two cool coronae in Coma requires that heat conduction must be
suppressed by a factor of 30 - 100 at the ISM-ICM boundary. S05 and SJJ05 derived
similar conclusions and the results also require viscosity to be suppressed.
The galactic magnetic field should be responsible for the suppression, and so
a better understanding of the galactic magnetic field evolution in rich
environments is required. We shall emphasize that these embedded coronae
in hot clusters are generally smaller than the mean free path of particles
in the ICM without magnetic field, which is:

\begin{eqnarray}
\lambda &=& 7.8 (\frac{kT_{\rm ICM}}{\rm 5 keV})^{2} (\frac{n_{\rm e, ICM}}{10^{-3} {\rm cm}^{-3}})^{-1} {\rm kpc}
\end{eqnarray}

Therefore, it is questionable whether we can apply hydrodynamics to the
ICM flow around the embedded coronae. The distribution function may not
be Maxwellian so ideally kinetic theory or at least rarefied gas dynamics
should be applied. This situation is different from coronae in cool groups
(e.g., NGC~1404 in Fornax, Machacek et al. 2005), where $\lambda$ is
generally smaller than 1 kpc while coronae are larger (e.g., $\sim$ 9 kpc for
NGC~1404). The inclusion of magnetic field in the ICM may further complicates
the problem, by making transport processes anisotropic. The details
are certainly complicated, largely depending on the magnetic field
configuration and evolution. We notice that simulations of the evolution
of X-ray coronae in clusters and groups all use hydrodynamics
(e.g., Toniazzo \& Schindler 2001), while the detailed transport processes
involving in the corona-ICM interaction were not incorporated.

S05 and SJJ05 also showed that many host galaxies of coronae are active in radio
and sometimes are luminous (e.g., NGC~1265). The radio emission generally
``turns on'' after traversing the dense coronae (i.e. morphologically anti-correlated,
e.g., NGC~3842, NGC~4874 and NGC~1265). In NGC~1265, since the jet power is
so strong ($>$ 1000 times the thermal energy of the corona), the narrow
jets must carry nearly all their energy away from the central SMBH and
release the energy outside the corona to preserve the small corona. Thus,
this fact implies that AGN heating may not always be able to heat the central
few kiloparsecs of a cooling core significantly.
For NGC~1265, we also found that the Bondi accretion luminosity of the detected
coronal gas is similar to the jet power. As the inner $\sim$ 2 kpc dense core of the
corona can survive both high-speed stripping and powerful
AGN feedback, the cooling of coronae can fuel the
central SMBH in rich environments where the amount of cold galactic
gas is at a minimum.

These embedded coronae are also ideal targets to examine the effects of
hot cluster environments on the properties of coronae (e.g., the
$L_{\rm X} - L_{\rm B}$ relation, the size and X-ray gas mass of coronae),
especially through a comparison with a sample of coronae in the field
or poor environments. These cool coronae may also be seeds for future
cluster cooling cores. Motl et al. (2004) presented a scenario for
the formation of cluster cooling cores via hierarchical mergers.
Although the cool cores in their simulations are larger than what we observed,
it is interesting to explore the evolution of these coronae and their
frequency in clusters, and to further compare with simulations. In principle,
the properties of the coronae also shed light on the evolution of
their early-type host galaxies, as the evolution processes (e.g., mergers)
can easily impact the small coronae.

Compared to the X-ray coronae of early-type galaxies, we know even less
about the X-ray coronae of late-type galaxies in hot clusters. One of the first
detailed studies with \chandra\ or \xmm\ is UGC~6697 in A1367 (Sun \& Vikhlinin
2005). We suggested
that the starburst in UGC~6697 was triggered by the interaction with the ICM.
During the course of this project, we have found a spectacular X-ray
tail behind a small starburst galaxy ESO~137-001 (SFR $\sim$ 10 M$_{\odot}$ / yr)
near the center of a massive cluster A3627 (Sun et al. 2006). The X-ray tail is
long (70 kpc) with a length-to-width ratio of $\sim$ 10. We interpret this tail
as the stripped ISM of ESO 137-001 mixed with the hot cluster medium, with this
blue galaxy being converted into a gas-poor galaxy.

These remarkable examples of coronae in early-type and
late-type galaxies are important to understand the galactic ISM in hot
clusters. However, the results based on just a few detections can be biased.
We do not know how common the embedded coronae are. We know nothing about
the properties of the whole population (e.g., luminosity, temperature, size
and gas mass). What are the environmental effects on the properties of
embedded coronae? What processes mediate the energy balance and transfer inside
coronae? What is the general connection between coronae and the AGN
activity of their host galaxies. Only a systematic analysis based on a
well-selected sample can answer these questions and characterize the general
properties of the whole population. Only with the knowledge of the whole
population, can we better understand the results from detailed investigations
of individual coronae. In this paper, we present the first systematic
study of X-ray coronae of cluster galaxies, in 25 nearby ($z <$ 0.05), hot
($kT >$ 3 keV) clusters using archival \chandra\ data. The cluster and
galaxy samples are defined in $\S$2. The data analysis is summarized
in $\S$3. The properties of the corona population are present in $\S$4
(for early-type galaxies) and $\S$5 (for late-type galaxies). We
summarize the fraction of luminous X-ray AGN in our sample in $\S$6.
$\S$7 is the discussion, while $\S$8 is the conclusions.
Notes of interesting coronae and galaxies in the sample are present
in the appendix.

\section{The sample}

\subsection{The cluster sample}

The cluster sample is drawn from the \chandra\ archive (as of March, 2006).
We limit our investigation to nearby ($z <$ 0.05) hot clusters to better resolve
embedded coronae and to detect faint coronae ($\lsim$ 10$^{40}$ ergs s$^{-1}$).
Clusters are selected with the following criteria: 1) $kT_{\rm ICM} \geq$
3 keV (from the X-Ray Galaxy Clusters Database, BAX); 2) $z <$ 0.05;
3) Combined exposure time of $>$ 10 ks for $z<$0.035 clusters, and $>$ 20 ks
for 0.035$<z<$0.05 clusters; 4) at least one known cluster member listed in the
NASA/IPAC Extragalactic Database (NED), excluding the cD if it is located in
a dense cluster cooling core. There are 24 clusters fulfilling these criteria
and all are included in the sample. As there are only 6 clusters hotter than
5 keV in these 24 clusters (including all known $>$ 5 keV clusters within $z$
of 0.035), we lower the threshold on exposure time to 14 ksec to add another
hot cluster A3558 ($kT$=5.4 keV). As the detection limits on coronae depend on
the \chandra\ exposure, the distance of the cluster, the surrounding ICM
background, the position of the galaxy in the detector and the Galactic
absorption, the resulting sensitivities are not uniform even in the
same cluster but
generally sufficient to detect X-ray coronae of $\gsim$ 2$\times10^{40}$
ergs s$^{-1}$ (0.5 - 2 keV). The sample includes 68 \chandra\ pointings
with a total exposure of 2.77 Ms and covers a total sky area of 3.3 deg$^{2}$.
The cluster temperatures range from 3 to 10.2 keV and the cluster 3D
velocity dispersion ranges from 900 km/s to 2300 km/s.
The clusters in the sample also span
a wide range of evolutionary stages, from dynamically young merging clusters
(e.g., A1367, ZW~1615+35 and A3627) to cooling core clusters (e.g., Centaurus,
Perseus and A496). The cluster distances are derived from their redshift
(from NED), assuming H$_{0}$ = 70 km s$^{-1}$
Mpc$^{-1}$, $\Omega$$_{\rm M}$=0.3, and $\Omega_{\rm \Lambda}$=0.7 (Table 1).

\subsection{The sample of cluster galaxies}

We mainly select galaxies by their near infrared (NIR) $K$ band
luminosities, as $K$ band
light is a more reliable measure of the stellar mass and is little affected
by extinction. Moreover, the Two Micron All Sky Survey (2MASS) measures the
NIR photometry of cluster galaxies homogeneously. As concluded
in S05, massive galaxies, more likely to have galactic cooling cores,
are much more likely to maintain their coronae in rich environments.
In this work, we selected a luminosity threshold of
10$^{10.95}$ L$_{\odot}$ ($L_{\rm Ks, cut}$ hereafter) in the
2MASS $K_{\rm s}$ band. All cluster galaxies more luminous than this threshold
have been included in the sample if they lie completely in the \chandra\ fields.
This $L_{\rm Ks, cut}$ is derived from an $B$ band luminosity of 2$\times10^{10}$
L$_{\odot}$ (about the \Ls\ in the $B$ band, e.g., Beijersbergen et al. 2002),
by assuming a $B - K_{\rm s}$ color index of
3.7 (Jarrett 2000) and $B_{\odot}$ = 5.47, $K_{\odot}$ = 3.39.
At $z$=0.05, this threshold corresponds to 12.8 mag, well within the 2MASS limit
magnitude of 13.5 mag. We have also confirmed that galaxies more luminous
than $L_{\rm Ks, cut}$ in the \chandra\ field have been well covered by
previous redshift surveys, except for galaxies in the 3C129.1 cluster that
is behind the Galactic plane.
Kochanek et al. (2001) measured the $K_{\rm s}$-band luminosity function of local
galaxies and found $M_{K*}$ = -24.30 and -23.75 for early-type and later-type
galaxies respectively (for H$_{0}$ = 70 km s$^{-1}$ Mpc$^{-1}$), which corresponds to
11.9$\times10^{10}$ L$_{\odot}$ and 7.2$\times10^{10}$ L$_{\odot}$.
However, as our analysis requires only a luminosity cut-off where
velocity and photometry data are nearly complete, we adopt a $K_{\rm s}$
band \Ls\ of 10$^{11.08}$ L$_{\odot}$ from Kochanek et al. (2001), and still
use a luminosity cut-off of 10$^{10.95}$ L$_{\odot}$ (0.74 \Ls) in this work.
As some regions of Centaurus and Perseus clusters were observed for $>$ 200
ks, we lower the threshold to 0.5 $L_{\rm Ks, cut}$ in these regions
to examine the X-ray properties of fainter cluster galaxies. For A3558, we
restrict our analysis to $>$ 2 \Ls\ galaxies because of the short
exposure time. We exclude cDs in large, dense cluster cooling cores
where there is no distinctive X-ray ISM component associated with the cD.
In cluster cooling cores like Perseus and Centaurus,
the ICM temperature decreases smoothly towards the
center and there are no distinctive $\sim$ 1 keV components with abrupt 
temperature gradient across the boundary. The cluster cooling cores,
sitting at the bottom of the cluster potential centered on the cD, have a
different gas origin from the X-ray coronae of early-type galaxies that
originates from the mass loss of evolved stars in galaxies. The X-ray ISM
of these cD galaxies may have already mixed with the surrounding dense ICM,
which is probably the end point of the evolution of cD galaxies' coronae.
In this work, 10 cDs in cooling core clusters (Centaurus, AWM7, Perseus,
Ophiuchus, A2199, A496, A2052, A3571, MKW3S and A4059) are excluded. We
further discuss the X-ray emission component centered on the cDs in $\S$7.

Besides early-type galaxies, we also selected late-type galaxies brighter
than $L_{\rm Ks, cut}$ or \Ls\ in the $B$ band (2$\times10^{10}$ L$_{\odot}$).
We define a galaxy as a late-type if its morphological type code from
HyperLeda is larger than 1.5. Thus, Sa galaxies are classified as early-types
in this work. We note that the HyperLeda classification for faint galaxies,
generally listed as early-types, may not be accurate.
$L_{\rm B}$ values of all galaxies are from HyperLeda, while the 2MASS $K_{\rm s}$
band luminosities are from NED. Galactic extinction (values from HyperLeda
and NED) has been corrected. Velocities of galaxies are from NED.
The resulting sample
includes 140 early-type galaxies brighter than $L_{\rm Ks, cut}$,
9 fainter early-type galaxies in Centaurus and Perseus clusters, and 19
late-type galaxies brighter than $L_{\rm Ks, cut}$ or $L_{\rm B*}$.

As shown in S05 and SJJ05, small coronae of early-type galaxies can be
associated with powerful radio sources without being destroyed. Thus, we also
examined all radio luminous ($L_{\rm 1.4 GHz}>10^{22.8}$ W Hz$^{-1}$) cluster
galaxies in the \chandra\ fields. The radio luminosities are generally from
The NRAO VLA
Sky Survey (NVSS), Faint Images of the Radio Sky at Twenty-cm (FIRST), Galaxy On
Line Database Milano Network (GOLDMine), and Sydney University Molonglo Sky 
Survey (SUMSS). All radio luminous galaxies are $> L_{\rm Ks, cut}$ galaxies,
except one in Ophiuchus.

We have also selected coronae associated with galaxies fainter than our
thresholds from the X-ray spatial and
spectral analysis (see $\S$3). The X-ray selection results in 7 additional
early-type galaxies that are fainter than $L_{\rm Ks, cut}$ and 3 more
late-type galaxies that are fainter than $L_{\rm Ks, cut}$ or $L_{\rm B*}$.
Therefore, our
final sample includes 157 early-type galaxies and 22 late-type galaxies.

\section{The data analysis}

\subsection{\chandra\ data reduction}

All observations were performed with the \chandra\ Advanced CCD Imaging
Spectrometer (ACIS). Standard data analysis was performed which includes
the corrections for the slow gain change
\footnote{http://cxc.harvard.edu/contrib/alexey/tgain/tgain.html}
and Charge Transfer Inefficiency (CTI). We examined data from all ACIS chips except
for chips I0, I1, S1 and S5 when ACIS-S was at the focus, since these chips lie
sufficiently far from the optical axis that the detection limits are high.
We investigated the light curve of
source-free regions to exclude time intervals with strong background flares.
As the local background is used for the analysis of small coronal sources,
any weak background flares do not affect our analysis, which we have verified
from the analysis of a sub-sample of embedded coronae.
The relevant information on the 64 \chandra\ pointings is listed in Table 1.
For the analysis of the cluster diffuse emission, we used the appropriate background
files\footnote{http://cxc.harvard.edu/contrib/maxim/bg/index.html}, and
excluded weak background flares.
We corrected for the ACIS low energy quantum efficiency (QE) degradation
due to the contamination on ACIS's Optical Blocking Filter
\footnote{http://cxc.harvard.edu/cal/Acis/Cal\_prods/qeDeg/index.html},
which increases with time and is positionally dependent.
The calibration files used correspond to \chandra\ Calibration Database
3.2.1 from the \chandra\ X-ray Center, released on December 2005.
In the spectral analysis, a lower energy cut of 0.5 keV is used to
minimize the effects of calibration uncertainties at low energy. The
upper energy cut is 7 keV unless sources are bright.
The solar photospheric abundance table by Anders \& Grevesse
(1989) is used in the spectral fits. Uncertainties quoted in this paper are 1 $\sigma$.

As we emphasized in $\S$2.2, most galaxies in our sample are selected in
the $K_{\rm s}$ or $B$ band. We also selected $< L_{\rm Ks, cut}$ galaxies (or
$<$ \Ls\ in the $B$ band for late-type galaxies) with
soft X-ray (0.5 - 2 keV) detections. This initial selection was not performed
by examining the spatial extent of the X-ray sources, as small coronae are
point-like at large (e.g., $> 7'$) off-axis angle. Moreover, the integrated
LMXB emission from nearby galaxies may also appear extended. The soft X-ray
detections come from the wavelet source detection tool in CIAO (WAVDETECT).
Any soft X-ray sources within 2$''$ of the nucleus of a known cluster galaxy
at a significance level of $\gsim$ 3 $\sigma$ are selected. Therefore, we have
a list of 160 $> L_{\rm Ks, cut}$ cluster galaxies (or $>$ \Ls\ in the $B$ band
for late-type galaxies) in the \chandra\ field, plus all $< L_{\rm Ks, cut}$
cluster galaxies (or $<$ \Ls\ in the $B$ band for late-type galaxies) with soft
X-ray detections (59 sources in total). Among these 160 sources, 101 have soft
X-ray detections. Embedded coronae will be identified from these 160 sources
(101 + 59) in the following spatial and spectral analysis.

\subsection{The identification of thermal coronae}

The most unambiguous evidence for the existence of 0.5 - 2 keV thermal
gas is the iron L-shell hump centered at 0.8 - 1 keV, which makes the 
spectra of coronae significantly different from those of AGN with a
featureless power law or integrated emission of LMXB with a 5 - 10 keV
bremsstrahlung. When X-ray sources are close to the optical axis,
coronae can also be identified by their spatial extents. Therefore,
either of these two criteria can be used to robustly identify thermal
coronae:

\begin{itemize}
\item[1)] A significant iron L-shell hump in the spectrum and a poor fit to
a power law
\item[2)] Extended soft X-ray sources
\end{itemize}

While the iron L-shell hump is very significant in the spectra of bright coronae
like those associated with two Coma central galaxies (V01) and two bright
galaxies in A1367 (S05), an objective measure is required to address
the significance of the iron L-shell hump presented in fainter coronae and the
departure of the fit from a power law. We identify a significant iron
L-shell hump in a spectrum in the following objective way. 
The X-ray spectrum was first fitted by a power law, which fits the hard spectra
of both AGN and LMXB. Then a MEKAL component
with an abundance fixed at 0.8 solar (the average abundance of coronae,
see $\S$4.7) is added to examine the improvement of the fit. Traditionally,
the F-test has been widely used to address the significance of an extra spectral
component. However, Protassov et al. (2002) has warned against the use of F-test
on testing a hypothesis that is on the boundary of the parameter space,
e.g., presence of an emission line in the spectrum. Although our test
is for the presence of a wide iron L-shell hump, we did not use the F-test but
instead performed Monte Carlo simulations to each spectrum to test
the significance of the iron L-shell hump (as recommended by Protassov et al. 2002).
Similar methods have already been widely used in X-ray spectral analysis,
mostly on the test of iron K emission or absorption lines (e.g., Markowitz
et al. 2006). The best-fit absorbed power law model was taken as the null
hypothesis model.
We used ``fakeit'' in XSPEC to simulate 3000 different fake spectra
according to the null hypothesis model. The same response and background
files, as well as the exposure time, have been used.
The simulated spectra have been binned
in the same way as that used to bin the real data. We fit each binned
fake spectrum with the null hypothesis model (an absorbed power law).
We then added a MEKAL component
with an abundance fixed at 0.8 solar. The improvement on $\chi^2$ has
been recorded for each fake spectrum. We calculate the frequency that
the improvement on $\chi^2$ is larger than what we observed for the
real data. We define a significant iron L-shell hump when the frequency is
smaller than 0.5\%, or the iron L-shell hump is significant at $>$ 99.5\%.
This analysis has been applied on all 160 X-ray candidates ($\S$3.1).
A significant iron L-shell hump was detected in 33 single-source spectra
so 33 coronae can be solely identified by the criterion 1 (Table 2 and 3).
As most X-ray sources are faint, spectra of 2 - 4 faint X-ray sources
(in the same cluster) were
stacked if they are close on the detector plane (usually within 4$'$).
A significant iron L-shell hump was also detected in 4 stacked spectra
of eight galaxies in A2634 and A2107 (two in each stacked group, Table 2).
The required significance cannot be achieved in the spectra of other
sources, although emission excess is present at the position
of the iron L-shell hump in some spectra.

We only examined the extent of X-ray sources of cluster galaxies if they
are within 7$'$ of the optical axis, where the \chandra\ angular resolution
is still good (50\% encircled energy radius of $\sim$ 3.5$''$ at 1 keV).
The 0.5 - 2 keV surface brightness profile of the source was compared
with the point spread function (PSF) model at 1 keV at the source location.
The PSF was first subtracted from the surface brightness profile with the
central brightness normalized to that of the source. We allowed the
normalization to scale up within the 1 $\sigma$ error of the central bin
of the source. Only sources with 4 $\sigma$ excess (after subtracting the
re-normalized PSF) are considered extended. We have also made sure that
these extended sources are all soft X-ray sources from their spectral fits,
as in principle, integrated emission from LMXB may also look extended.
Thirty-four sources were found to be extended, with 13 of them identified
as coronae only by their spatial extents (Table 2 and 3).
For these 34 resolved coronae, their sizes can be estimated.
The size is estimated from the fit to the 0.5 - 2 keV surface brightness
profile centered on the corona, with a model composed of a truncated 
$\beta$ model plus the local background (see detail in S05 for NGC~3842 and
NGC~3837, SJJ05 for NGC~1265). The local PSF at 1 keV is also included in the
modeling. However, the contribution from the LMXB emission is not included
in the fit as most galaxies do not have \hst\ imaging data. The uncertainty
of the size is generally 10\% for bright coronae and up to 20\% for
faint ones. For unresolved coronae or soft X-ray sources within 7$'$ of the
optical axis (32 sources in total), we can also derive the upper limit of
their sizes from their surface brightness profiles, with a similar way as did
for resolved coronae (but without convolving the local PSF).

While the coronae identified by these two criteria (generally with $>$ 80
net counts in the 0.5 - 3 keV band) are robust, the data reveal many fainter
soft X-ray sources in cluster galaxies. The identification of faint coronae
($\lsim$ 10$^{40}$ ergs s$^{-1}$) is indeed tricky, because most \chandra\
pointings of nearby hot clusters are not deep enough to study the coronal
emission from a single galaxy in detail. Nevertheless, the existence of faint
coronae in hot clusters is unambiguous, e.g., the faint, soft extended sources
associated with NGC~3841 and CGCG~97090 in A1367 (S05), several coronae in
the nearby Centaurus and Perseus clusters with significant iron L-shell hump
(NGC~4709, NGC~4706, NGC~1277 and CGCG~540-101) revealed from deep exposures
as analyzed in this work. These faint coronae are small ($\lsim$ 2 kpc in radius),
if resolved. They have been unambiguously identified only because: 1) they
happen to be very close to the optical axis (e.g., NGC~3841 and CGCG~97090);
or 2) the observations are deep enough to reveal the iron L-shell hump in nearby
cluster galaxies (e.g., in Centaurus and Perseus with $z <$ 0.02).
In our list of galaxies with soft X-ray detections, there are
indeed some sources resolved. But many
soft X-ray sources remained unresolved, especially the faint ones. 
Moreover, the iron L-shell hump in faint X-ray sources may not be significant.
For example, the stacked spectrum of NGC~3841 and CGCG~97090 in A1367,
two confirmed faint coronae, does not show a significant iron L-shell hump, and
the 90\% upper limit on the abundance is 0.13 solar. The iron abundance
of early-type galaxies' coronae may indeed be low in some cases. Detailed work with
\chandra\ and \xmm\ data on coronae of early-type galaxies in poor
environments shows that iron abundances can be lower than 0.1 solar
in some coronae (e.g., O'Sullivan \& Ponman 2004; Humphrey \& Buote
2006). The low iron abundance, combined with low statistics,
can make a seemingly featureless spectrum. Moreover, many cluster
galaxies have X-ray AGN (e.g., 3C~264 and IC~310 in this work). When
the X-ray AGN are bright, it is very difficult to have a significant detection
of a faint corona. The coronae of 3C~264 and IC~310 are only detected
because they are close to the optical axis so the extended coronal
emission can be resolved. Even when the X-ray AGN are not as bright
as those of 3C~264 and IC~310, it requires a deep exposure to distinguish
the thermal spectral component from the hard AGN component when the source is
unresolved. Therefore, the problem is how to identify faint coronae,
when they are spatially unresolved or mixed with a central X-ray AGN.

Unfortunately, with poor data statistics there is no a good way to
reliably identify faint coronae. However, we can use the different
spectral shapes of $<$ 1 keV thermal sources from hard AGN sources
or X-ray binaries to explore the population of faint coronae.
In this work, we have taken the following approach to study the faint
unresolved X-ray sources and define a sample of soft X-ray sources.
We identify a soft X-ray source if the following criterion is met:

\begin{itemize}
\item[3)] For faint unresolved X-ray sources, if the spectrum is fitted
with a power law, the 1 $\sigma$ lower bound of the photon index ($\Gamma$)
is larger than 2.4.
\end{itemize}

This criterion comes from the fact that the spectra of AGN are flatter
or harder with $\Gamma$ of 1.5 - 2.0 (e.g., Kim et al. 2004; Shemmer et al.
2005), compared with the much steeper spectra of coronae when the iron
L-shell hump is insignificant because of poor statistics or low abundance
(e.g., $\Gamma$=2.9 for the stacked spectrum of NGC~3841 and CGCG~97090 in
A1367). As the uncertainty on the photon index for
faint X-ray sources is large, the lower limit (2.4) is set on the 1 $\sigma$
lower bound of the photon index, instead of the best fit of the photon index.
Therefore, the best fits of the photon index are larger than 3.0 in most cases
that sources are identified as soft X-ray sources. The photon index of 2.4
was chosen conservatively from XSPEC simulations.
We have run a series of XSPEC simulations (with background added) to understand 
the power law fits to weak X-ray sources (50 net counts in the 0.5 - 7 keV
band). If a photon index of 2.2 is assumed for AGN (steeper than the usual
value), 1000 XSPEC simulations with different responses and absorption
show that in at most 12\% of the simulations, a photon index with 1 $\sigma$ lower
bound $>$ 2.4 is derived from the spectral fits. On the contrary,
for genuine coronae with a high temperature ($>$ 1 keV) and a low
abundance ($<$ 0.1 solar), the chance that criterion 3 is not
met is substantial even when no hard component is included. If a temperature
of 1 keV and an abundance of 0.05 solar
are assumed, 1000 XSPEC simulations with different responses and absorption
show that in up to 45\% of simulations criterion 3 is not met. When a hard
component is added into the model, the chance that criterion 3 is not met
becomes much higher.
Therefore, the criterion 3 is a conservative way to select faint coronae. 
In this work, we have examined the spectra of all 114 sources that do not
meet the criteria 1 and 2.
Fifteen single spectra and six stacked spectra meet the criterion
3 and are selected (31 galaxies in total).
Representative spectra of a corona identified by the criterion 1
and a soft X-ray source identified by the criterion 3 are shown in Fig. 1.
In Table 2 and 3, we give the codes (1 or 2: robust detections of
coronae; 3: soft X-ray sources) on how these sources are identified. In $\S$3.4,
we will discuss more on the nature of the faint soft X-ray sources. We are
confident that most of such sources are genuine coronae. Nevertheless, their
nature does not affect our conclusion that coronae of massive cluster galaxies
($>$ 2 \Ls\ galaxies) are ubiquitous in hot clusters.

\subsection{The spectral analysis}

This section details the spectral analysis to identify coronae and to derive
the spectral properties of the X-ray sources.
X-ray sources with $\gsim$ 60 net counts
and a $\gsim 5 \sigma$ detection in the 0.5 - 2 keV band are analyzed
individually. The source spectrum was extracted within the region
determined from the spatial analysis (a 0.5 - 2 keV surface brightness
profile centered on the source). The region source for the spectral extraction
is then the same as the spatial extent of the source (without PSF
correction). The background spectrum was extracted locally (an annulus between
the source radius and twice the source radius).
The background spectra are generally
normalized by the ratio of the geometric area of the source and background
regions, unless the local background profile is not constant. In such
cases (generally only for cD galaxies), we extrapolated the background
profile into the source region and this extra factor was included in
the normalization of the local background. Because of small sizes of
embedded coronae, the local background is generally still within the
optical radius of the galaxy. Each spectrum has been examined for a
significant iron L-shell hump, based on the quantitative test stated
in $\S$3.2.

However, there are many fainter X-ray sources associated with cluster
galaxies. To fully explore the current data, we have stacked spectra
of fainter sources in the same cluster (2 - 4 sources stacked in this
work), if they are close to each other in the same FI or S3 chips and
have similar net counts.
The choice of sources to combine depends on the source location and
we did not specifically combine sources with similar X-ray color, although
almost all faint sources are only significantly detected in the soft
band. In some clusters, there are faint X-ray sources that cannot find
others to stack so they can only be examined individually. Stacked galaxies
are listed adjacently in Table 2 and only temperatures derived from the
stacked spectra are showed. Stacked spectra with
a significant iron L-shell hump are given a code of 1 (four such cases in
Table 2), while those with a steep spectrum to meet the criterion 3 are
given a code of 3. The X-ray luminosities of stacked galaxies are derived
by dividing the total X-ray luminosity according to the relative 0.5 - 2
keV flux of the combined sources.
The stacking analysis and the classified soft X-ray sources by the
criterion 3 allow us to explore faint unresolved coronae as a population,
although the results may not always be accurate individually. These faint
soft X-ray sources have been discussed more in the next section, where
we argue that most of them should be genuine thermal coronae.

Because of the LMXB emission and the possible nuclear hard source,
we always include a hard component (a power law)
in the spectral model of coronae or soft X-ray sources. For bright coronae
($> 10 \sigma$ detections
in the 0.5 - 2 keV band), we left the power law index free. For fainter
coronae, we fixed the power law index at 1.7,
which is a good approximation for both LMXB and AGN. 
When a significant iron L-shell hump is present, the determination of the
coronal gas temperature is robust and changes little with the variation of
abundance and the spectral shape of the hard component.
However, the X-ray bolometric luminosities (unabsorbed 0.01 - 100 keV
luminosities in this work) can vary by a factor of up to 2 with the change of
the abundance (e.g., 0.03 - 2 solar).
Lowering the abundance increases the bolometric luminosity
while the rest-frame 0.5 - 2 keV luminosity changes little.
In this work, as the abundance of the coronal gas is constrained through
the joint fit in $\S$4.7, we fix the abundance of coronae at 0.8 solar
when the X-ray luminosities are derived, unless the coronal abundance is
smaller than 0.8 solar at the 95\% confidence level (only for NGC~3308, NGC~3841
and CGCG~97090 in this work). The 1 $\sigma$ confidence level of the rest-frame
0.5 - 2 keV luminosities of coronae are also estimated, by examining the
allowed luminosity values when the coronal properties are changed within
their allowed 1 $\sigma$ range. For faint coronae, the
uncertainty is close to the statistical uncertainty. For bright coronae,
the uncertainty from the contribution of the hard component becomes dominant.
The uncertainty on the bolometric luminosities, at least as high as that
for the rest-frame 0.5 - 2 keV luminosities, is not given because of the
unknown spectrum of coronae below 0.5 keV. We emphasize that the bolometric
luminosities are derived by assuming an abundance of 0.8 solar.
Nevertheless, the uncertainties of bolometric luminosities are mainly on
the upper bound. The bolometric luminosity at most decreases by 8\% when
abundance is increased from 0.8 solar to 2 solar. The increase brought by
the decrease of abundance from 0.8 solar to 0.05 solar is however 15\% for
1.5 keV gas and 100\% for 0.4 keV gas.
The 0.5 - 2 keV luminosity of the hard component is generally 10\% - 30\% of
the total 0.5 - 2 keV luminosity, except for sources with significant X-ray AGN.
We cannot determine the absolute luminosity contributed by LMXB in the
aperture used to analyze coronal spectra, since some LMXB contribution is
contained in the local background, which is generally still within the galaxy.
Nevertheless, we did fit any residual LMXB component and thus our parameter
estimates for the coronal emission are uncontaminated by LMXB emission.

Besides the parameter estimates of identified thermal coronae, we also
want to determine the upper limits of coronal emission for galaxies
without coronal detection. The upper limits need to be derived in a
homogeneous way for both galaxies without X-ray detection and galaxies
detected in X-rays but without coronal detection.
For NIR/optical or radio selected galaxies without X-ray detections, a
3-$\sigma$ upper limit ($\sigma$ from Poisson distribution) was derived
from the 0.5 - 2 keV emission at the
position of the galaxy. We fix the aperture at 3 kpc radius.
To address the smearing of the PSF (especially at large off-axis angle),
we added the 90\% encircled energy average radius of the local PSF at 1 keV
to the 3 kpc radius. Although there are some coronae with radii larger
than 3 kpc, the enclosed X-ray light within 3 kpc radius is always
$>$ 90\%, except for NGC~7720 and IC~1633, which are all cDs with luminous
coronae. 
For the purpose of upper limits, we assume a coronal spectrum with a temperature
of 0.7 keV and an abundance of 0.8 solar. These temperature and abundance
values are typical for embedded coronae of $L_{\rm Ks} < 10^{11.8}$
$L_{\rm K\odot}$ early-type galaxies (see $\S$4), while all galaxies with
upper limits in our sample are indeed fainter than this threshold.
These temperature and abundance values are also fair assumptions for coronae
of late-type galaxies (see $\S$5). The response files at the position of
the source are generated to convert count rates to fluxes. 

Compared with the galaxies without X-ray detection,
It is trickier to determine the upper limits for galaxies with X-ray
sources that do not meet any of criteria in $\S$3.2 (e.g.,unresolved hard
X-ray sources without a significant iron L-shell hump). The concern
is how to determine the upper limits of the soft component in the spectra
and how to define the aperture. 
There are some soft X-ray sources that do not meet criterion 3. However,
their spectra are also fitted well by a thermal model plus a power law
with the limited statistics.
There are also unresolved hard X-ray sources close to the optical axis.
As the aperture for the spectral extraction may be smaller than 3 kpc
(especially for lower $z$ clusters, e.g., Perseus), the possible
coronal emission beyond the region of the central point-like source
(but still within a 3 kpc radius) needs to be considered.
In this work, we have taken the following approach to estimate upper
limits in a consistent way as we did for galaxies without X-ray detections.
We first added a 0.7 keV thermal component with an abundance of 0.8 solar
to the source spectra. The upper limit of the coronal emission in the
source spectrum is set at the best-fit of the soft component plus 1 $\sigma$
error. This approach allows the possible coronal flux in the source
spectrum to enter the upper limit. If the radius of the spectral
extracting region is smaller than 3 kpc (after PSF correction),
a 3-$\sigma$ upper limit was further derived between the the 3 kpc radius
and the spectral extracting region as did for galaxies without X-ray detection.
A coronal spectrum with a temperature of 0.7 keV and an abundance of 0.8 solar
is still assumed. The final upper limit is then the sum of the limits
in these two regions.

We treat late-type galaxies in a different way as thermal gas of the
late-type galaxies can be more extended and fill large portions of the
galaxy ($\S$5 and examples in the Appendix). We used an elliptical aperture with a semi-major
axis of 10 kpc. The size of the semi-minor axis is determined from the
ellipticity of the galaxy and the assumed semi-major axis. We still
added the 90\% encircled energy average radius of the local PSF at 1 keV
to the aperture. All other procedures and assumption to estimate upper
limits are the same as those for early-type galaxies.

\subsection{The faint soft X-ray sources}

Our analysis has generated a list of robust detections of coronae
(by criteria 1 or 2) and faint soft X-ray sources (single or stacked).
For soft X-ray sources identified by the criterion 3, the simulations
($\S$3.2) indicate that the contamination from the pure power
law AGN should be small ($<$ 12\%). Therefore, the most natural
explanation of these soft X-ray sources are faint coronae.
In fact, many sources show emission excess at the energy band of the iron
L-shell hump, but not significantly enough to meet the 99.5\% significance level
we set. If the four sources in A3376, and the eight sources in MKW3S
identified by the criterion 3 (Table 2) are stacked respectively,
a significant iron L-shell hump is measured in both stacked spectra.
A significant iron L-shell hump is also measured if the four faint
soft X-ray sources identified by the criterion 3 in A496, A576, A3571
and A4059 are stacked. Therefore, a significant portion of these 16
soft X-ray sources should be genuine coronae.

One contaminating source to faint coronae identified by the criterion 3
could be the faint AGN with soft X-ray excess, which dominates below $\sim$
0.5 keV (e.g., Arnaud et al. 1985; Walter \& Fink 1993). The soft excess
generally behaves as a steep power-law (photon index up to 3.4,
Walter \& Fink 1993) at the soft band (e.g., 0.1 - 2.4 keV \rosat\ band).
While the soft X-ray excess is especially strong in narrow-line Seyfert 1
galaxies, the galaxies we are mainly interested are early-types. In fact,
none of the early-type galaxies identified by the criterion 3 are known optical
AGN in the literature. Moreover, the usual hard component with a photon
index of 1.4 - 2.2 is always significant in the 2 - 10 keV spectra of AGN
with soft excess.
Page et al. (2004) analyzed \xmm\ spectra of seven QSOs with soft excesses.
If the soft excess is fitted with a MEKAL model, the temperature is between 0.1 -
0.5 keV and the abundance is zero. The photon index of the hard component
is 1.7 - 2.2. Piconcelli et al. (2005) investigated
\xmm\ data of 40 QSOs. They presented fits to the soft excess with bremsstrahlung,
which is close to MEKAL model with zero abundance. The average temperature of
the soft excess is about 0.38 keV, while the average photon index of the hard
component is $\sim$ 1.8. We have fitted the spectra of soft X-ray sources we 
identified with the criterion 3 with the same model, power law + bremsstrahlung.
Although the temperature range, 0.3 - 0.6 keV, is not too far from the typical
values for the soft excess in AGN, the soft-to-hard flux ratio at the 0.5 - 2 keV
band is much higher, 2.1 - 27 with a median of 4-5, compared with 0.06 - 3.0 with
a median of 0.65 in Piconcelli et al. (2005). In other words, the soft component
is dominant in sources identified by criterion 3, different from typical
soft excess seen in AGN. Moreover, the fits with ``power law + bremsstrahlung''
are always worse than the fits with ``power law + MEKAL'' (with abundance fixed
at 0.8 solar, the same degree of freedom), which implies the emission excess
at the energy band of the iron L-shell hump. Based on all these facts,
we consider that the contamination from the AGN with soft excess is small.

Therefore, we conclude that most of soft X-ray sources identified by the
criterion 3 are genuine thermal coronae. They just cannot be unambiguously
identified individually as the available \chandra\ exposures are usually
not deep enough. Nevertheless, when the properties of embedded coronae are
investigated in $\S$4, we always labeled the soft X-ray sources identified
by criterion 3 differently from robust detections and always
discuss the effects of adding them into the coronal population.

\section{Coronae of early-type galaxies in hot clusters}

For the 3.77 deg$^{2}$ sky coverage of these 25 clusters in the sample,
46 coronae (identified by the criteria 1 or 2) and 30 soft X-ray sources
(identified by the criterion 3) are detected from a sample of 157
early-type galaxies. Their properties are summarized in Table 2
and the color-luminosity relation of all galaxies in our sample
is shown in Fig. 2. The properties of the population
of embedded coronae can be investigated.

\subsection{The L$_{\rm optical}$ - L$_{\rm X}$ relation}

The $L_{\rm B} - L_{\rm X}$ and $L_{\rm Ks} - L_{\rm X}$ relations for
embedded coronae were investigated. $K_{\rm s}$ band luminosity is homogeneously
measured by 2MASS and is a more accurate measure of the stellar mass for
early-type galaxies, but $L_{\rm B}$ has been widely used in previous
studies. We plot both relations in Fig. 3, where $L_{\rm 0.5 - 2 keV}$
is used as it is more robustly measured than $L_{\rm bol}$. The relations are
complete for $L_{\rm Ks} >$ 0.74 \Ls\ galaxies, and for $L_{\rm B} >$
0.88 \Ls\ galaxies (assuming $B - K_{\rm s}$ = 4.0) in the \chandra\ field.
Dispersion in both plots
is large (greater than one order of magnitude). The dispersion should
arise from the combined effects of the internal dispersion and the variation
caused by environmental effects. Cluster galaxies with low density coronae
(in the wind phase) may have already been fully stripped and
are currently ``naked'' with little galactic gas. Cluster galaxies with dense
galactic cooling cores can maintain their gas cores even in dense environments,
and therefore are still X-ray luminous. Therefore, the $L_{\rm optical}$ -
$L_{\rm X}$ relation for embedded coronae in hot clusters is really an
envelope instead of a tight relation. 

We perform fits with a function of Log $L_{\rm 0.5-2 keV}$ = $\alpha + \beta$
Log ($L_{\rm op} / 10^{11} L_{\odot}$). For robust detections with criteria
1 or 2, we measured: $\alpha = 40.91, \beta = 1.30\pm0.03$ at $L_{\rm B} >$
10$^{10.18}$ L$_{\rm B\odot}$ (or 0.88 \Ls\ at the $B$ band); $\alpha = 39.65,
\beta = 1.55\pm0.03$ at $L_{\rm Ks} > L_{\rm Ks, cut}$.
If soft X-ray sources identified by criterion 3 are added, we measured:
$\alpha = 40.87, \beta = 1.27\pm0.03$ in the $B$ band and $\alpha = 39.78,
\beta = 1.50\pm0.03$ in the $K_{s}$ band.
We have also applied the survival analysis (Feigelson \& Nelson
1985) to compute linear regression for both detections and upper limits.
The survival analysis requires that the censoring distribution about the
fitted line is random. We consider it is a fair assumption for most of
our data, as upper limits are randomized by many factors (exposure,
cluster background, Galactic absorption and the offset of the galaxy to
the optical axis) and not tightly related to the real luminosities of the coronae.
Moreover, almost all \chandra\ observations we analyzed were intended to study
the ICM, rather than cluster galaxies. The Buckley-James algorithm in
STSDAS, with the Kaplan-Meier estimator, has been used. This algorithm is
considered the most reliable one in the survival analysis package. With
the addition of upper limits, we measured:
$\alpha = 40.50, \beta = 1.45\pm0.15$ in the $B$ band and $\alpha = 39.40,
\beta = 1.63\pm0.13$ in the $K_{s}$ band.

We then compared the $L_{\rm B} - L_{\rm X}$ and $L_{\rm Ks} - L_{\rm X}$ 
relations of our sample with those of galaxies in poor environments by 
O'Sullivan et al. (2001) and Ellis \& O'Sullivan (2006) from the \rosat\
data, as shown in Fig. 4. Although Fig. 4 may imply that the coronae in hot
clusters are systematically X-ray fainter than coronae in poorer environments,
this comparison needs to be examined more carefully and the \rosat\ results
need to be used with caution. The \rosat\ X-ray luminosities of galaxies
actually include all X-ray emission components in galaxies.
There are ten galaxies in both our sample and O'Sullivan
et al.'s sample: ESO~137-006, ESO~137-008, ESO~137-010, IC~1633, NGC~3311,
NGC~3842, NGC~3862, NGC~4709, NGC~4889 and IC~5358. The \rosat\ luminosities
derived by O'Sullivan et al. (2001) are 3.5 - 79 times the \chandra\ luminosities
of thermal coronae
for the first nine galaxies, and $>$ 3300 times higher than the \chandra\ value
for IC~5358 (only a \chandra\ upper limit for the thermal coronal component).
While the \chandra\ results are robust and the hard X-ray components (AGN and
X-ray binaries) have been subtracted, the \rosat\ data
suffer from the contamination of AGN, X-ray binaries and ICM emission, which
are all included in the luminosities of galaxies.
Besides these contaminants, the O'Sullivan et al. (2001) sample also includes
many cD galaxies at the centers of the cooling core clusters (or groups)
where there is no distinctive cool galactic component associated with the cD.
\rosat\ is unable to resolve these cores and derives temperature distribution,
while \chandra\ can easily do.
The X-ray gas around these cD galaxies is mostly of ICM origin, and should
be discussed separately from coronae with the stellar origin.
We have also examined several overlapping galaxies in detail.
For IC~5358 (cD of A4038), the \rosat\ luminosity is the X-ray luminosity
of the whole cluster gas core (3.2 keV) around IC~5358, as IC~5358's AGN
(revealed by \chandra) is still 500 times fainter than the \rosat\ luminosity.
For NGC~3862, the \rosat\ luminosity (55 times the \chandra\ value), is the 
contribution of the bright nucleus. For NGC~4889 (\rosat\ value 79 times
larger) and NGC~3311 (\rosat\ value 32 times larger), the luminosities are
dominated by the contribution from the surrounding ICM as the PSPC apertures
for the sources are large. All these factors contribute to the large difference
we observe in other six galaxies. We have also estimated the luminosity of a large,
luminous corona with the following parameters, n$_{\rm e, center}$ = 0.4 cm$^{-3}$,
$\beta$=0.5, r$_{0}$=0.8 kpc, r$_{\rm cut}$=20 kpc, $kT$=0.6 keV and Z=1.0 solar.
All these parameters make $L_{\rm X}$ of the assumed corona close to the
maximal value for a corona in the field or poor environments.
The resulting X-ray bolometric luminosity is 2.7$\times10^{42}$ ergs
s$^{-1}$, and the corresponding X-ray gas mass is 6.1$\times10^{9}$
M$_{\odot}$. We have examined 19 galaxies more luminous than this
threshold in the O'Sullivan et al. sample. IC310 and NGC3516 are AGN,
while the other 17 are all cD galaxies in cluster cooling cores.
Therefore, we should be aware of the limitation
of the \rosat\ data and the $L_{\rm optical} - L_{\rm X}$ relation of
coronae in poor environments needs to be re-examined by \chandra. 

The upper limit on the soft coronal emission gives constraints on the
coronal gas density and the size of the corona. The coronal gas density of
a cluster galaxy needs to be high enough to survive ram-pressure stripping
with a velocity of $\approx \sqrt{3} \sigma_{\rm clu}$ (1000 - 2300 km/s
for this sample). The X-ray luminosity of a corona is:

\begin{eqnarray}
L_X&=& \int_0^{r_{\rm cut}} n_{\rm e}^2(r) \varepsilon (T,Z) dV \\
&=& 3.1\times10^{39} (\frac{x_{0}}{2})^{-3} (\frac{r_{\rm cut}}{\rm 3 kpc})^{3} (\frac{n_{\rm e, cut}}{10^{-2} {\rm cm}^{-3}})^2 f(\beta,x_{0}) {\rm ergs/s} \nonumber
\end{eqnarray}

\begin{eqnarray}
( f(\beta,x_{0}) = \int_0^{x_{0}} [\frac{(1+x^2)}{(1+x_{0}^2)}]^{-3\beta} x^2 dx )  \nonumber
\end{eqnarray}

where $\varepsilon$(T,Z) is the emissivity of the coronal gas, which depends on
temperature and abundance of the gas. We here assume a typical temperature
and abundance value for coronae
(0.7 keV and 0.8 solar). The emissivity can vary by $\gsim$ 50\% with
different temperatures and abundances.
r$_{\rm cut}$ is the size of the corona (pressure-confined by the ICM) and
n$_{\rm e}$ is the electron density of the coronal gas, for which we assume:
n$_{\rm e}$(r) = n$_{\rm e, 0}$ [1+(r/r$_{0}$)$^{2}$]$^{-3\beta/2}$. 
x$_{0}$=r$_{\rm cut}$/r$_{0}$, while n$_{\rm e, cut}$ is the electron density
of the coronal gas at r$_{\rm cut}$. n$_{\rm e, cut}$ is determined by the
ICM pressure (thermal pressure + ram pressure) applied on the corona. We have
found that the ratio of the average ram pressure to the thermal pressure
is 1.5 - 4 in the clusters of our sample. Therefore, roughly we have:

\begin{eqnarray}
n_{\rm e, cut} &\approx& \frac{n_{\rm e, ICM} v_{\rm gal}^2 \mu m_{\rm p}}{kT_{\rm ISM}} \\
&=& 0.018 (\frac{n_{\rm e, ICM}}{10^{-3} {\rm cm^{-3}}}) (\frac{kT_{\rm ISM}}{\rm 0.7 keV})^{-1} (\frac{v_{\rm gal}}{1400 {\rm km/s}})^2 {\rm cm^{-3}} \nonumber
\end{eqnarray}

In the simplest scenario, when a corona is moving into a denser region, r$_{0}$
is constant while x$_{0}$ decreases with
the increasing ram pressure. The X-ray luminosity of an embedded corona depends
mainly on its density distribution and size. We have computed the X-ray luminosities
for some sets of ($\beta$, x$_{0}$), assuming r$_{0}$ = 0.5 - 1 kpc (S05, SJJ05)
and n$_{\rm e, cut}$=0.01 cm$^{-3}$.
If $\beta$ = 0.8 (similar to what we found for some luminous coronae, S05, SJJ05),
r$_{\rm cut}$ has to be small to make the luminosity small, e.g.,
8.7$\times10^{39}$ ergs s$^{-1}$ for x$_{0}$=3.0 and r$_{0}$ = 0.5 kpc,
1.1$\times10^{40}$ ergs s$^{-1}$ for x$_{0}$=2.0 and r$_{0}$ = 1 kpc.
If $\beta$ = 1/3 (e.g., for some X-ray faint elliptical galaxies, NGC~4697, Sarazin
et al. 2001), r$_{\rm cut}$ can be larger, e.g.,
1.9$\times10^{40}$ ergs s$^{-1}$ for x$_{0}$=6 and r$_{0}$ = 0.5 kpc,
1.6$\times10^{40}$ ergs s$^{-1}$ for x$_{0}$=3 and r$_{0}$ = 1 kpc.
The upper limits we derived are for a fixed aperture of 3 kpc radius.
However, if galaxies only with upper limits really have coronae with a
dense core (n$_{\rm e, 0} \sim 0.2$ cm$^{-3}$), the size of the corona
must be small and $\beta$ must be large (e.g., 0.8).
By repeating the estimates on upper limits and the predictions on
X-ray luminosities, we conclude that the derived average
upper limits on the coronal emission ($\sim 10^{40}$ ergs s$^{-1}$)
constrain the possible faint coronae: either low density ones
(n$_{\rm e, 0} \lsim 0.08$ cm$^{-3}$) with a radius of $<$ 3 kpc, or
high density ones (n$_{\rm e, 0} \sim 0.2$ cm$^{-3}$) with a radius
of $<$ 1.6 kpc.

For BCGs in the bottom of the cluster potential, the thermal pressure
of the coronal gas should mainly be
balanced by the thermal pressure of the ICM, as the residual motion of these
central galaxies should generally be small. Therefore, we have:

\begin{eqnarray}
n_{\rm e, cut} = 0.03 (\frac{n_{\rm e, ICM}}{5\times10^{-3} {\rm cm^{-3}}}) (\frac{kT_{\rm ICM}/kT_{\rm ISM}}{6}) {\rm cm^{-3}}
\end{eqnarray}

We can apply this further constraint to the cD galaxy of A4038, one of the
most massive galaxies without a coronal detection in our sample. The ICM
density at the center of A4038 is $\sim$ 0.022 cm$^{-3}$, while the ICM
temperature is $\sim$ 3.2 keV from our analysis. Even if we assume a
temperature of 1.5 keV for the coronal
gas of A4038's cD, $n_{\rm e, cut}$ should not be smaller than 0.045 cm$^{-3}$.
The upper limit we set for the coronal emission of A4038's cD is for a fixed
aperture of 3 kpc radius. However, for a corona with this high value of
$n_{\rm e, cut}$ at r$_{\rm cut}$=3 kpc, we estimate that the expected
coronal luminosity is over 5 times
higher than the derived upper limit, even by assuming small $\beta$ and x$_{0}$
(1/3 and 3 respectively). Therefore, the corona of A4038's cD, if exist,
must be smaller. By repeating the analysis of the upper limit and the
estimate of the expected luminosity for a smaller coronal size, we conclude
that the corona of A4038's cD must be smaller than 1.5 kpc in radius and the
upper limit on its coronal emission can be reduced by a factor of 2.8.
Therefore, with the extra constraint from the pressure equilibrium, tighter
upper limits on the coronal emission can in principle be set. However, this
approach does not work for most galaxies as the external pressure is generally
uncertain.

\subsection{The L$_{\rm optical} \sigma_{\ast}^{2}$ - L$_{\rm X}$ relation}

We also examined the $L_{\rm optical} \sigma_{\ast}^{2} - L_{\rm X}$
relation, for galaxies with known $\sigma_{\ast}$. The
quantity $L_{\rm optical} \sigma_{\ast}^{2}$ is proportional to the
total energy released by stars in the galaxy, and also proportional
to the gravitational heating energy (Canizares et al. 1987). Based
on the \xmm\ observations of the Coma cluster, Finoguenov \& Miniati
(2004) used detections and claimed a positive effect of
environment on the X-ray luminosities of coronae, when compared
with the \rosat\ results in poorer environments (Matsushita 2001).
We have examined the $L_{\rm B} \sigma_{\ast}^{2} - L_{\rm 0.5 - 2 keV}$
relation of our sample, with upper limits included (Fig. 5).
The data show a large dispersion. We fit all data points at
$L_{\rm B} \sigma_{\ast}^{2} > 10^{14.85} L_{B\odot}
(km/s)^{2}$ with a function of Log $L_{\rm X}$ = $\alpha + \beta$
Log ($L_{\rm B} \sigma_{\ast}^{2}$ / 10$^{16} L_{B\odot} (km/s)^{2}$).
This threshold comes from an \Ls\ galaxy in the B
band with a velocity dispersion of 190 km/s, where the optical
data are quite complete and the X-ray upper limits are on average
lower than detections. As we are testing the environmental effects
on the X-ray luminosities of embedded coronae, both detections and
upper limits need to be included in the analysis. We used the
Buckley-James algorithm to fit this censored data and obtained:
$\alpha = 40.68, \beta = 1.03\pm$0.12. As shown in Fig. 5, our
line of the fit is below both the lines of Finoguenov \& Miniati (2004)
and Matsushita (2001). Furthermore, we have examined the
$L_{\rm B} \sigma_{\ast}^{2} - L_{\rm X}$
relations in two sub-samples, one with galaxies in
15 low $\sigma_{\rm clu}$ clusters ($<$ 880 km/s in this work, see
Table 1), another with galaxies in 10 high $\sigma_{\rm clu}$ clusters
($>$ 880 km/s). There is no $\sigma_{\rm clu}$ value available for
the 3C~129.1 cluster. We attribute it to the high $\sigma_{\rm clu}$
clusters based on the $\sigma_{\rm clu}$ - $kT_{\rm ICM}$ relation.
Therefore, all $>$ 5 keV clusters are in the high $\sigma_{\rm clu}$
cluster sample. As shown in Fig. 5 (blue: galaxies in low
$\sigma_{\rm clu}$ clusters; red: galaxies in low $\sigma_{\rm clu}$
clusters), there is no indication that coronae in high $\sigma_{\rm clu}$
clusters are systematically more luminous than coronae in low
$\sigma_{\rm clu}$ clusters. The fits with the Buckley-James algorithm
show that $\alpha = 40.69, \beta = 1.27\pm$0.16 for high $\sigma_{\rm clu}$
clusters, and $\alpha = 40.62, \beta = 0.79\pm$0.23 for low
$\sigma_{\rm clu}$ clusters. At $L_{\rm B} \sigma_{\ast}^{2} < 10^{15.7}$
$L_{\rm B\odot}$ (km/s)$^{2}$, coronae in low $\sigma_{\rm clu}$
clusters are on average $\sim$ 2.5 times more luminous than coronae in high
$\sigma_{\rm clu}$ clusters, although dispersion is large.
Thus, we conclude that there is no
evidence for a positive effect of environment on the X-ray luminosities
of coronae. Instead, the data favor a negative effect. However, as we
cautioned in the last section, a good sample of coronae in poor
environments is required to better compare with our results.

Fig. 5 clearly indicates that most coronae in massive galaxies with
deep potentials (e.g., $L_{\rm Ks} \sigma_{\ast}^{2} > 10^{16.3}$
L$_{\rm K\odot}$ (km/s)$^{2}$) will survive, presumably because
galactic cooling cores can form and are sustained in these galaxies.
For detected coronae, $L_{\rm X, bol} \propto$  ($L_{\rm Ks}$
$\sigma_{\ast}^{2}$)$^{1.05}$ and $L_{\rm X, bol} \propto$  ($L_{\rm B}$
$\sigma_{\ast}^{2}$)$^{0.94}$. The closeness of the logarithmic slope to
one may imply that the energy released from the stellar mass loss balances
the X-ray cooling (or radiation). However, the kinetic energy from
the stellar mass loss, 3/2 $\dot{\rm M}_{*}\sigma^{2}_{*}$ (the factor of 3
comes from the fact that $\sigma_{*}$ is just the radial stellar velocity
dispersion), inside the small coronae (enclosing 20\% - 40\% of the total
stellar mass), is on average 3.5 times smaller than the X-ray bolometric
luminosities in the $L_{\rm Ks} \sigma_{\ast}^{2} - L_{\rm X, bol}$ plot
(Fig. 5). The enclosed stellar light fraction of 20\% - 40\% is estimated from a
sample of $\sim$ 20 galaxies with \hst\ data or good ground imaging data
(not from DSS). The enclosed optical light fractions of previously known coronae
are also in this range (V01; S05; SJJ05).
We used the stellar mass loss rate by Faber \& Gallagher (1976),
$\dot{M}_*$ = 0.15 M$_{\odot}$ yr $^{-1}10^{10}$ L$_{\rm B\odot}^{-1}$, and
assumed $B - K_{\rm s}$ = 4.0. The difference is a bit larger ($\sim$ 4 times)
in the $L_{\rm B} \sigma_{\ast}^{2} - L_{\rm X, bol}$ plot.
As we emphasized in $\S$3.3, the lower bound uncertainty of $L_{\rm X, bol}$ is
small ($\sim$ 10\%) so the observed difference is very significant.
The NOAO fundamental plane survey shows that
the age of an early-type galaxy is correlated with the stellar velocity
dispersion (Nelan et al. 2005). Therefore, less massive galaxies may have
a larger stellar mass loss rate, as $\dot{M}_*$ = $\dot{M}_{\rm *, FG}$
(t/13 Gyr)$^{-1.3}$ (Ciotti et al. 1991). Almost all galaxies with coronal
detections in Fig. 5 have $\sigma_{\ast} >$ 200 km/s, which corresponds
to an age of $>$ 10 Gyr (Nelan et al. 2005). Thus, at the low end of
the relations, the expected kinetic energy injection rate from stellar
mass loss is still on average 2.3 (in the $K_{\rm s}$ band) or 3.5
(in the $B$ band) times smaller than the X-ray bolometric luminosities
of coronae, while the difference is unchanged in
the high end. We conclude that for almost all luminous coronae ($> 10^{40}$
ergs s$^{-1}$) the energy release by the stellar mass loss 
is too small to balance cooling of the coronal gas.
This conclusion is further supported in $\S$4.5 where we find that the
coronal temperatures are generally higher than the stellar kinetic temperatures.

\subsection{The detection rate of X-ray coronae}
 
We would like to understand which factors govern the survival and destruction of
coronae, by examining the properties of their host galaxies and clusters.
However, we should be aware that the detection limit of embedded coronae
varies even in the same cluster. As shown in Fig. 3, below $L_{\rm Ks}$ of
$\sim 10^{11.2} L_{\rm K\odot}$, the upper limits become comparable
to the detections. Indeed, identifications of coronae and soft X-ray sources
are found to be distance-dependent. For 12 clusters at $z \leq 0.03$, 24 coronae
(by the criteria 1 or 2) and 8 soft X-ray sources (by the criterion 3) are
identified from 66 galaxies that are more luminous than $L_{\rm Ks, cut}$.
For 12 clusters at $z > 0.03$ (excluding A3558), 18 coronae and 17 soft X-ray
sources are identified from 71 $> L_{\rm Ks, cut}$ galaxies. While the rate
of coronae + soft X-ray sources in each group changes little, there is a trend
that more soft X-ray sources are identified in more distant clusters in the sample.
Besides distance, exposure, ICM background, galaxy's angular distance from the
optical axis of the observations, and Galactic absorption all affect the achieved
sensitivity. Nevertheless, our discussion can only be based on the current data
sample. This ``bias'' factor for X-ray faint coronae has to be kept
in mind when results are interpreted.

The detection rate of X-ray coronae varies significantly from cluster to
cluster. There are no coronae detected in A2147, A2199 and A4038 (17
galaxies brighter than 0.74 \Ls\ in total), while the detection rate is over
50\% for Perseus, A1367 and A2634 ($\gsim$ 7 galaxies per cluster). It
is unclear what causes the variation from cluster to cluster and why coronae in
A2147, A2199 and A4038 are faint or no longer exist.
We also do not find any significant changes of detection rate with
the cluster temperature or velocity dispersion, no matter whether the soft
X-ray sources identified by the criterion 3 are included.

We have also examined the relation of the detection rate with $L_{\rm Ks}$
or $L_{\rm B}$ of the host galaxy. We took a conservative measure
to count only half of soft X-ray sources identified by criterion 3 as genuine
coronae. For $>$ 2 \Ls\ galaxies ($L_{\rm Ks} > 10^{11.38} L_{\rm K\odot}$;
$L_{\rm B} > 10^{10.60} L_{\rm B\odot}$), the detection rate is 60\% and
64\% respectively (out of 44 galaxies in the $K_{\rm s}$ band and 37 galaxies
in the $B$ band). In fact, some $>$ 2 \Ls\ galaxies without coronal detections
have small $\sigma_{*}$. Thus, we also examined the detection rate from the
$L_{\rm optical} \sigma_{\ast}^{2} - L_{\rm X}$ plot. We set the thresholds
for a 2 \Ls\ galaxy with a $\sigma_{*}$ of 250 km/s. Above that thresholds,
the detection rates of coronae are 68\% and 72\% respectively (out of 38
galaxies in the $K_{\rm s}$ band and 30 galaxies in the $B$ band).
As there are still upper limits higher than detections, we conclude that
$>$ 60\% of $>$ 2 \Ls\ galaxies have their coronae survived.
When fainter galaxies are examined, the average
level of upper limits becomes comparable to that of detections. For galaxies
with $L_{\rm Ks}$ between \Ls\ and 2 \Ls\ (54 in total in this sample),
we conclude that at least 40\% of galaxies have their coronae survived,
as there are 18 robust detections and 12 identified as soft X-ray sources.
Even for $<$ \Ls\ (10$^{11.08} L_{\rm K\odot}$) galaxies, a significant
number of coronae (5 robust detections and 11 identified as soft X-ray
sources, Fig. 3) have been revealed although many upper limits are higher
than detections.
The detection rate of corona also increases with the stellar
velocity dispersion of the host galaxy, from $\sim$ 20\% at
$\sigma_{*} <$ 200 km/s, to $\sim$ 50\% at 200 kms/ $< \sigma_{*} <$
300 km/s, to $\sim$ 80\% at $\sigma_{*} >$ 300 km/s. We have also
examined the relation of the detection rate with the $B - K_{\rm s}$
color (Fig. 2) and failed to find any clear evidence for correlation. This may
not be surprising as the color-luminosity map shows that the average
$B - K_{\rm s}$ color is nearly constant for $> L_{\rm Ks, cut}$ galaxies.

We have also examined the distributions of the relative velocity of galaxies
to that of the cluster for galaxies with and without coronal detections.
No significant difference was found, for all galaxies or for only $<$ 2 \Ls\
galaxies. The relation between the detection rate and galaxy position in clusters
is also examined. The distance of galaxies to the cluster center was normalized
with the size of the cluster dense core defined as the radius
where the ICM electron density reaches 0.005 cm$^{-3}$. However, we did not
find any significant difference between detections and non-detections, for
all $>$ 0.74 \Ls\ galaxies or only for 0.74 \Ls\ $< L_{\rm Ks} <$ 2 \Ls\
galaxies. The failure to find a significant difference may not be surprising
as the difference may only exist for faint galaxies generally with low
density coronae, for which the true detection rate is confused by high
upper limits. Moreover, little knowledge of the complete stripping history
of cluster galaxies (as they move around the clusters), as well as the
projection effect, also dilutes correlations.

In summary, our results show that coronae are very common for $\gsim$ \Ls\
galaxies. The increase of the detection rate of coronae with the optical/NIR
luminosity and the stellar velocity dispersion of the galaxy is also
implied by the current data, but how much is due to the observational
bias for X-ray faint coronae is unknown. The X-ray coronae of $\lsim$ \Ls\
galaxies are generally indeed faint and have low density (e.g., David
et al. 2006), which should make them less likely to survive in high
pressure environments compared with massive galaxies with galactic
cooling cores. The detection rate can vary a lot from cluster to cluster
(at least for bright coronae). The detection rate is not apparently correlated
with the cluster temperature or velocity dispersion.

\subsection{L$_{\rm X}$ - L$_{\rm radio}$ relation}

We have also examined the $L_{\rm X} - L_{\rm radio}$ relation for coronae
in our sample (Fig. 6). The generalized Kendall's tau and the generalized
Spearman's rho tests in STSDAS have been used to examine the correlation
coefficient. Both tests are able to handle the data with censoring in both
independent and dependent variables. A moderate correlation (96.9\%
significance) is found from both tests. If the luminosity of the nuclear
hard source is added (the right panel of Fig. 6), the correlation becomes
more significant (99.99\% from both tests).
Many coronae in our sample are associated with powerful radio sources. 
In fact, for 16 galaxies with an 1.4 GHz luminosity of $> 10^{22.8}$ W Hz$^{-1}$
(our selection threshold for radio bright galaxies, $\S$2.2),
10 have robust coronal detections, while PGC018297 in A3376 has a soft
X-ray source. Four other sources are detected in X-rays (generally with hard
spectra), while only PGC3097068 in Ophiuchus is not detected. Even for these
sources without coronal detections, the upper limits are generally high
($> 1.2\times10^{40}$ ergs s$^{-1}$ in the 0.5 - 2 keV band) except for
NGC~4869 in Coma, which has a soft X-ray spectrum that however does not
meet the criterion 3. The presence of a significant gas component
($\lsim 10^{8} M_{\odot}$) potentially provides the fuel for the
central super-massive black hole (SMBH), in environments where the
amount of the galactic cold gas is at a minimum.
Therefore, our results suggest a general connection between
the cooling of the coronal gas and the AGN activity.

We can classify the X-ray emission of radio luminous galaxies into
three classes: corona-dominated (e.g., ESO~137-006, the most luminous
radio galaxy in our sample, see Appendix), X-ray-nucleus-dominated (e.g., ESO~137-007,
IC~310 and 3C~264, see Appendix) and mixed (e.g., NGC~1265). All seven galaxies with
$>$ 10$^{41}$ ergs s$^{-1}$ (0.5 - 2 keV) coronae are corona-dominated,
while both galaxies with X-ray AGN more luminous than 10$^{41}$ ergs s$^{-1}$
are X-ray-nucleus-dominated. There is however no a galaxy with both a
luminous corona ($> 10^{41}$ ergs s$^{-1}$) and a luminous X-ray AGN
($> 10^{41.5}$ ergs s$^{-1}$).
Studies of the X-ray emission of luminous radio sources actually
extend beyond those in hot clusters. We note that dense coronae
have also been found around the nuclei of luminous radio sources in cooler
groups or clusters, e.g., 3C~31 (Hardcastle et al. 2002), 3C~296 (Hardcastle
et al. 2005), cD galaxies of A160 and A2462 (Jetha et al. 2005).

S05 and SJJ05 reported an anti-correlation of X-ray morphology with
radio surface brightness distributions in NGC~3842, NGC~4874 and
NGC~1265, and concluded that the jets traverse the coronae with little
energy dissipation. Just outside the coronae, the radio emission ``turns on''.
However, not many coronae in our sample are well resolved and not many
radio galaxies in our sample have high-resolution radio images that allow
us to examine this morphological anti-correlation
in a much larger sample. Anti-correlation of the radio emission with the
coronal gas is also found in WEIN~51 (3C129.1, Krawczynski et al. 2003),
ESO~137-006 (Jones \& McAdam 1996) and NGC~7720 (3C~465, Hardcastle et al. 2005).
However in NGC~3862, such an anti-correlation is not observed (3C~264,
Lara et al. 2004).

\subsection{Temperature of coronae}

The temperatures of coronae and soft X-ray sources are plotted with the $K_{\rm s}$
band luminosities of their host galaxies (Fig. 7). Temperature values from
stacked spectra are not included. The coronal temperature is not correlated
with the $K_{\rm s}$ band luminosities of the galaxy (or the total stellar mass),
especially for $< 10^{11.8}$ L$_{\rm K\odot}$ galaxies. The temperatures of coronae
are generally in the range of 0.4 - 1.1 keV. Only three coronae (NGC~4889,
IC~1633 and PGC020778) are significantly hotter than 1.1 keV and they are
all cDs. The $L_{\rm X} - T$ relation of embedded coronae also does not show a
significant correlation, as coronae that are less luminous than 10$^{41}$ ergs
s$^{-1}$ coronae all have a similar temperature of 0.5 - 1 keV.
The coronal temperatures are also compared with the stellar kinetic
temperature of their host galaxies. We calculate $\beta_{\rm spec}$ ($\mu
m_{\rm p} \sigma_{\ast}^{2} / kT$) for galaxies with measured stellar velocity
dispersion and coronal temperatures. As shown in Fig. 8,
$\beta_{\rm spec}$ = 0.2 - 1.1, implying generally hotter X-ray gas than
stars. NGC~6107 has the smallest $\beta_{\rm spec}$ ($\sim$ 0.18).
We have also examined the relationship between $\beta_{\rm spec}$ and $kT_{\rm ICM}$
(or $kT_{\rm ICM}^{3/2}$ to mimic the saturated evaporation by the ICM), and
find no correlation. This implies that the ICM heat flux is not
the reason for the over-heating of the coronal gas relative to stars.
We notice that small $\beta_{\rm spec}$ has also been found for coronae
in the field or poor environments with the \chandra\ data (e.g., 0.3 - 1
by David et al. 2006).

One may notice that the bottom left of Fig. 8 is not populated with
any coronae and almost all coronae are hotter than 0.4 keV. Is this
real or due to the observational bias? For low temperature plasma
($<$ 0.35 keV), the iron L-shell hump becomes not significant any
more. Therefore, these coronae cannot be well identified by the
criterion 1. However, they should still be identified by the criteria
2 and 3. The soft X-ray sources identified by the criterion 3 are also
generally hotter than 0.4 keV. Galaxy coronae with temperature of
$<$ 0.4 keV are indeed reported in poor environments (e.g., $\sim$ 0.25 keV
gas in NGC~4697, Sarazin et al. 2001; $\sim$ 0.3 keV gas in Centaurus A,
Kraft et al. 2003), but the X-ray gas luminosities of these galaxies
are all low. Recently, David et al. (2006) studied
18 low-luminosity early-type galaxies in poor environment. Seven coronae
they examined just fill the bottom left of Fig. 8 with temperatures
of 0.2 - 0.4 keV. However, these coronae are too faint ($\sim 10^{39}$
ergs s$^{-1}$) to be detected with our data and their low gas density
(central electron density of 0.003 - 0.02 cm$^{-3}$) makes them hard
to survive in dense environment. Thus, we conclude, above our detection
limit ($L_{X} \sim 10^{40}$ ergs s$^{-1}$), embedded coronae are almost
all hotter than 0.4 keV.

For seven luminous coronae within $5'$ of the optical axis,
the superior spatial resolution of \chandra\ allows us to derive temperature
values in 2 - 4 radial bins.
We summarize the results in Fig. 9, which includes the published profiles
of NGC~3842, NGC~3837, NGC~4874 and NGC~1265 (V01; S05; SJJ05). The temperature
profiles of two larger X-ray sources (IC~1633 and NGC~7720) are discussed separately
in Appendix (Fig. 21 and 23).
There is clearly a large temperature gradient across the
coronal boundary. Inside the coronae, temperature profiles can have a positive
gradient (NGC~3842, NGC~4874, NGC~1265, ESO~137-006, NGC~3309 and NGC~1265) or
be flat (NGC~6109 and NGC~3837).
Within the very center ($\lsim$ 0.6 kpc), the gas temperature is 0.5 - 0.8 keV.

\subsection{Abundance of the coronae}

Although the temperature and luminosity of coronae can be robustly determined,
the abundance is generally poorly constrained as the data statistics
are not very good. For the two largest and brightest coronae in this sample,
NGC~7720 and IC~1633, their abundances are fairly well determined,
1.08$^{+0.62}_{-0.18}$ solar (from MEKAL) and 1.37$^{+0.51}_{-0.29}$ solar
(iron abundance from VMEKAL). As
constraints for other coronae are poor, we selected 20 coronae with
temperatures determined better than 13\% (NGC~1265, NGC~1270, NGC~1277,
CGCG~540-101, NGC~3309, NGC~3311, NGC~3842, NGC~3837, NGC~4874, NGC~4889,
NGC~6107, NGC~6109, PGC020767, ESO~137-006, NGC~4706, IC~5342, NGC~5718,
ESO~444-046, PGC047197 and PGC073007) to perform a joint fit. MEKAL model
was adopted and only the abundance was linked in the joint fit. 
The deep observations of Centaurus and Perseus
clusters allow us to include three faint coronae (NGC~1277, CGCG~540-101
and NGC~4706, all with $L_{X} <10^{40}$ ergs s$^{-1}$) in the joint fit.
The derived abundance from the MEKAL model is 0.79$^{+0.23}_{-0.13}$ solar,
which is consistent with the derived abundances of all these coronae within 95\%. 
As this abundance value is only an average value for the whole corona and an
abundance gradient may exist inside coronae (S05; SJJ05), the coronal gas
is indeed more enriched in iron than the ICM ($\sim$ 0.3 solar), which is
consistent with a stellar origin of the coronal gas (e.g., Humphrey \& Buote 2006).

\subsection{Size and gas mass of the coronae}

There are 27 coronae of early-type galaxies resolved in this work. 
Their sizes are estimated and shown in Fig. 10. Besides these 27 resolved
sources, we can also put upper limits on the sizes of 32 unresolved
coronae or soft X-ray sources (see $\S$3.2). Thirty one sources are
smaller than 4.0 kpc in radius, while one is smaller than 4.3 kpc
in radius. Thus, most coronae have radii of 1 - 4 kpc.
Coronae that are significantly larger than 4 kpc in radius are:
NGC~6109 (4.6$\pm$0.4 kpc), NGC~3860 ($\sim$ 6.2 kpc),
NGC~7720 (9.6$\pm$0.8 kpc), IC~1633 (9.0$\pm$0.7 kpc),
IC~310 and PGC~018313. 
NGC~6109 is the central galaxy of a 3 keV cluster ZW1615+35 with the second
lowest radial velocity dispersion (584 km/s) in the sample. The cluster also has
a low X-ray surface brightness. Thus, although NGC~6109 is certainly moving
as implied by its long radio tail, the stripping may not be very strong.
Both IC~310 and PGC~018313 have an X-ray tail (Appendix, Fig. 19 and 24), while
their stripping fronts are only $\sim$ 4 kpc from the galactic
center. NGC~3860 is a Sa galaxy with a low-density X-ray corona (Sun \&
Murray 2002a). H$\alpha$ observations implied it may still have a substantial
star formation activity (e.g., Cortese et al. 2006), which may help to
explain its larger low-density corona than the typical ones of early-type
galaxies (see $\S$5). Both NGC~7720 and IC~1633 are cD galaxies so the
residual motion of coronae may be small, which allows the retention of the
outskirts of the galactic cooling core. Nevertheless, $\sim$ 90\% of embedded
coronae have a radius of $\lsim$ 4 kpc, which is expected as stripping can
quickly remove the gas on the outskirts in dense environments.
We should be aware that the LMXB emission is not subtracted from the fits of
surface brightness profiles. Including the LMXB component may reduce
the measured sizes (e.g., S05).

One may argue that any low surface brightness outskirts of X-ray coronae
may hide in the ICM background. However, as these embedded coronae are
pressure confined in the high-pressure environments, the coronal density
at the boundary cannot be low from the argument of pressure equilibrium
(equ. 3 and 4). For three coronae that we have done detailed analysis
(NGC~3842, NGC~3837 and NGC~1265, from S05 and SJJ05), we find that the
coronal densities at the boundary derived from surface brightness fits
are consistent with the expected values from pressure equilibrium (note
we normalized the densities of NGC~3842 and NGC~3837 to lower values as
S05 used low coronal abundances of 0.32 - 0.56 solar). Moreover, we use
the density models of NGC~1404 (r$_{0}$=0.31 kpc and $\beta$=0.481 from
our own analysis) and NGC~4697 (r$_{0}$=0.16 kpc and $\beta$=0.335, Sarazin
et al. 2001) to estimate the expected sizes of coronae in high-pressure
environments. For a luminous NGC~1404-like corona with a central density
of 0.3 cm$^{-3}$, the gas density at 4 kpc radius is 7.5$\times10^{-3}$ cm$^{-3}$.
Assuming a local ICM density of 5$\times10^{-4}$ cm$^{-3}$ and a $kT_{ICM} /
kT_{ISM}$ ratio of 6, the required ambient pressure (thermal pressure +
ram pressure) can be achieved with a velocity of 1000 km/s for a 0.7 keV corona.
With the same assumptions for a faint NGC~4697-like corona with a central density
of 0.08 cm$^{-3}$, the coronal boundary has to be as small as 1.7 kpc for
the pressure equilibrium. Thus, the derived size distribution is consistent
with expectation in dense environments.

The X-ray gas mass of coronae is generally in the range of 10$^{6.5}$ - 10$^{8}$
M$_{\odot}$, for coronal sizes of 1.5 - 4 kpc. The most massive coronae are those of
NGC~7720 and IC~1633, with gas mass of 0.64 - 1.0 $\times 10^{9}$ M$_{\odot}$,
because of their larger sizes. Therefore, the coronae of NGC~7720 and IC~1633
really stand out in this sample. Although we have argued that large sizes of their
coronae may be related to the small residual motion of the galaxies
relative to the surrounding ICM, it is also possible that the X-ray sources of
NGC~7720 and IC~1633 have a different origin from other coronae. We note that
central X-ray sources with the similar size and mass as the ones associated
with NGC~7720 and IC~1633 are also presented in some 2 - 3 keV clusters
(e.g., ESO~306-017 by Sun et al. 2004, ESO~5520200 and NGC~6269). We have
argued that the small cool core of ESO~306-017 may be the relic of a previous
cluster cooling core, after heating by the central AGN. This picture may
also apply for the X-ray sources of NGC~7720 and IC~1633, especially
as the radio source associated with NGC~7720 is the second most luminous
radio galaxy in our sample. 

\subsection{Scarcity of X-ray tails}

Although we have detected 76 embedded coronae or soft X-ray sources of
early-type galaxies in hot clusters, only PGC~018313 in A3376 has a
significant long tail (Appendix, Fig. 24). NGC~1265's corona is asymmetric (SJJ05)
and IC~310's corona has a faint 9 kpc tail (Appendix, Fig. 19). Therefore,
even excluding coronae detected at large off-axis angles (unresolved),
the coronae of early-type  galaxies with a significant stripped tail
in hot clusters are $<$ 5\%, which implies that the period with a
high density X-ray tail is either rare or short ($\lsim$ several 10$^{8}$
yr). The stripped ISM clumps may quickly mix with the ICM and only has a small
density enhancement.

\section{Coronae of late-type galaxies in hot clusters}

There are 22 late-type galaxies in our sample and at least 8 of them have
thermal coronae (Table 3, Fig. 22), with $kT$ = 0.3 - 0.9 keV. UGC~6697 and
ESO~137-001 were discussed in Sun \& Vikhlinin (2005) and Sun et al.
(2006). Both show features indicative of stripping and ESO~137-001 has
a dramatic 70 kpc X-ray tail. Other coronae of late-type
galaxies do not show significant features indicative of stripping.
Among the three most X-ray luminous ones, ESO~137-001 and UGC~6697 are
starburst galaxies, while ESO~349-009 may also have substantial star
formation activity from its high NUV brightness and the likely
ultra-luminous HMXB associated with the galaxy.
A substantial galactic cooling core has
never been detected in a late-type galaxy, since galactic winds are
strong in late-type galaxies. Indeed, the X-ray thermal emission of
late-type galaxies in our sample is generally much more extended
than that of early-type galaxies and is not in high surface brightness.
The X-ray coronae in late-type galaxies usually have low density and
are thus difficult to sustain in high-pressure environments, unless
there is significant gas supply from active star formation.

Sun \& Vikhlinin (2005) and Sun et al. (2006) presented evidence
for star formation triggered by the ICM pressure in UGC~6697 and
ESO~137-001. Interestingly, evidence for enhanced star formation
in other two late-type galaxies (NGC~4921 and ESO~349-009) has also
been found (Appendix, Fig. 21 and 26).

The abundances of the thermal gas in late-type galaxies were also
examined, though good constraints cannot be obtained. The best fits
of the abundance range from 0.1 to 1.2 solar. The low abundance derived in some
galaxies may be due to the mixing of the multi-phase emission components
(Fabbiano et al. 2004), as we can only examine the global spectra.
Because of the mixing effect, we did not examine the abundance from a joint
fit to all detections as we did for coronae of early-type galaxies.

\section{X-ray AGN in our sample}

We have also examined the fraction of luminous X-ray AGN in our sample,
to compare with the recent results by Martini et al. (2006) in eight
0.05 $< z <$ 0.31 clusters. As we are studying the galactic nuclear
component, all cD galaxies are included in this study. Assuming a
$B - R$ color of 1.3 (e.g., Martini et al. 2006), our luminosity
threshold corresponds to $M_{R}$ of -21.30 mag. A3558 is excluded for
this analysis as its \chandra\ exposure is the shallowest. For 163 galaxies
(including both early-type and late-type) that are more luminous
than our luminosity threshold, we found nine AGN with a 0.3 - 8 keV intrinsic 
luminosity of $> 10^{41}$ ergs s$^{-1}$, IC~310, NGC~1275, 3C~264, 
NGC~3861, NGC~4911, IC~5358, NGC~7720, GIN~190 and the cD of A2052. 
This implies a fraction of $\sim$ 5\% for $M_{R} <$ -21.30 mag galaxies
with luminous AGN. Martini et al. (2006) derived a fraction of 5$\pm$1.5\%
for $M_{R} <$ -20 mag galaxies with $L_{0.3 - 8 keV} > 10^{41}$ ergs s$^{-1}$
AGN. We notice that none of AGN detected by Martini et al. are in BCGs and
over half of luminous
AGN in their sample are $<$ \Ls\ galaxies, while all nine galaxies
from our sample are $>$ 1.5 \Ls\ galaxies and five of them are BCGs.
As five clusters in Martini et al. sample are at $z >$ 0.15 (two at
$z >$ 0.30), the AGN activity in Martini et al. sample may
be stronger than that in our sample ($z$ of 0.01 - 0.05 clusters). A better
comparison requires more work on galaxies fainter than our threshold
and likely optical follow-ups for completeness, which is beyond
the scope of this paper. Fainter AGN ($L_{0.3 - 8 keV} \sim 10^{40}$
ergs s$^{-1}$) are indeed detected in many galaxies in our sample as
about two third of galaxies are detected by \chandra\
and at least 20\% galaxies in our sample have nuclear hard component
(Table 2 and 3).

Another uncertainty on the AGN fraction is the fraction
of absorbed AGN as they are faint. In the nine luminous AGN in our
sample, the one in the big spiral NGC~3861 (in A1367) has an absorbed X-ray
spectrum ($N_{H} = 4 - 9 \times 10^{22}$ cm$^{-2}$) and has a similar
optical spectrum as the ``optically dull'' galaxy 3C~264 (Elvis et al. 1981;
Sun \& Murray 2002b). This kind of absorbed AGN is more difficult to
be identified at higher $z$ than the Type I AGN.

\section{Discussion}

A cartoon of the structure of embedded corona is shown in Fig. 12.
A boundary layer lies between the corona and the ICM. As strong
gradients of temperature and velocity are present in the boundary layer,
transport phenomena become important there. We next discuss
the relevant physics of these embedded coronae.

\subsection{Stripping of the coronal gas in hot clusters}

The coronae in hot clusters are subject to stripping by the ICM
as the velocity dispersion of galaxies in hot clusters is high
(520 - 1100 km/s along the line of sight in this sample). The
internal thermal pressure of coronae is generally high enough to
overcome the ram pressure as shown in $\S$4.1. Gravitational restoring
force can also help to sustain the coronal gas if the ram pressure
moves the coronae off-center. Nevertheless, nearly all resolved coronae
are centered on the galactic nuclei. 
Besides ram pressure, coronae also suffer stripping
through transport processes (viscosity or turbulence) at the corona-ICM
interface (or a boundary layer) (Nulsen 1982). The typical mass-loss
rate of the corona by stripping through transport processes is approximately
(e.g., Nulsen 1982):

\begin{eqnarray}
\dot{M}_{\rm strip} &\approx& \pi r^2 \rho_{\rm ICM} {\rm v_{gal}} \\
&=& 0.69 (\frac{n_{\rm e, ICM}}{10^{-3} {\rm cm}^{-3}}) (\frac{r}{2.5 {\rm kpc}})^2
(\frac{{\rm v}_{\rm gal}}{1200 {\rm km/s}}) {\rm M_{\odot} / yr} \nonumber
\end{eqnarray}

The typical timescale of complete stripping is:

\begin{eqnarray}
t_{\rm strip} &=& \int \frac{d M}{\dot{M}} = \frac{4}{n_{\rm ICM} v_{\rm gal}} \int n(r) dr \\
&=& 0.224 \hspace{0.13cm}g_{1} (\frac{n_{e, 0}}{\rm 0.2 cm^{-3}}) (\frac{n_{\rm ICM}}{\rm 10^{-3} cm^{-3}})^{-1} (\frac{r_{0}}{\rm 0.4 kpc}) \nonumber \\
&& (\frac{v_{\rm gal}}{\rm 1400 km/s})^{-1} {\rm Gyr} \nonumber
\end{eqnarray}

\begin{eqnarray}
( g_{1} = \int (1+x^2)^{-1.5\beta} dx , x = r / r_{0}  ) \nonumber
\end{eqnarray}

where a $\beta$ model is still assumed for the coronal density profile and
all symbols have the same meaning as those in $\S$4.1. We have
calculated $g_{1}$ for several coronae with different starting and ending
coronal sizes. For NGC~1404, $g_{1}$(4-10 kpc) = 0.24 and $g_{1}$(0-4 kpc) = 2.16
(r$_{0}$=0.306 kpc, $\beta$=0.48). For NGC~3842, $g_{1}$(2-4.1 kpc) = 0.23
and $g_{1}$(0-4.1 kpc) = 1.69 (r$_{0}$=0.56 kpc, $\beta$=0.56). For NGC~1256,
$g_{1}$(1-2.3 kpc) = 0.16 and $g_{1}$(0-2.3 kpc) = 1.29 (r$_{0}$=0.40 kpc,
$\beta$=0.73). Thus, the removal of the small dense core of a corona may
not be very quick ($\sim$ 0.5 Gyr), but the outer layers ($\gsim$ 4 kpc radius)
of large coronae can be removed in much shorter time ($\sim$ 50 Myr). Therefore,
stripping is very efficient at reducing the size of coronae
to $\lsim$ 4 kpc and should be the main reason for the small sizes of embedded
coronae ($\S$4.8). Even for coronae with a size of $\sim$ 4 kpc, the
stripping is still efficient to reduce its size to 1 - 2 kpc quickly. The
complete stripping of the whole coronae may take up to 1 Gyr, but the
stripping beyond 1.5 kpc radius region is $>$ 5 times faster. As the prevalence
of coronae of massive galaxies implies a lifetime of about at least several Gyr
and many coronae have a size of 2 - 4 kpc, the coronal gas must be replenished.
 
As we discussed in S05 and SJJ05, the stellar mass loss can replenish the
coronal gas lost through stripping. Although the enclosed stellar light in
embedded coronae is only $\sim$ 30\% of the total light, the stellar mass
loss outside the corona (region III in Fig. 12) may also help to reduce the
impact of the ICM wind momentum and ``shield'' the small corona at the center. This
``shielding'' effect has been examined for three galaxies with known
optical light profiles, NGC~3842, NGC~3837 and NGC~1265
(S05; SJJ05). The stellar mass loss rate by Faber \& Gallagher (1976)
was used. We assume that the stellar mass loss rate is proportional to
the stellar light and follows the stellar light distribution. Therefore,
the percentage of the stellar light ahead of the corona (within the
cross section of the corona) can be derived by deprojecting the measured
two-dimensional stellar light profile. The result is shown in Fig. 13 as
a function of the assumed coronal size (r). The direct ``shielding'' ahead of
small coronae is however small, only 3\% - 5\% of the total stellar mass loss,
for coronae of NGC~3842, NGC~3837 and NGC~1265 (2.3 - 4.1 kpc in radius).
Therefore, the mass loss through stripping (equ. 5) needs to be balanced
by the stellar mass loss in the boundary layer of the corona.

We have examined this scenario for coronae of NGC~3842, NGC~3837 and 
NGC~1265 by assuming a width of the boundary layer of 0.5 kpc. We assume
that only stellar mass loss in the boundary layer can replenish the gas
lost from stripping, while the stellar mass loss interior to the boundary
layer cannot contribute. The properties of the coronae and the ICM are
adopted from S05 and SJJ05. We use velocities of 500 km/s, 1400 km/s and
3000 km/s for NGC~3842, NGC~3837 and NGC~1265 respectively (S05 and SJJ05).
It is easy to find out that stripping estimated from equ. 5 is too strong
and must be suppressed. As shown in Fig. 14, stripping needs to be
suppressed by a factor of 13 - 22 from the rate predicted by equ. 5, to
allow the gas balance to achieve in the boundary layers of these coronae.
Since the stellar mass loss at the back side of coronae may not help to
balance the gas loss from stripping (acting on the front and the side),
the required suppression factors may be even larger. There are certainly
some uncertainties in this simple analysis, e.g., the size of the boundary
layer and the velocity of coronae. Nevertheless, we conclude that stripping has
to be suppressed by at least a factor of ten from that predicted by
equ. 5 to allow the gas balance in the boundary layer. Thus, turbulence
or viscosity has to be suppressed by at least a factor of ten at the
coronal boundary. Obviously, magnetic field, responsible for the suppression
of heat conduction ($\S$7.2), should also be responsible for suppressing
the instabilities at the boundary. As long as the stripping is largely
suppressed, the balanced radius is stabilized as implied by Fig. 14.
If stripping overwhelms at a point, the size of the
corona decrease so the stellar mass loss rate in the boundary layer
increases to return to the balance point. Therefore, the size distribution
of embedded coronae (generally 2 - 4 kpc in radius) can be explained
by this scenario with reduced stripping balanced by the stellar mass
loss in the boundary layer.

Another good example to examine is the cD of A3558 (ESO~444-046, see Appendix
and Fig. 27 and 28),
which is located in a dense non-cool core with a central electron density
of $\sim$ 0.015 cm$^{-3}$ (Rossetti et al. 2006). The gas core of A3558
is ``sloshing'', which may be triggered by the past merger with a nearby
cluster SC~1327-312 (e.g., Rossetti et al. 2006). Using the pressure
difference across the northwestern cold front (Fig. 27), Rossetti et al.
(2006) derived a velocity of $\sim$ 940 km/s for the gas core. However,
internal velocity field may be complicated, as implied by the eastern
sharp edge (Fig. 27) and simulations (e.g., Ascasibar \& Markevitch 2006). 
As SC~1327-312 is 1.4 Mpc from the gas core of A3558 in the plane of sky,
the ``sloshing'' of A3358's gas core must have proceeded for over 1 Gyr.
Over the course of ``sloshing'', the corona of ESO~444-046 should be
subject to the stripping of the dense ICM even the galaxy may stay
at the bottom of the potential. Repeating the exercise that we did for
the three previous coronae, we find that stripping has to be suppressed by a factor
of over 20 from equ. 5, even assuming a small relative velocity of
ESO~444-046's corona to the surrounding ICM (200 km/s).
Therefore, the very existence of ESO~444-046's corona in the dense core
of A3558 either puts strong constraint on stripping, or on the central
velocity field during the core ``sloshing'', or both.

\subsection{Heat conduction and magnetic field}

The prevalence of cool (0.5 - 1.1 keV) coronae in massive galaxies
in hot clusters indicates that their lifetime is long (at least several
Gyr) and that they are in approximate energy balance, which contrasts
with the short timescale of evaporation as previously noticed (V01; S05;
SJJ05). The saturated conductive heat flux across the ISM-ICM interface
is (Cowie \& McKee 1977):

\begin{eqnarray}
q_{\rm sat} &=& 0.4 (\frac{2kT_{e}}{\pi m_{e}})^{1/2} n_{e}kT_{e} \\
&=& 6.8 \times 10^{-4} (\frac{T}{1 {\rm keV}})^{3/2}
(\frac{n_{e}}{10^{-3} {\rm cm^{-3}}}) {\rm \mbox{ }ergs\mbox{ }s^{-1}\mbox{ }cm^{-2}} \nonumber
\label{eq:cond_sat}
\end{eqnarray}

Thus, the total heat flux across the whole surface of the corona is:
8.2$\times10^{42}$ (T / 5 keV)$^{3/2}$ (n$_{\rm ICM} / 10^{-3}$ cm$^{-3}$)
(r / 3 kpc)$^{2}$ ergs s$^{-1}$, which is much larger than the typical X-ray
luminosity of embedded coronae. From equ. 63 of
Cowie \& McKee (1977), the saturated mass loss rate by evaporation is:
$\dot{M} \propto$ n$_{\rm ICM}$ T$_{\rm ICM}^{1/2}$ r$^{2}$ F($\sigma_{0}$),
where F($\sigma_{0}$) is a weak function of the saturation
parameter $\sigma_{0}$ (F($\sigma_{0}$) $\propto$ $\sigma_{0}^{\sim 0.5}$,
equ 32 and Fig. 4 of Cowie \& McKee 1977). Therefore, the evaporation time
of the layer (r, r+dr) is $\propto$ n$_{\rm ISM}$ r$^{\sim 1/2}$ dr, assuming
constant n$_{\rm ICM}$ and T$_{\rm ICM}$ in the process of evaporation.
As $\beta$ of luminous coronae is always $>$ 1/3 and the core radius of
the coronal gas is generally $<$ 0.5 kpc, the evaporation time is largest
at small radii,
e.g., 0.5 - 3 kpc radius. The times for the current coronae of NGC~4874,
NGC~4889, NGC~1265, NGC~3842 and NGC~3837 to fully evaporate are 3.4 -
35 Myr, with the properties of the coronae and the ICM derived in V01,
S05 and SJJ05, if the Spitzer conductivity is assumed. The total evaporation
time at most increases by $\sim$ 20\% if an original coronal size of 10 kpc
is assumed for these coronae. Therefore, the coronal boundary cannot be
the conduction front and heat conduction across the coronal
boundary has to be greatly suppressed. The best constraints on the suppression of
heat conduction are from coronae in the hottest clusters (e.g., $>$ 5 keV).
We have not found that the detection rate of coronae in $>$ 5 keV clusters
differs from that in 3 - 5 keV clusters. By balancing the evaporating flux
and the cooling flux of the corona, it is required that heat conduction has
to be suppressed by a factor of at least 70 - 500 for coronae with $\gsim$ 2 kpc
radii in six $kT >$ 5 keV clusters in our sample. The limit on the
suppression factor is a little weaker in 3 - 5 keV clusters, but still
$>$ 30 - 300.

The natural way to inhibit heat conduction is magnetic field. What is the
origin of magnetic field at the coronal boundary? Lyutikov (2006)
argued that the draping of the ICM magnetic field is responsible for
the suppression of heat conduction across the boundary of cold fronts
in clusters. When a plasma cloud (e.g., a cluster gas core) moves through
a weakly magnetized ICM, a thin, strongly magnetized boundary layer will
form in the front, where transport processes across the layer may be
strongly suppressed.
However, it is unclear whether magnetic draping can occur
at scales as small as a few kpc around coronae, as the scale is smaller than the
mean free path of particles in the ICM and likely smaller than the coherence
length of magnetic fields in the ICM. Moreover, although magnetic draping
may explain the suppression of heat conduction at the front, the transport
processes at the back and probably the sides are not addressed. Magnetic
draping also should not be a factor for coronae at the cluster centers
with little residual motion (e.g., cD of A3558, two cDs in Coma). 
Therefore, the longevity of embedded coronae in hot clusters may be strong
evidence that the coronal gas of early-type galaxies is magnetized. Moreover,
the galactic magnetic field is disconnected from the magnetic field in the
ICM, keeping the heat conduction across the corona boundary very low.
This raises a general question on the galactic magnetic field evolution
when the galaxy is falling into a dense environment.
Current simulations ignore the effects of magnetic field, although
it is shown to be important for energy transfer in embedded coronae.
Our knowledge of the magnetic field in elliptical galaxies is also poor
(reviewed by Widrow 2002). Galactic magnetic
fields may also dissipate if turbulent diffusion is strong. This
requires continuous amplification to maintain the galactic magnetic field.
It is unclear how the galactic magnetic field remains disconnected from
the field in the ICM over cosmological timescales.

\subsection{The effects of rich environments on galaxy coronae}

Since the first derivation of the $L_{\rm X} - L_{\rm B}$ relation,
the effects and importance of environment have been discussed. 
Different observational results have been obtained supporting both
positive and negative environmental effects on coronae.
Brown \& Bregman (2000) claimed a positive (but weak) correlation between
$L_{X}$ / $L_{B}$ and local galaxy density, based on \rosat\
results for 34 galaxies. They suggested positive effects arise from
accretion of the surrounding ICM and/or stifling galactic winds by the ICM
pressure. However, O'Sullivan et al. (2001) found no such correlation in a
\rosat\ analysis of 401 galaxies. Recently, Finoguenov \& Miniati (2004)
claimed a positive effect of rich cluster environments on galaxy coronae
based on \xmm\ data of the Coma cluster. Their main argument is that
the coronae are more luminous because of the compression by the
ICM pressure.

Our sample includes 157 early-type galaxies in 25 nearby hot clusters
and unambiguously reveal the negative effects of environments on the size and
gas mass of embedded coronae, as embedded coronae are smaller and
contain less X-ray gas than coronae in poor environments. The negative
effect on the $L_{\rm X} - L_{\rm optical}$ relation of embedded coronae
is also suggested by our results ($\S$4.1 and $\S$4.2). In
Fig. 5, we also derived a best fit lower than those from 
Finoguenov \& Miniati (2004) and Matsushita (2001). However, we
caution that reliable relations for coronae in poor environments
need to be derived using \chandra\ data for better comparison.
If we examine the high end of the $L_{\rm Ks} - L_{\rm X}$
relation in detail (Fig. 3), we find that cD galaxies in high pressure
environments (e.g., NGC~4874 and NGC~4889 in Coma, the cDs of A3558 and 
A4038) have systematically lower $L_{\rm X} / L_{\rm Ks}^{1.5}$ (or
the normalization of the best fit line in Fig. 3) than cD galaxies in
lower pressure environments (e.g., NGC~6107, IC~1633, NGC~7720, ESO~137-006
and NGC~3842). This trend is also against the positive environmental
effects on the luminosities of coronae.

The limitations of \rosat\ prevented any previous analysis from accurately
determining the properties of thermal coronae in rich environments (e.g.,
as shown in $\S$4.1). The accretion of the surrounding ICM is only likely
important for cDs, while the ICM around other cluster galaxies is not
dense. In fact, even for cDs of non-cooling core clusters in our sample,
the cooling time of the surrounding ICM is always longer than 3 Gyr.
The ICM may stifle the galactic wind. However, any low density
galactic winds will be easily stripped out of the galaxy as the
velocity dispersion in hot clusters is very high (1000 - 2300 km/s
in our sample). The argument of coronal density enhancement by ICM
pressure (Finoguenov \& Miniati 2004) also does not hold for luminous
embedded coronae, as the internal thermal pressure gradient inside
these coronae is high. For example, the central gas pressures in NGC~3842,
NGC~3837 and NGC~1265 are 20 - 30 times the gas pressures at their
boundaries. The ICM pressure has little effect on the dense cores of
coronae, while the outskirts contribute little to the total luminosity.
Even for coronae with flat density profiles, the compression will only
happen on the leading edge, while the sides and likely the back will be
stripped. The effect of ambient pressure on the coronal luminosity thus
depends on the initial internal pressure gradient of coronae and the
detailed stripping process. Although a moderate pressure may enhance
density at the front (e.g., NGC~4552 in Virgo, Machacek et al. 2006;
David et al. 2006), as the gas content and the size of coronae have
been significantly reduced by the stripping, the X-ray luminosity 
should eventually decrease, as we have observed.

We have also examined the enclosed X-ray luminosity as a function
of radius for two representative nearby coronae in the poor
environments, NGC~4697 and NGC~1404 (Fig. 15). The properties
of NGC~4697 are from Sarazin et al. (2001), while those of NGC~1404
are from our own analysis. For flat coronae with small $\beta$ like
NGC~4697, most X-ray emission is from large radii. When such
coronae move into hot clusters, the outskirts, where most of the X-ray
gas mass resides, will be quickly removed. Although density
enhancement may occur along the leading edge, the X-ray luminosity
eventually should greatly decrease.
On the other hand, the luminosities of luminous coronae with
a dense cooling core are less affected by stripping. If NGC~1404
falls into a hot cluster and stripping reduces its size to
2 kpc, the X-ray bolometric luminosity is only reduced by 50\% to
be $\sim 1.0\times10^{41}$ ergs s$^{-1}$, which will make NGC~1404
($L_{Ks}=10^{11.27} L_{\odot}$) similar to PGC~018313 in A3376
(Table 2).
We also show the profile of NGC~1265's corona (extrapolated to 10 kpc
radius) in Fig. 15. It is very concentrated because NGC~1265's corona
is small ($\sim$ 2 kpc radius) and its surface brightness profile
is steep close to the boundary (SJJ05).

\subsection{The X-ray component centered on the cD galaxy}

We observed a variety of X-ray components centered on the cD
galaxies in our cluster sample, as shown in Fig. 16. There are
clearly at least four classes:

\begin{itemize}
\item cD filled with hot diffuse ICM, without a cooling core or a cool corona (A2147)
\item a small cool corona of the cD embedded in the hot ICM (Coma, A1060,
A3627, 3C129.1, A1367, MKW8, ZW1615+35, A576, A3376 and A3558)
\item a larger distinctive cool X-ray component associated with the cD, embedded
in the hotter ICM (A2877 and A2634)
\item a big cluster cooling core (Centaurus, AWM7, Perseus, Ophiuchus,
A2199, A496, A2052, MKW3S and A4059)
\end{itemize}

A2147 is a non-cooling core cluster. Over the course of the lifetime of its
cD galaxy, over 10$^{10}$ M$_{\odot}$ ISM gas should have been injected into
the galaxy. The corona of its cD galaxy must have been destroyed and mixed
with the ICM. On the other hand, from the large number
of clusters in the second class it seems that most coronae of the cD galaxies
can survive until probably a cluster cooling core begins to form.
The central X-ray component in A2107 is bigger and hotter than those
of A2877 and A2634. A2107 may then lie between the class three and four.
Both A4038 and A3571 have a high density gas core and a short central
cooling time ($\sim$ 1.3 Gyr). However, no temperature decline towards the
center has been observed in both clusters. We consider that their gas
cores may either have been re-heated, or still in the early stage of cooling
core formation. Therefore, they may be more similar to the class four clusters
than other classes. We also noticed that the luminosities of the central
radio sources do not appear to relate to the X-ray luminosities of the central
cool component. The radio source in A496, embedded in a big cluster cooling core,
is 80 times fainter than the radio source associated with the small corona
of ESO~137-006.
Most clusters in our sample either belong to the class two or the class
four, which implies that the cD galaxies spend most of time
in these two stages. This variety of X-ray components centered on the cD
galaxy needs to be fitted into our understanding of cluster evolution,
especially the evolution of the cluster gas core.

\subsection{The energy \& gas balance in embedded coronae}

Our results indicate that coronae of massive galaxies are very common
in hot clusters, which implies approximate energy and gas balance inside
coronae. However, these embedded dense coronae also have short cooling
time. The mass deposition rate from cooling is
$\dot{M} \approx 2\mu$$m_{\rm p}L_{\rm X, bol}/5kT = 0.44$
($L_{\rm bol}/10^{41}$ ergs s$^{-1}$) ($kT$/0.9 keV)$^{-1}$ M$_{\odot}$/yr.
Thus, the cooling timescale for a typical corona is: 0.22
($M_{\rm gas}$/10$^{8}$ M$_{\odot}$)
($kT$/0.9 keV) ($L_{\rm bol}$/10$^{41}$ ergs s$^{-1}$)$^{-1}$ Gyr, while
the expected lifetime is an order of magnitude higher. 
The stellar mass loss rate within coronae of massive galaxies can be
substantial, 0.3 - 0.6 ($L_{\rm B}/10^{11} L_{\odot}$) M$_{\odot}$/yr,
assuming an enclosed stellar light fraction of 20\% - 40\%. As long as
stripping is largely suppressed at the boundary ($\S$7.1), the fraction of
the stellar mass loss rate required to balance the mass lost by stripping
is not large ($\sim$ 20\%). However,
as shown in $\S$4.2, the total energy release from the stellar
mass loss is too small to balance cooling. Moreover,
as the stellar kinetic temperature is generally lower than the gas
temperature ($\S$4.5), the coronal gas cannot be heated by the stellar
mass loss. Therefore, another heat source is required to offset cooling
for coronae of massive galaxies to explain their ubiquity and other
properties.

The problem for these embedded mini cooling cores is similar to that in
big cluster cooling cores, where cooling needs to be offsetted by heating.
While AGN has generally be considered as the primary heat source in cluster
cooling cores, AGN heating may be too powerful for these embedded coronae. 
The central SMBHs do not know where they are, if the accretion
is determined by the gas properties close to the SMBHs (e.g., within the
Bondi radius, which is generally smaller than 0.1 kpc). Indeed, many
radio sources, as powerful as those in cluster cooling cores, have
been found to originate from embedded coronae ($\S$4.4).
As we argued in S05, SJJ05 and $\S$4.4, coronae associated with strong radio
sources should not be much heated
by central AGN as the jet power can easily destroy the coronae and unbind
the gas. Jets may just penetrate the coronae with little energy dissipation.
We indeed notice that some radio sources can disturb the surrounding gas
within several kpc radius significantly and gas distribution is very
asymmetric (e.g., M84, Finoguenov \& Jones 2001;
M87, Forman et al. 2006). However, if this happens for a small corona in
a high pressure environment, the radio source may easily destroy the corona
or alter the density distribution to make coronae vulnerable to stripping
in certain directions.
Even if we assume weak AGN can offset cooling and not destroy coronae
in some cases, it may be unavoidable to conclude that some coronae have
already been destroyed by AGN as the radio power of the AGN spans a very
wide range. It is then difficult to reconcile the facts
of ubiquity of embedded coronae and the possible high frequency of AGN
bursts (with a duty cycle of e.g., $\sim 10^{8}$ yr).

Therefore, a gentle heat source is required. There are
at least two possible such heat sources. The first one is the
reduced ICM heat flux through the coronal boundary. This was first
suggested for NGC~4874 (V01), as there is a temperature
gradient inside the corona. This mechanism requires a largely suppressed
but not zero heat conductivity across the coronal boundary. The suppression
factor can be estimated from the ratio of the heat flux and the X-ray
luminosity of the corona (e.g., $\S$7.2). As the reduced
heat flux (proportional to the surface area of the corona) drops much
more quickly than the cooling with radius, a balance
radius at 1 - 4.5 kpc can always be achieved for heat conductivity
reduced by a factor of 50 - 400. However, as examined in V01 and S05,
this mechanism requires a heat conductivity as high as the Spitzer
value interior to the coronae. As we argued in $\S$7.2, the coronal gas is
magnetized. Therefore, it is unclear whether such a high heat
conductivity can be achieved inside coronae. Moreover, as shown in
Fig. 9, the temperature profiles of NGC~6109 and NGC~3837 are
flat. Any ICM heat flux crossing the corona/ICM boundary may not be
re-distributed in at least these two coronae. 

Another heat source is the kinetic energy of SNe. As shown in Fig. 4,
the SNe inside coronae (enclosing $\sim$ 30\% of the total stellar mass)
generally have enough
energy to balance cooling, if the SN kinetic energy can be efficiently
coupled into the coronal gas. We have used the SN rate in early-type galaxies
by Cappellaro et al. (1999), 0.166 h$_{70}^{2}$ per 100 yr per 10$^{10}$
L$_{\rm B \odot}$, and assumed a kinetic energy of 10$^{51}$ ergs per SN.
The requested efficiency ranges from $\sim$ 20\% for 0.7 \Ls\ galaxies, to
$\sim$ 40\% for 2 \Ls\ galaxies, to 100\% for most luminous galaxies.
However, studies of the $L_{\rm X} - L_{\rm B}$ relation for coronae in poor
environments (e.g., Canizares et al.
1987; Brown \& Bregman 1998) indicate that the SN kinetic energy is
poorly coupled into the coronal gas, especially for X-ray faint coronae. 
Most of the SN kinetic energy is probably used to drive galactic winds,
which have a low density and are X-ray dim. The question is then where
the SN energy can go in embedded coronae.
The energy transfered to cosmic rays may not be strong as
the SN blastwaves in 10$^{7}$ K coronal gas can only be weak shocks. 
The evolution of Supernova remnants (SNRs) in
the X-ray coronae of early-type galaxies has been examined by Mathews (1990).
With the typical gas properties of luminous coronae in our sample,
the SN-blown bubbles are generally $\lsim$ 10 pc in radius. The lifetime
of these buoyant bubbles is also short as Rayleigh-Taylor and Kelvin-Holmholtz
instabilities grow quickly. Therefore, SNe in dense coronae may only
be able to deposit their kinetic energy locally, to offset cooling.
However, if a single SN can only heat locally, the integrated SN heating
should follow the optical light profile, which is flatter than the X-ray
light profile of embedded luminous coronae (Fig. 4 of S05; Fig. 2 of SJJ05).
Therefore, even if the SN kinetic energy is sufficient
to balance cooling globally, gas cooling in the center may be unavoidable.
Buoyant bubbles from SNe transfer more energy to the outskirts and cannot help
to heat the center. Strong cooling at the very center may not be
a problem as long as heating can offset a large portion of cooling
globally. Moreover, strong cooling at the very center may be required to
fuel the central SMBH and trigger radio sources.

Now we have a toy model set to explain the longevity of these embedded
coronae and the energy balance of these mini cooling cores in hot clusters.
The strong evaporation is greatly suppressed by the magnetic field
at the boundary, which may also help to significantly suppress
stripping through surface instabilities. Stellar mass loss in the
boundary layer can balance the mass lost from the suppressed stripping
and the balance is stable.
Inside the corona, radiation loss is largely offsetted by heating from
the kinetic energy of SNe. Cooling may still happen at the very center
and is responsible for triggering the activity of the central SMBH.
The gas mass dropped out from the X-ray phase can be replenished by
the stellar mass loss. The detailed modeling of the internal energy
balance of embedded coronae is however beyond the scope of this paper.

What is the fate of these embedded coronae of early-type galaxies?
Although they can survive even in environments as dense as the center
of the Coma cluster, it is unlikely that they can survive if
passing through a dense cluster cooling core ($>$ 0.01 cm$^{-3}$) with
a high velocity.
However, the volume of such a dense cluster core (generally $<$ 100 kpc
radius) is tiny, compared with the volume of the whole cluster
(with a radius of several Mpc). The fate of coronae then depends
on the orbit of the galaxy. The coronae of cD galaxies may eventually
mix with the ICM when cluster cooling cores begin to form.

\section{Conclusions}

We have systematically searched for and investigated thermal coronae of
179 galaxies (both
early and late types) in 25 nearby ($z <$ 0.05), hot ($kT >$ 3 keV)
clusters, based on 68 \chandra\ archival pointings with a total
exposure of 2.77 Msec and a sky coverage of 3.3 deg$^{2}$. 
The galaxy sample is complete for all NIR luminous ($>$ 0.74 \Ls\
in the $K_{\rm s}$ band) or radio luminous ($L_{\rm 1.4 GHz} > 10^{22.8}$
W Hz$^{-1}$) galaxies in the \chandra\ field. This work
represents the first systematic study of X-ray thermal emission of galaxies
in rich environments. The main observational results and conclusions of
our study are:

1) We find a new population of embedded X-ray coronae of early-type
galaxies in hot clusters. Despite the effects of ICM stripping,
evaporation, rapid cooling, and powerful SMBH bursts, X-ray coronae
of massive early-type galaxies (excluding cDs in cluster cooling cores)
are very common ($>$ 60\% of $>$ 2 \Ls\ galaxies in the $K_{\rm s}$ band)
in hot clusters, although their properties have been
significantly modified by the dense ICM. Significant number of coronae
have also been found in less massive galaxies ($>$ 40\% of \Ls\ $<$
$L_{\rm Ks} <$ 2 \Ls\ galaxies; $>$ 15\% of $<$ \Ls\ galaxies),
although the \chandra\ data of most clusters are not deep enough to
unambiguously identify faint coronae ($L_{\rm 0.5 - 2 keV} \lsim 10^{40}$
ergs s$^{-1}$). These embedded coronae are smaller (generally 1.5 - 4 kpc
in radius) and contain less gas (10$^{6.5} - 10^{8}$ M$_{\odot}$) than their
counterparts in poor environments. The negative effect on the coronal
luminosity is also suggested by the data. These embedded coronae may
correspond to the dense cores of coronae in poor environments. 
Therefore, our work has demonstrated the negative environmental effects
of rich environments on galaxy coronae. The ubiquity of embedded coronae
in massive galaxies implies a lifetime comparable to that of clusters
(or at least several Gyr).

2) The temperatures of embedded coronae range from 0.3 to 1.7 keV.
The gas temperature is generally higher than the stellar kinetic temperature
of the galaxy ($\beta_{\rm spec}$ = 0.2 - 1.1, Fig. 8). Internal temperature
profiles of embedded coronae generally show a decrease of gas temperature
towards the center. The abundance of the coronal gas, constrained from the
joint data of 20 coronae with best statistics, is $\sim$ 0.8 solar, which
implies a stellar origin of the coronal gas.

3) For the cool coronae of early-type galaxies to survive in the hot ICM,
heat conduction across the boundary of the coronae has to be suppressed by
a factor of $\gsim$ 100. We argue that this fact implies the X-ray gas in
early-type galaxies is magnetized. Magnetic field plays an important role
in energy transfer.
However, it is unclear how the galactic magnetic field can remain disconnected
from the field in the ICM for the lifetime of embedded coronae. 

4) The embedded coronae of early-type galaxies are subject to ICM stripping.
The internal thermal pressure of coronae is generally high enough to
overcome the ram pressure. However, the stripping through transport processes
(viscosity or turbulence) at the coronal boundary is too strong and has
to be suppressed by at least a factor of ten. The stellar mass loss
in the boundary layer can balance the mass lost from the suppressed
stripping and the balance is stable.

5) The embedded coronae of early-type galaxies have high gas densities
($\sim$ 0.3 cm$^{-3}$ at the center of luminous coronae) so they are
the mini version of big
cluster cooling cores, with a boundary. As the prevalence of the coronae
of massive galaxies implies a long lifetime ($\gsim$ several Gyr), there
must be a heat source inside these embedded mini cooling cores. While we
argue that both AGN and the stellar ejecta cannot be the major heat source,
SN heating inside coronae is a good candidate if its kinetic energy can be
efficiently (20\% - 100\%) couple into the coronal gas.

6) We have observed a connection between these mini-cooling-cores and the
radio activity of their host galaxies.
Radiative cooling of the coronal gas may provide fuel for the central
SMBH in environments where the amount of galactic cold gas is at a minimum.
We have also found a general morphological anti-correlation of the radio
jet emission and the X-ray coronae. The radio jets generally ``turn on''
after traversing the dense coronae. The embedded dense coronae may also
provide cold gas for nuclear star formation, while the star formation
in ``naked'' galaxies exposed to the hot ICM may have been truncated.

7) We have observed a variety of X-ray components associated with or centered on
the cD galaxies in our sample, ranging from no cool component,
to small cool coronae, to large X-ray cool cores, to big cluster cores.
Most clusters in our sample either have a small cool corona associated with
the cD galaxy, or a big cluster cooling core centered on the cD galaxy.

8) We also detected thermal coronae of at least 8 late-type galaxies (Sb or later)
from 22 galaxies in our sample. Late-type galaxies with luminous X-ray
thermal emission usually have substantial star formation activity.
In four galaxies with the most X-ray luminous coronae in the sample,
evidence for enhanced star formation triggered by the ICM pressure has
been found.

9) Nine luminous X-ray AGN ($L_{0.3 - 8 keV} > 10^{41}$ ergs s$^{-1}$)
are found from 163 galaxies (both early-types and late-types) brighter
than 0.74 \Ls\ (in the $K_{s}$ band), which indicates a not small
fraction ($\sim$ 5\%) of X-ray luminous AGN in local cluster galaxies. Fainter
nuclear hard sources are also found in at least 20\% of galaxies in our
sample.

\acknowledgments

We thank Ewan O'Sullivan for providing the \rosat\ results and
Alexey Voevodkin for providing the density profiles of some clusters.
We thank the referee for prompt reading and comments.
The financial support for this work was provided by the NASA Grant AR6-7004X
and NASA LTSA grant NNG-05GD82G.

\begin{appendix}

\section{Notes on interesting coronae and galaxies}

\paragraph{A1060} is a non-cooling-core cluster. Two coronae associated
with NGC~3311 and NGC~3309 were first reported by Yamasaki, Ohashi \&
Furusho (2002). We performed a new analysis with the updated \chandra\
calibration. The NGC~3311 corona is double-peaked
and a dust filament runs through the nuclear region (Fig. 17).

\paragraph{A3627} is a merging massive cluster in the core of the
Great Attractor (Kraan-Korteweg et al. 1996; B\"ohringer et al. 1996).
It hosts two luminous radio
sources, the NAT source B1650-605 with a 600 kpc tail and a powerful
Wide Angle Tail (WAT) source PKS~1610-608 (Jones \& McAdam 1996).

A luminous corona was found to be associated with the WAT source
PKS~1610-608 (cD galaxy ESO~137-006, Fig. 18). PKS~1610-608 is among the brightest
radio source in the nearby universe (4.4 times more luminous than M87
at 1.4 GHz) and its giant power is expected to have significant impacts
on the cluster medium. Nevertheless, the survival of this small corona
again indicates that small coronae can survive powerful AGN outbursts.
Similar to the radio sources associated with NGC~1265, NGC~4874 and NGC~3842,
the radio jets
of PKS~1610-608 significantly brighten after they transverse the dense
corona (Jones \& McAdam 1996). If we examine the coronal morphology in
detail (Fig. 18), the corona of ESO~137-006 is double-peaked.
The surface brightness and temperature profiles of the corona
are also derived (Fig. 18). The central 2$''$ (in diameter) region is
indeed abnormal. What causes the surface brightness decrement at the nucleus
is unknown.
If we fit the surface brightness profile of ESO~137-006's corona with a truncated
$\beta$ model as we did for NGC~3842 and NGC~1265 (S05; SJJ05), the derived
central gas electron density is 0.43 cm$^{-3}$, assuming an abundance of 0.8 solar.
Such a high gas density implies a central gas cooling time as short as 15 Myr.
The exact boundary of the corona is uncertain as the emission in the flat region
(2.5 kpc - 4.2 kpc, PSF corrected) may be largely contributed by the LMXB (Fig. 18).
The total gas mass is 1.0$\times10^{8}$ M$_{\odot}$ within 2.5 kpc radius, and
2.1$\times10^{8}$ M$_{\odot}$ within 4.2 kpc radius.

ESO~137-006 is $\sim 5.5'$ (or 106 kpc) west of A3627's X-ray peak and still
in the 10$'$ radius cluster core (at least in projection) revealed by \rosat\
(B\"ohringer et al. 1996). The radial ICM light profile within 100$''$ radius
of the galaxy is
flat (Fig. 18). The lack of ICM enhancement around ESO~137-006 implies
that its corona is subject to ICM stripping. We also examined the
temperature of the ICM around ESO~137-006. As A3627 is behind the Galactic
plane, the Galactic soft X-ray foreground is strong. This is supported by
measurements of the PSPC flux in the R4-R5 band of the RASS image around the
position of A3627: 3.1 - 5.1 $\times 10^{-4}$ cts s$^{-1}$ arcmin$^{-2}$,
3 - 5 times the typical flux of fields in the \chandra\ blank sky background
dataset (0.9 - 1.5 $\times 10^{-4}$).
This soft Galactic component indeed shows up in the ICM spectra that are
largely absorbed by Galactic absorption, as a single MEKAL component
plus the Galactic absorption ($N_{\rm H}$ = 2$\times10^{21}$ cm$^{-2}$)
underestimates the observed flux below 0.7 keV. This soft excess may be the
primary reason that the absorption from the analysis of the \asca\ data (Tamura
et al. 1998) is lower than the Galactic absorption, as Tamura et al. used
\asca\ blank sky background. There are no other \chandra\ pointings within
1 degree of A3627 and A3627 emission fills these three \chandra\ pointings.
Thus, we used the blank-sky background and added a 0.25 keV MEKAL component
(unabsorbed) to mimic the soft X-ray excess. The temperature of the soft component
comes from the study of the soft X-ray background with \chandra\ (Markevitch et al.
2003). We allowed the temperature of the soft X-ray background to change within
0.1 - 0.4 keV, and have counted the resulting error into the final error of
the ICM temperatures.
The derived temperature of the surrounding ICM is $\sim$ 6 keV (Fig. 18).

We have also detected a long and narrow X-ray tail
associated with a small starburst galaxy (ESO~137-001). The detailed
analysis of this source has been present in Sun et al. (2006).

\paragraph{Perseus} is a well-known cooling-core cluster. We have analyzed
all 18 observations of Perseus, with a total exposure of 1.15 Msec.
The detailed work on NGC~1265 was presented in SJJ05.
The deep exposure allows us to unambiguously detect some
faint coronae. However, as the deep exposure is on the bright cluster center
and many cluster galaxies are projected on the bright cluster core, the
upper limits are generally not very tight (Table 2).
IC~310 is another NAT radio source in Perseus (e.g., Sijbring \& de Bruyn 1998).
It is $>$ 50 times brighter than NGC~1265 in X-rays as the nuclear source of
IC~310 is luminous ($L_{\rm 0.5-10 keV}$=2.3$\times10^{42}$ ergs s$^{-1}$).
The \chandra\ 0.5 - 1.5 keV surface brightness
profile (Fig. 19) clearly reveals excess emission in the 2.5$'' - 12''$ annulus
bin, which is confirmed to be the coronal emission from the follow-up
spectral analysis. There are clumps
12$'' - 25''$ from the southwest of the galaxy in the 0.5 - 1.5 keV image
(3.4-$\sigma$ feature). These clumps may be the stripped gas of the corona
as IC~310 is moving towards the northeast.

\paragraph{A1367} is a dynamically young cluster with multiple merging
and infalling activity (e.g., Cortese et al. 2004).
Two A1367 pointings have been investigated in detail and galaxy coronae of
both early-type and late-type galaxies are discussed in Sun \& Murray (2002a),
S05 and Sun \& Vikhlinin (2005). In addition to these, we analyzed the
new 1/8 sub-array observation on the luminous radio galaxy 3C~264 (NGC~3862).
The X-ray nucleus of NGC~3862 is luminous ($L_{\rm 0.5-10 keV}$
= 2$\times10^{42}$ ergs s$^{-1}$). The 0.5 - 1.5 keV \chandra\ surface
brightness profile clearly reveals the excess soft emission between 1.5$''$
and 6$''$ from the nucleus, while the 2 - 6 keV X-ray emission is well
fitted by the PSF (Fig. 20). The subsequent spectral
analysis at that region confirms the coronal emission nature of the excess
and constrain its temperature at 0.65$^{+0.29}_{-0.09}$ keV.

\paragraph{Coma} is a well-known rich cluster. V01
discovered two luminous coronae associated with central dominant galaxies
NGC~4874 and NGC~4889. We re-analyzed all five central pointings of Coma.
Two additional off-center pointings were also analyzed.
Thermal diffuse emission is detected in two giant spiral galaxies
NGC~4921 and NGC~4911. Most of soft X-ray emission of NGC~4921 comes
from the bulge and an arm-like feature to the west (Fig. 21).
The HI depletion
was revealed in the northwest of the galaxy (Bravo-Alfaro et al. 2000),
which implies the motion of the galaxy to somewhere west.
Interestingly, enhanced star formation is also revealed in the west by the
\xmm\ OM UVW1 - V color image (Fig. 21) and the enhancement is just
outside of the X-ray arm-like feature. We consider this as an evidence
of induced star formation by ICM pressure (see $\S$5).
Unlike small coronae of early-type galaxies, X-ray emission in these
two spirals can be traced to the radii of 12 - 23 kpc. Spectra of these
two sources were extracted from circles with a radius of 0.5$'$ for
NGC~4911 and a radius of 0.9$'$ for NGC~4921. NGC~4911 also hosts a
bright nuclear source ($L_{\rm 0.5-10 keV}$ = 2$\times10^{41}$ ergs s$^{-1}$),
while the nuclear source of NGC~4921 is about 8 times fainter.

\paragraph{A2877} is a poor cluster. The cD galaxy IC~1633 is $\sim$ 40 kpc
to the south of the cluster centroid defined by the cluster X-ray emission
at large radii, which may imply relative motion of the galaxy to the
surrounding ICM. As shown by the 0.5 - 3 keV surface brightness profile
(Fig. 22), there is an X-ray extended source associated with
IC~1633. The X-ray source of IC~1633 seems distinctive of other coronae
by its size, $\sim$ 9 kpc, which is $\gsim$ 2 times the other coronae
in our sample (except NGC~7720 in A2634). Its temperature, $\sim$ 1.5 keV,
is also high. The IC~1633 source is enriched with heavy elements compared
to its surroundings. Iron abundance is 1.37$^{+0.51}_{-0.29}$ solar for
the IC~1633 source and 0.44$^{+0.14}_{-0.13}$ solar for the surrounding
ICM (within 70 kpc). The X-ray source of IC~1633 can be fitted by a $\beta$
model, r$_{0}$=2.3$\pm$0.2 kpc and $\beta$=0.91$\pm$0.06, with a cut-off
radius of 9.0$\pm$0.7 kpc. Assuming the same emissivity (from the spectral
best fit) for the gas within
9 kpc radius, the central electron density is 0.174 cm$^{-3}$ and the total
gas mass within 9 kpc is 1.0$\times10^{9}$ M$_{\odot}$. Thus, the IC~1633
X-ray source is larger and much more massive than all coronae in hot clusters
(except NGC~7720).
Its properties are similar to the central X-ray source in the fossil group
ESO~306-017 (Sun et al. 2004), which was suggested to be the relic of a
group cooling core. For completeness, we include this source in the sample.

\paragraph{A4038} is a poorly studied cluster without a very dense gas core
(n$_{\rm e}\sim$0.022 cm$^{-3}$ and t$_{\rm e}\sim$1.3 Gyr at the center).
An X-ray point source is detected at the center of the cD galaxy IC~5358, which
is only 20$''$ from the optical axis of the observation. Its spectrum is well
fitted by a power-law with a photon index of $\sim$ 1.9 plus a soft excess
with a temperature of $\sim$ 0.3 keV and a metallicity of zero. The soft-to-hard
flux ratio in the 0.5 - 2 keV band is about 0.6. As the properties of the soft
component are consistent with those of soft excess in AGN, we only put an upper
limit on the corona emission of IC~5358.

\paragraph{ZW~1615+35} is a poor and irregular cluster. The cluster X-ray emission
is elongated on the north-south direction and concentrated on two brightest
galaxies, NGC~6107 and NGC~6109 (Feretti et al. 1995). Both galaxies have
coronae. The 0.5 - 2 keV surface brightness profile of the NGC~6109's corona
(Fig. 23) shows a sharp boundary at r $\sim 7''$, which indicates the
confinement by the surrounding ICM.
Temperatures of its coronal gas were also derived in two radial bins.
The surrounding ICM was used as the background, and a $\beta$-model is fitted to
the surface brightness profile within 10$'' - 50''$ annulus to get the proper
estimate of local background inside 8$''$.

\paragraph{A2634} hosts the second luminous radio source 3C~465 (cD NGC~7720) in
this sample. The \chandra\ data
of 3C~465 were first discussed by Hardcastle et al. (2005).
We present here our own analysis, focusing on thermal coronae. A luminous X-ray
corona is associated with NGC~7720, and a central nuclear source is also
detected. The temperature and abundance of the NGC~7720 corona is 1.01$\pm$0.03 keV
and 1.08$^{+0.62}_{-0.18}$ solar respectively, from the fit to the global spectrum.
The surface brightness profile at the 0.5 - 2 keV band is
shown in Fig. 24. We fit the profile with a
$\beta$-model with a cut-off radius, plus a $\beta$-model for the surrounding
ICM emission. The best-fit parameters for the corona are: r$_{0}$=0.43$\pm$0.04 kpc,
$\beta$=0.58$\pm$0.01 and r$_{\rm cut}$=9.6$\pm$0.8 kpc. Thus, the NGC~7720 corona
is bigger than all other embedded coronae and has
similar properties as those of IC~1633's corona. If we subtract the nuclear and LMXB
emission (derived from the spectral fits) from the 0.5 - 2 keV surface brightness
profile, r$_{0}$=0.79$\pm$0.08 kpc, $\beta$=0.66$\pm$0.02, while r$_{\rm cut}$
is the same. Assuming the same emissivity for the X-ray gas of NGC~7720 and an abundance
of one solar, the
central electron density is 0.33 cm$^{-3}$ and the total X-ray gas mass
is 6.4$\times10^{8}$ M$_{\odot}$. The temperature profile of the corona
was also derived (Fig. 24). The temperature gradient inside the corona
is large. The ICM temperature is found to be $\sim$ 4.5 keV. There is an
extra absorption column associated with the
central AGN (1.6-6.0$\times10^{21}$ cm$^{-2}$).

\paragraph{A3571} has a dense and elongated X-ray gas core, which is also
very asymmetric with extension to the north. The gas core of A3571 is not as
dense as other cooling cores in this sample (n$_{\rm e} \sim$ 0.032 cm$^{-3}$).
The temperature of the gas core is high, 8.0$^{+1.0}_{-0.7}$ keV within the
central 8 kpc from the \chandra\ data, and there is no sign of gas cooling in
the gas core. It was suggested to be a very advanced merging cluster from its
radio properties (Venturi et al. 2002), which is also supported by the X-ray
morphology of the gas core. If that is true, the dense cluster cooling core
may have been heated and disrupted by the ongoing merger. For this reason,
we exclude the giant cD galaxy, ESO~383-076, from our analysis, as it may
have hosted a dense cluster cooling core prior to the merger.

\paragraph{A2107} is a relaxed cluster. The \chandra\ data have been
first analyzed by Fujita et al. (2006). The surface brightness profile of
A2107's gas core shows two components (Fig. 6 of Fujita et al. 2006),
but their separation is much smoother and subtler than what we observe
for IC~1633 and NGC~7720. The central component is too hot ($\sim$ 2.7 keV)
and too big ($\sim$ 18 kpc in radius), compared to all other coronae of BCGs in our sample.
It is more likely the central component of the cluster cooling core, rather
than the ISM only from the stellar mass loss from the cD. For this reason,
we exclude it in our analysis of embedded coronae.

\paragraph{A3376} is a merging cluster with a very elongated gas core running
from the west to the east. A 30 kpc X-ray tail of
the S0 galaxy PGC~018313 is detected to the east of the galaxy (Fig. 25),
just opposite to the moving direction of the cluster gas core.

\paragraph{A4059} is a cooling core cluster. A luminous X-ray corona
is detected from a large spiral, ESO~349-009, which is only 265 kpc
from the cD in the plane of sky (Fig. 26). Its corona is among the most luminous
for a late-type galaxy in our sample, along with UGC~6697 and ESO~137-001.
Most of the thermal X-ray emission is from the red bulge, while the soft
X-ray emission is also enhanced in the western arm with active star formation.
The 0.5 - 2 keV X-ray morphology may imply a motion of the galaxy to the west,
though the case is not as strong as those in UGC~6697 and ESO~137-001.
Without a significant X-ray nucleus, the hard X-ray component (measured by a
power law) is still luminous in ESO~349-009, $L_{\rm 2-10 keV}$ = 0.8 - 2.7
$\times10^{41}$ ergs s$^{-1}$, depending on the nature of three bright point
sources at the end of the south spiral arm. Interestingly, the southern
one of these three point sources was not detected in the \chandra\ observation
taken four years ago, which implies the large flux variation of the source.
In any sense, if we attribute the hard X-ray emission to emission from
high mass X-ray binaries (HMXB), the estimated SFR from
the $L_{\rm X}$ - SFR correlation by Grimm et al. (2003) is $>$ 12 M$_{\odot}$/yr,
which makes ESO~349-009 a starburst galaxy. Indeed, the galaxy is luminous
in the NUV band, as revealed by the \xmm\ OM image at the UVW1 band (Fig. 26).

\paragraph{A3558} is at the core of Shapley Supercluster. It was observed
as a cooling-flow cluster in cycle 2 with the GTO time (PI: Fabian).
However, the \chandra\
data show no temperature decline towards the center and the ICM remains
nearly isothermal ($\sim$ 5.9 keV) within 250 kpc from the X-ray peak.
As the exposure is only 11\% - 40\% of other clusters at $z$ = 0.04 - 0.05,
we restrict optically selected galaxies to $>$ 2 \Ls\ galaxies.
Two of three $>$ 2 \Ls\ galaxies in the field have coronae, including
the cD galaxy ESO~444-046 (Fig. 27). The corona of ESO~444-046 is especially
interesting, as it is located in a dense core (n$_{\rm e} \approx$
0.015 cm$^{-3}$ at the center) and the cooling time of the surrounding
ICM is shorter than the Hubble time ($\sim$ 4 Gyr). For all coronae known,
the ICM density around ESO~444-046's is the highest. The surface brightness
and temperature profiles around the cD galaxy are shown in Fig. 28. This
detection also brings an interesting question on the relation of these mini
cooling cores with large cluster cooling cores. The existence of small
corona of ESO~444-046 may imply that the cluster cooling core of A3558
has not formed so the corona has not yet mixed with the ICM.

\end{appendix}

\setlength{\voffset}{-20mm}
\begin{table}
\begin{scriptsize}
\begin{center}
\caption{Nearby hot clusters in the sample}
\vspace{6mm}

\begin{tabular}{ccccccccc}
\hline \hline
 Cluster & z\tablenotemark{a} & D\tablenotemark{b} & N$_{\rm H}$ & Area\tablenotemark{c} & ObsID & Exposure\tablenotemark{d} & kT\tablenotemark{e} & $\sigma_{\rm r}$\tablenotemark{f} \\
 & & (Mpc) & (10$^{21}$ cm$^{-2}$) & (deg$^{2}$) & & (ks) & (keV) & (km/s) \\
\hline

Centaurus & 0.0114 & 49.4 & 0.81 & 0.225 & 504, 505, 4190, 4191 & 31.7, 10.0, 34.3, 34.0 & 3.7 & 863 \\
          &        &      &      &       & 4954, 4955, 5310     & 87.0, 44.7, 49.3 &     &   \\
A1060 & 0.0126 & 54.6 & 0.49 & 0.093 & 2220 & 26.6 & 3.4 & 647 \\
A3627 & 0.0157 & 68.2 & 2.0 & 0.258 & 4956, 4957, 4958 & 14.5, 14.1, 14.1 & 5.6 & 897 \\
AWM7  & 0.0172 & 74.8 & 0.92 & 0.112 & 908  & 47.8 & 3.7 & 740 \\
Perseus & 0.0179 & 77.9 & 1.2-1.6 & 0.405 & 502, 503, 1513, 3209 & 5.3, 9.0, 22.2, 95.8 & 6.4 & 1324 \\
        &        &      &   &       & 3237, 4289, 4946, 4947 & 65.7, 95.4, 23.7, 29.8 & &  \\
        &        &      &   &       & 4948, 4949, 4950, 4951 & 111.6, 29.4, 76.8, 96.1 & & \\
        &        &      &   &       & 4952, 4953, 5597, 6139 & 149.4, 30.1, 25.2, 56.4 & & \\
        &        &      &   &       & 6145, 6146 & 85.0, 45.0 & & \\
3C129.1 & 0.0210 & 91.6 & 7.1 & 0.098 & 2218, 2219 & 26.6, 9.6 & 5.6 & - \\
A1367 & 0.0220 & 96.1 & 0.22 & 0.152 & 514, 4189, 4916 & 37.8, 43.2, 34.8 & 3.6 & 879 \\
Coma & 0.0231 & 100.9 & 0.092 & 0.341 & 555, 556, 1086, 1112 & 8.7, 9.6, 9.5, 9.7 & 8.2 & 1008 \\
     &        &       & &       & 1113, 1114, 2941, 4724 & 9.6, 9.1, 61.9, 59.7 & & \\
A2877 & 0.0247 & 108.1 & 0.21 & 0.093 & 4971 & 24.8 & 3.5 & 898 \\
MKW8  & 0.0270 & 118.3 & 0.28 & 0.093 & 4942 & 23.1 & 3.0 & 518 \\
Ophiuchus & 0.0291\tablenotemark{g} & 127.7 & 2.2 & 0.093 & 3200 & 47.1 & 10.2 & 1101\tablenotemark{g} \\
A4038 & 0.0300 & 131.8 & 0.16 & 0.093 & 4188, 4992 & 6.1, 33.5 & 3.2 & 882 \\
A2199 & 0.0302 & 132.7 & 0.087 & 0.151 & 497, 498 & 19.5, 18.9 & 4.3 & 733 \\
ZW1615+35 & 0.0310 & 136.3 & 0.14 & 0.068 & 3985, 3340, 3341 & 19.4, 4.8, 5.1 & 3.2 & 584 \\
A2634 & 0.0314 & 138.1 & 0.50 & 0.093 & 4816 & 49.5 & 3.5 & 886 \\
A496  & 0.0329 & 144.8 & 0.47 & 0.125 & 931, 3361, 4976 & 17.7 (FI), 8.2, 57.3 & 4.1 & 714 \\
A2147 & 0.0350 & 154.3 & 0.34 & 0.093 & 3211 & 17.9 & 4.3 & 821 \\
A2052 & 0.0355 & 156.6 & 0.28 & 0.093 & 890  & 36.8 & 3.1 & 751 \\
A576  & 0.0389 & 172.0 & 0.57 & 0.074 & 3289 & 27.6 & 3.8 & 977 \\
A3571 & 0.0391 & 172.9 & 0.33 & 0.074 & 4203 & 23.0 & 7.7 & 988 \\
A2107 & 0.0411 & 182.0 & 0.45 & 0.093 & 4960 & 35.6 & 3.9 & 672 \\
MKW3S & 0.0450 & 199.9 & 0.30 & 0.107 & 900  & 57.3 & 3.2 & 617 \\
A3376 & 0.0456 & 202.6 & 0.49 & 0.093 & 3202, 3450 & 44.3, 19.8 & 4.4 & 641 \\
A4059 & 0.0475 & 211.4 & 0.11 & 0.154 & 897, 5785 & 24.5 (BI), 38.3 (FI), 92.1 & 3.9 & 628 \\
A3558 & 0.0480 & 213.7 & 0.39 & -     & 1646 & 14.4 & 5.4 & 977 \\

\hline\hline
\end{tabular}
\tablenotetext{a}{Cluster redshifts from NED}
\tablenotetext{b}{Luminosity distance of the cluster derived from their redshift}
\tablenotetext{c}{Sky area covered by the \chandra\ pointing(s) and used in the analysis.
Some CCD chips far away from the optical axis are not included in the analysis
($\S$3.1). The A3558 pointing is not included as we only studied galaxies complete to
$>$ 2 \Ls. The total sky area covered is 3.3 deg$^{2}$.}
\tablenotetext{d}{Effective total exposure after excluding time intervals of strong background
flares. Small difference on the effective exposure of BI and FI chips are
not listed. The total effective exposure is 2.50 Ms out of a total 
observation time of 2.77 Ms from 68 pointings.}
\tablenotetext{e}{The average emission-weighted temperature from BAX,
which is generally within 10\% of the \chandra\ values.
Note that there is generally temperature variation across each cluster (cooling-core
or merging). The temperatures of MKW8, ZW1615+35 and A3571 are from our \chandra\
analysis, as the BAX's values are too uncertain or $\gsim$ 10\% different from
the \chandra\ results.}
\tablenotetext{f}{Cluster radial velocity dispersion from Struble \& Rood (1999),
Miller et al. 2002 (for ZW1615+35), Koranyi \& Geller 2002 (for AWM 7 \& MKW 8),
Pinkney et al. 1993 (for A2634), Mohr et al. 1996 (for A576) and Mahdavi
et al. 2000 (for MKW3S). There are only redshift measurements on the brightest
two galaxies in 3C129.1, so the velocity dispersion is unknown.}
\tablenotetext{g}{There are 27 cluster galaxies within 80$'$ (about the virial radius)
of the cD galaxy from NED. We derive an average velocity of 8731$\pm277$ and a
velocity dispersion of 1101$\pm$191 km/s.}
\end{center}
\end{scriptsize}
\end{table}
\clearpage

Table 2 (void); Table 3 (void) \\
\\
Please check the complete version at: 
http://www.pa.msu.edu/\~{}sunm/coronae\_all\_v1.5\_emuapj.ps.gz

\clearpage

\begin{figure}
\vspace{-1.5cm}
 \centerline{\includegraphics[height=0.6\linewidth]{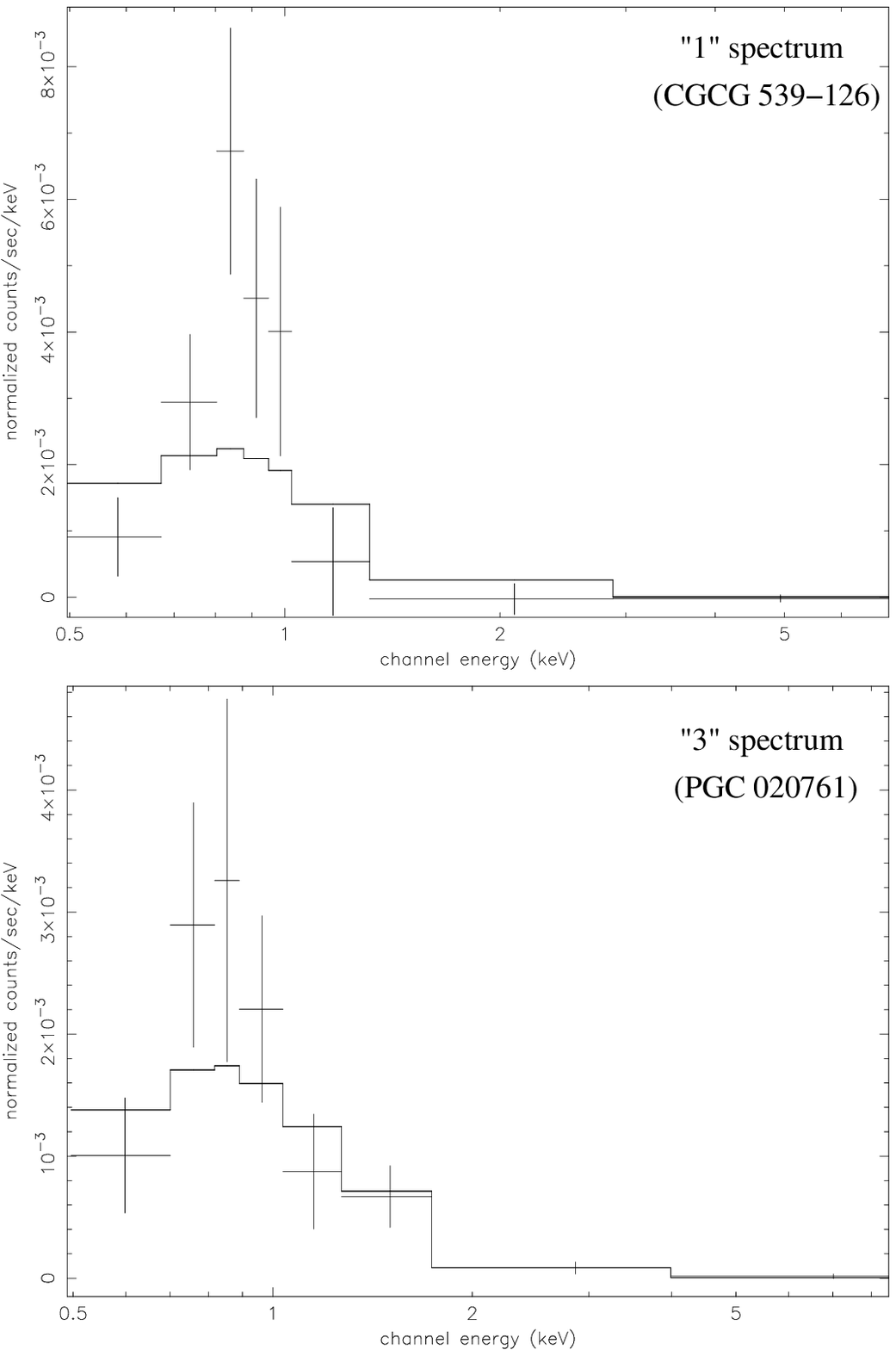}}
\caption{Representative spectra of a corona identified by the criterion
1 (see $\S$3.2) in AWM7 and a soft X-ray source identified by the criterion 3 in A576.
The null hypothesis models are also plotted (a power law). The power index
in PGC~020761 is 3.22$^{+0.42}_{-0.40}$.
}
\end{figure}

\begin{figure}
\vspace{-3.3cm}
 \centerline{\includegraphics[height=0.65\linewidth,angle=270]{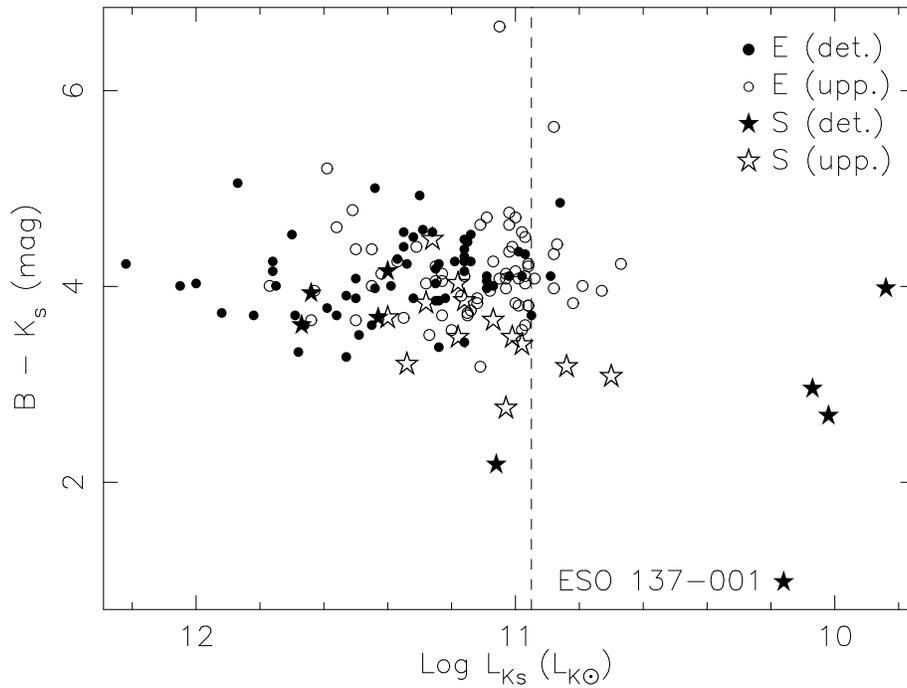}}
\vspace{-0.1cm}
\caption{The color-luminosity relation for the galaxies (with and without
detections of coronae) in our sample. The soft X-ray sources identified by
the criterion 3 are included in the detections. The
$B - K_{\rm s}$ color of early-type galaxies is around 4 with small dispersion
(i.e. the red sequence). Note that only cluster galaxies more luminous
than $L_{\rm Ks, cut}$ (the dashed line) are complete in our sample.
ESO~137-001, a starburst galaxy with a long X-ray tail (Sun et al. 2006),
is marked. It is now completely off the red sequence and is located in the
so-called ``blue cloud''.
}
\end{figure}
\newpage

\begin{figure}
\vspace{-2.5cm} 
 \centerline{\includegraphics[height=0.9\linewidth,angle=270]{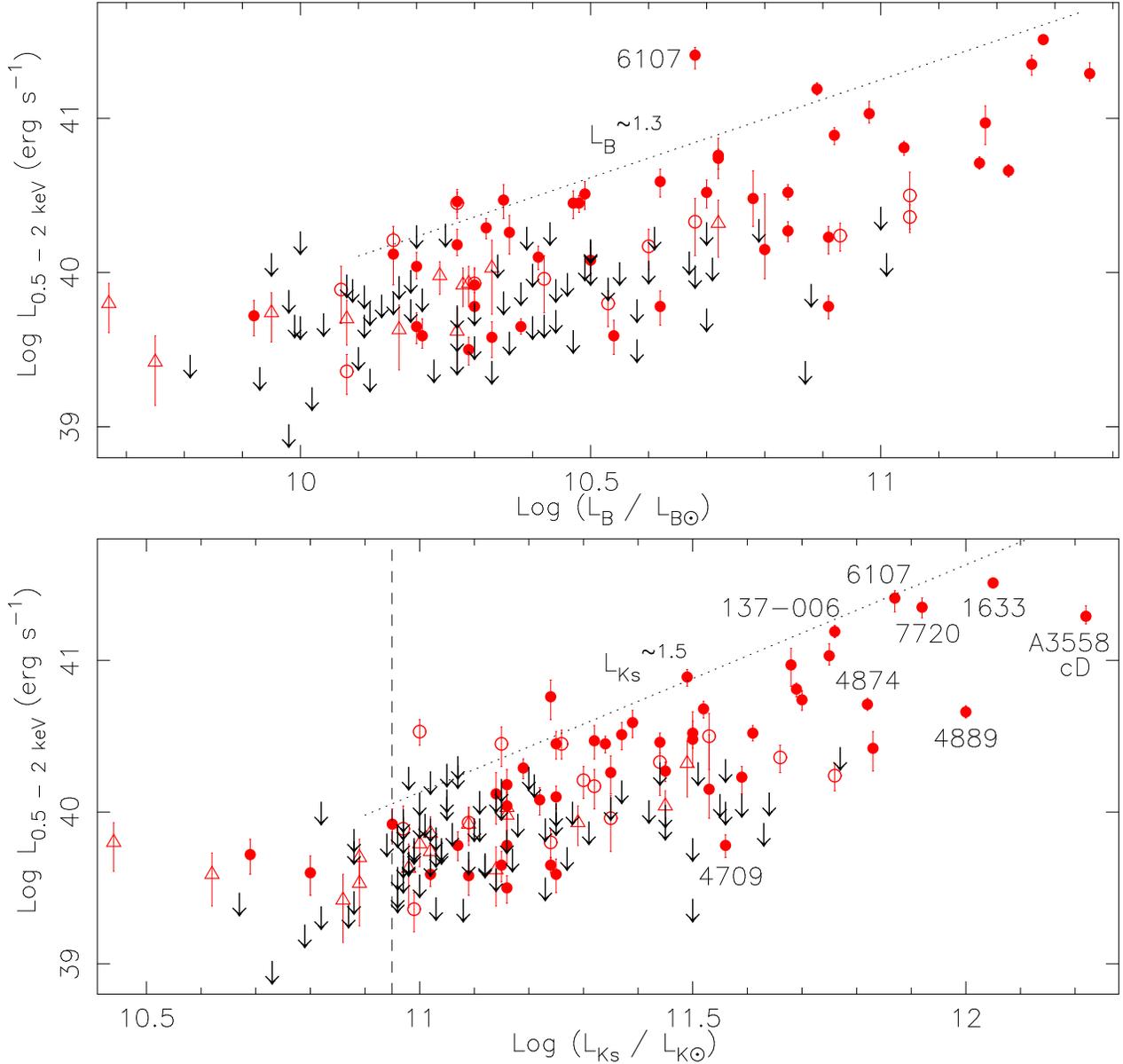}}
\vspace{-0.1cm}
\caption{The stellar luminosity $L_{\rm B}$ (upper) or $L_{\rm Ks}$ (lower)
vs. the rest-frame 0.5 - 2 keV thermal luminosity of the coronae ($L_{\rm 0.5 - 2 keV}$)
for 157 early-type galaxies in 25 nearby hot clusters from our work.
The detections are in red, while the upper limits are in black.
The data points with filled circles are coronae identified by the criteria 1 or 2.
The data points with empty circles are soft X-ray sources individually identified
by the criterion 3 (see $\S$3, hereafter in Fig. 4 - 8), while the data points
with empty triangles are soft X-ray sources identified by the criterion 3 from
stacked spectra. Fits to all detections
for $L_{\rm B} > 10^{10.18} L_{\rm B\odot}$ or $L_{\rm Ks} > L_{\rm Ks, cut}$
(the vertical line) are shown, with the normalizations increased to delineate
the approximate envelopes in each relation (see $\S$4.1 for more detail).
Some interesting galaxies are marked (prefix NGC, IC or ESO omitted).
}
\end{figure}
\clearpage

\begin{figure}
\vspace{-1cm}
 \centerline{\includegraphics[height=0.75\linewidth,angle=270]{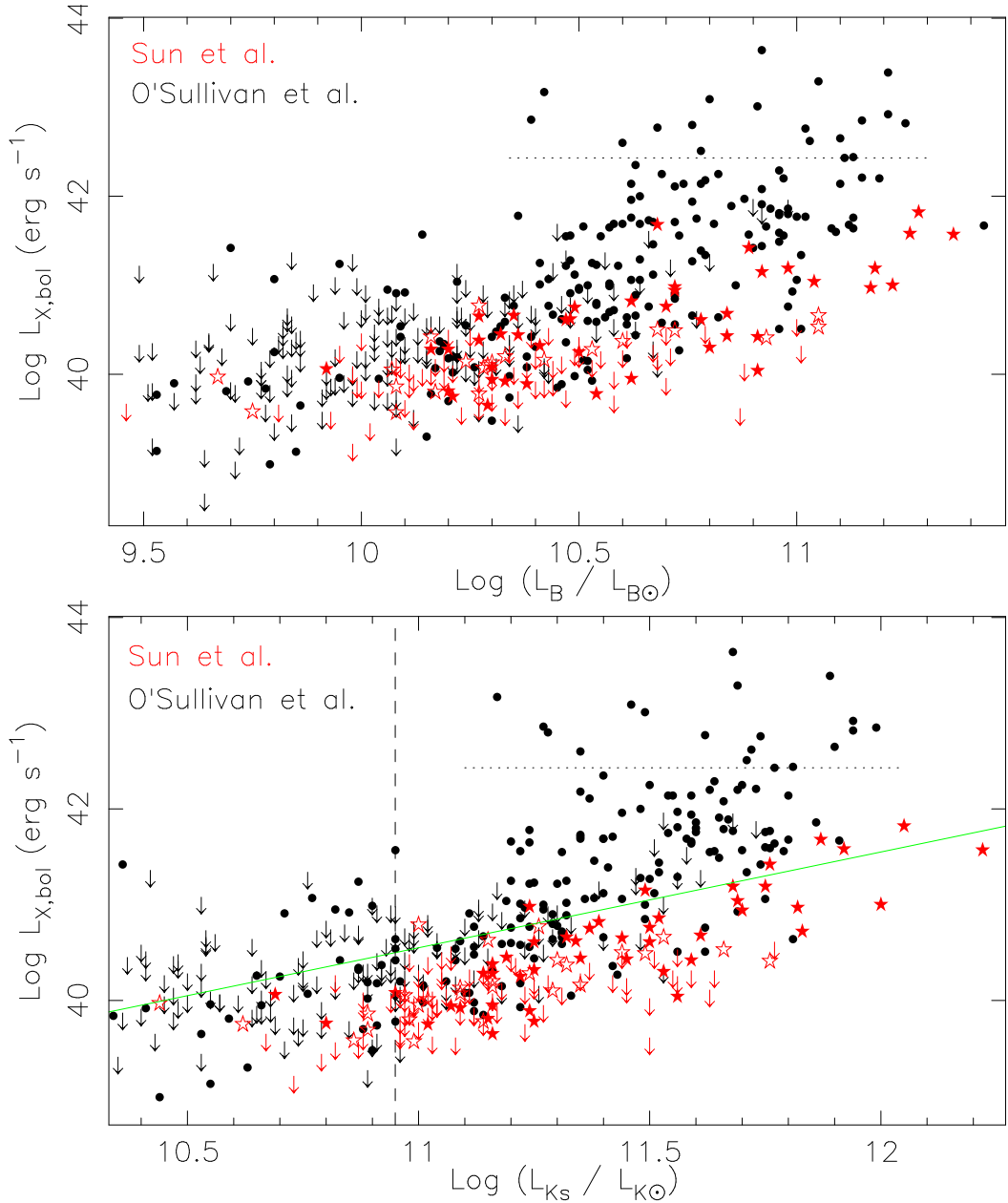}}
\vspace{-0.1cm}
\caption{The stellar luminosity $L_{\rm B}$ (upper) and $L_{\rm Ks}$ (lower)
- the X-ray bolometric luminosity of the thermal coronae ($L_{\rm X, bol}$) for
157 early-type galaxies in 25 nearby hot clusters from our work (red), compared
with the relation for the total X-ray luminosity of galaxies (black) in
poorer environments based on \rosat\ data (O'Sullivan et al. 2001;
Ellis \& O'Sullivan 2006). The data points with filled stars are coronae
identified by the criteria 1 or 2, while those with empty stars are soft
X-ray sources identified by the criterion 3.
The vertical dashed line marks the position of
$L_{\rm Ks, cut}$. The \rosat\ luminosities include emission from AGN,
LMXB and the surrounding ICM for some galaxies, while the \chandra\
luminosities for galaxies in our sample are only from the thermal gas.
The horizontal dotted line marks a representative luminosity of a corona
(2.7$\times10^{42}$ ergs s$^{-1}$) that is close to the maximal value
in the field or poor environments (see $\S$4.1). We have confirmed that
any points above the line in the O'Sullivan et al. sample are either
cluster cooling cores or AGN. The green line represents the energy release
from SNe if the SN kinetic energy can be fully coupled into the coronal gas
(assuming an enclosed stellar mass of 30\% of the total mass, see $\S$7.5).
}
\end{figure}
\clearpage

\setlength{\voffset}{-20mm}
\begin{figure}
\vspace{0.1cm}
 \centerline{\includegraphics[height=0.75\linewidth,angle=270]{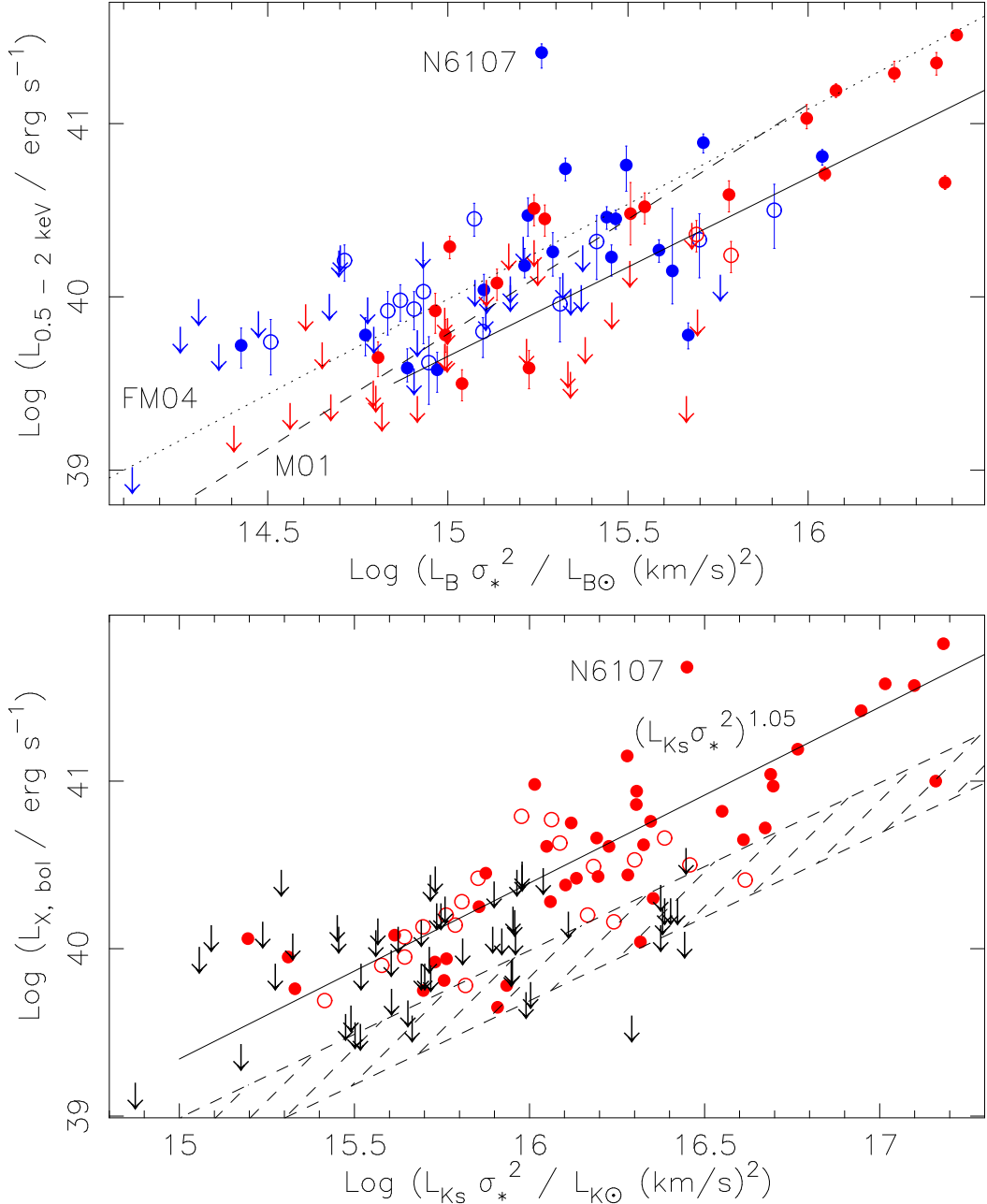}}
\caption{{\bf Upper}: $L_{\rm B} \sigma_{\ast}^{2}$ vs. $L_{\rm 0.5 - 2 keV}$
for early-type galaxies in our sample. The blue data points represent galaxies in
low $\sigma_{\rm clu}$ clusters ($<$ 880 km/s in this work), while the
red data points represent galaxies in high $\sigma_{\rm clu}$ clusters
($>$ 880 km/s). The data points with filled circles are coronae identified by
the criteria 1 or 2, while those with empty circles are soft X-ray sources
identified by the criterion 3.
The dotted line is the best fit from Finoguenov \& Miniati (2004)
who claimed a positive effect of environments on the X-ray luminosities of
coronae based on the \xmm\ observations of Coma (only detections), while
the dashed line is the best fit from Matsushita (2001) for galaxies in poorer
environments based on \rosat\ data. The solid line is the fit to
all our data (detections + upper limits) at $L_{\rm B} \sigma_{\ast}^{2} >$
10$^{14.5} L_{\rm B\odot}$ (km/s)$^{2}$ (see $\S$4.2).
Our results suggest a negative effect of environments on the X-ray
luminosities of coronae, but a more reliable sample of coronae in poor
environments is required for a better comparison with our results.
{\bf Lower}: $L_{\rm Ks} \sigma_{\ast}^{2}$ vs. $L_{\rm X, bol}$ for cluster
galaxies in our sample (detections in red and upper limits in black).
Our results again show that most coronae of optical
luminous galaxies with a deep potential can survive in hot clusters.
The solid line is the best fit to all detections. If soft sources
identified by the criterion 3 are removed, the fit is identical on the plot.
As shown by the shaded area, the expected total energy release rate by the
stellar mass loss in embedded coronae is on average 3.5 times smaller
than the X-ray luminosities of coronae ($\S$4.2). An ``outlier'' is marked
(NGC~6107), which also has the smallest $\beta_{\rm spec}$ (Fig. 8).
Empty data points have the same meaning as in Fig. 3.
}
\end{figure}
\clearpage
\setlength{\voffset}{0mm}
\begin{figure}
 \centerline{\includegraphics[height=0.95\linewidth,angle=270]{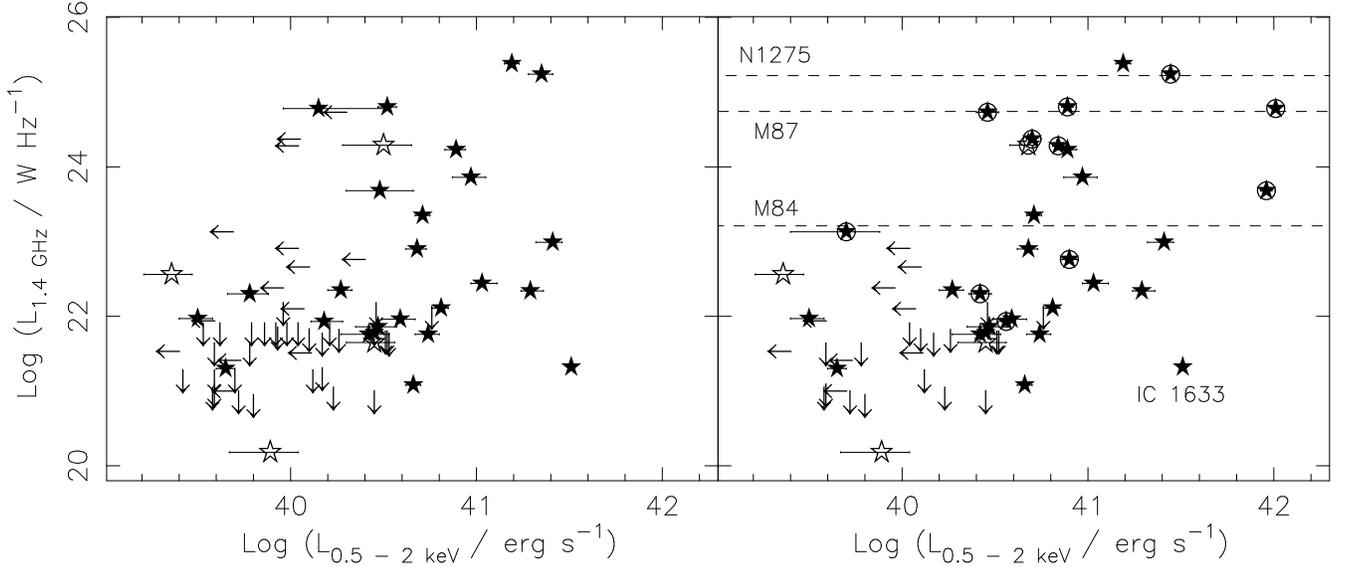}}
\caption{
{\bf Left}: The 0.5 - 2 keV luminosity of the corona vs. the 1.4 GHz luminosity
of the host galaxy.
{\bf Right}: The 0.5 - 2 keV luminosity of the corona (+ nucleus) vs. the 1.4 GHz
luminosity of the host galaxy. The points in a circle are
galaxies with a bright X-ray nucleus ($> 10^{40.4}$ ergs s$^{-1}$),
while the nuclei in other galaxies are not detected or faint.
A correlation between $L_{\rm X}$ and $L_{\rm radio}$ is present, which
may imply a connection between coronal gas cooling and the activity of the
central SMBH. At least ten of sixteen radio luminous galaxies
($> 10^{22.8}$ W Hz$^{-1}$ at 1.4 GHz) are detected to have X-ray coronae, while
five others have X-ray detection. For comparison, the 1.4 GHz luminosities
of Perseus, M87 and M84 are also shown.
}
\end{figure}

\begin{figure}
\vspace{-1.5cm}
 \centerline{\includegraphics[height=0.6\linewidth,angle=270]{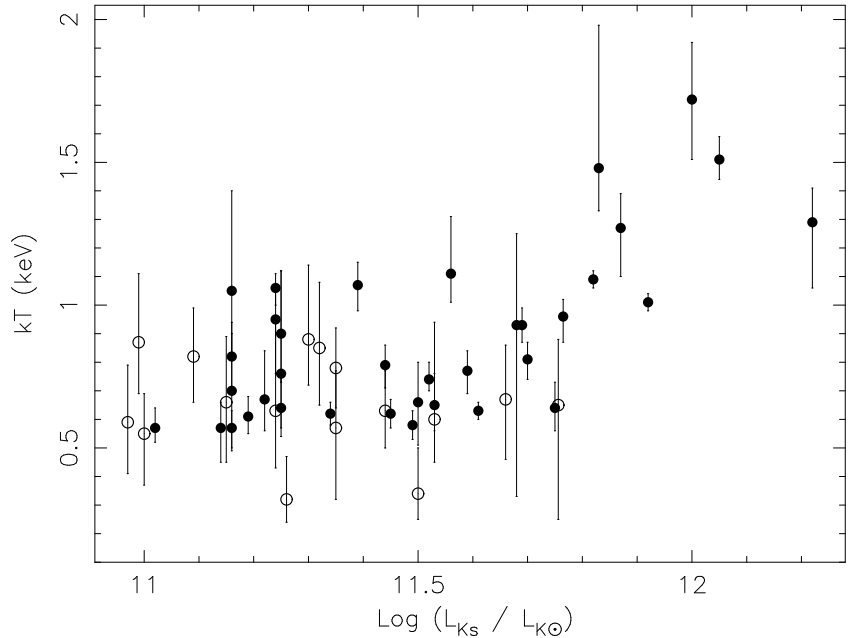}}
\caption{Temperatures of coronae vs. $L_{\rm Ks}$ of the host galaxy.
Temperature values from stacked spectra are not included. 
The coronal temperatures of $< 10^{11.8} L_{\rm K\odot}$ galaxies
are generally 0.4 - 1.1 keV and are not correlated with L$_{\rm Ks}$,
but coronae of very massive galaxies ($> 10^{11.8} L_{\rm K\odot}$,
all cDs) are systematically hotter (1.0 - 1.8 keV). Empty data points
have the same meaning as in Fig. 3.
}
\end{figure}
\clearpage

\begin{figure}
 \centerline{\includegraphics[height=0.6\linewidth,angle=270]{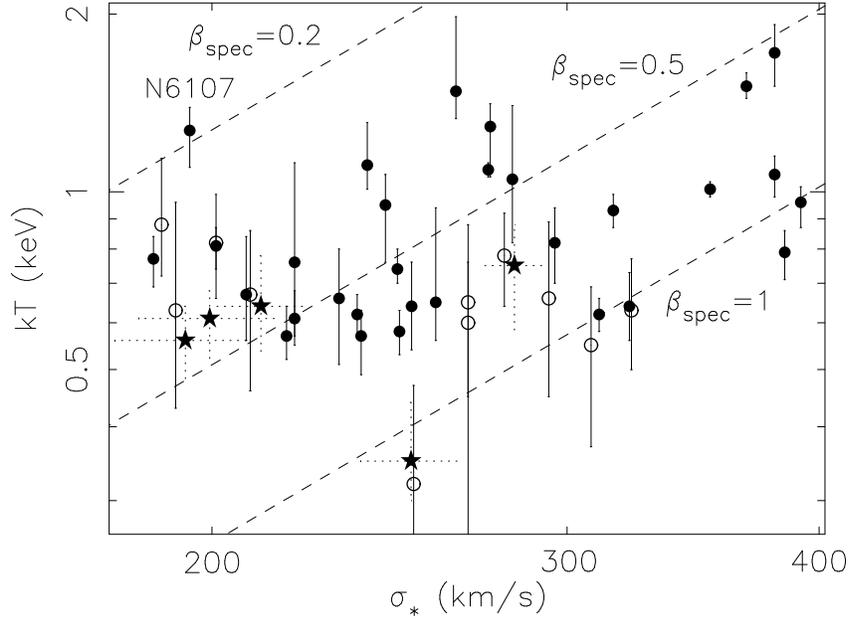}}
\caption{Temperatures of coronae vs. the central velocity dispersion of the stars.
Temperature values from stacked spectra with a significant iron
L-shell hump are also included (stars with dotted lines), as stacked galaxies
have similar stellar velocity dispersion.
The corresponding $\beta_{\rm spec}$ is between 0.2 and 1.1, which implies
that the coronal gas is almost always hotter than stars of the host galaxy.
The corona with the smallest $\beta_{\rm spec}$ is marked (NGC~6107).
Empty data points have the same meaning as in Fig. 3.
}
\end{figure}

\begin{figure}
\vspace{-0.3cm}  
 \centerline{\includegraphics[scale=0.68,angle=270]{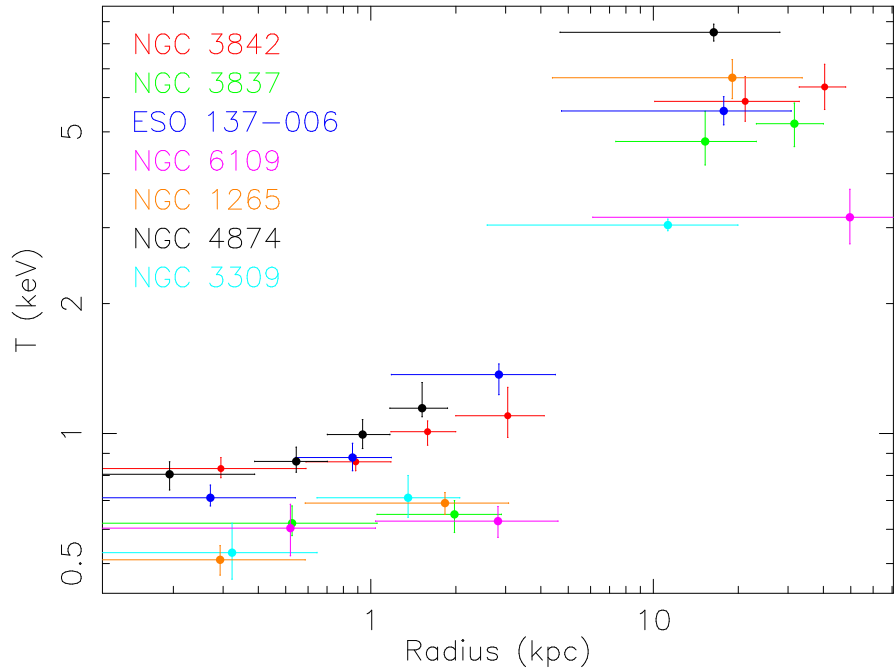}}
\caption{The temperature profiles of seven luminous galaxy coronae and the
surrounding hot ICM. The temperature profiles of two bigger ones, IC~1633 and NGC~7720,
are shown in Fig. 22 and 24.
}
\end{figure}
\clearpage

\begin{figure}
\vspace{-2.5cm}
 \centerline{\includegraphics[scale=0.75,angle=270]{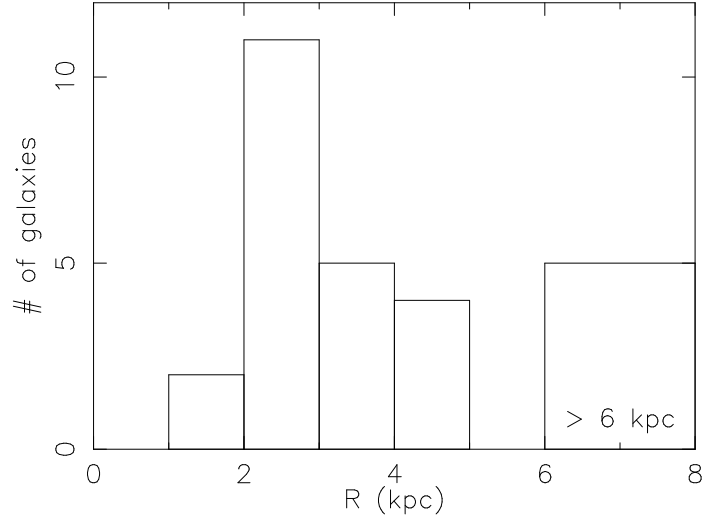}}
\caption{The histogram of the radius of the resolved embedded coronae
(see $\S$4.7 for detail). Besides these 27 coronae, we can also put upper
limits on the sizes of other 32 unresolved coronae or soft X-ray sources.
They are all smaller than 4.3 kpc in radius.
}
\end{figure}

\begin{figure}
\vspace{-5cm}
 \centerline{\includegraphics[height=0.55\linewidth,angle=270]{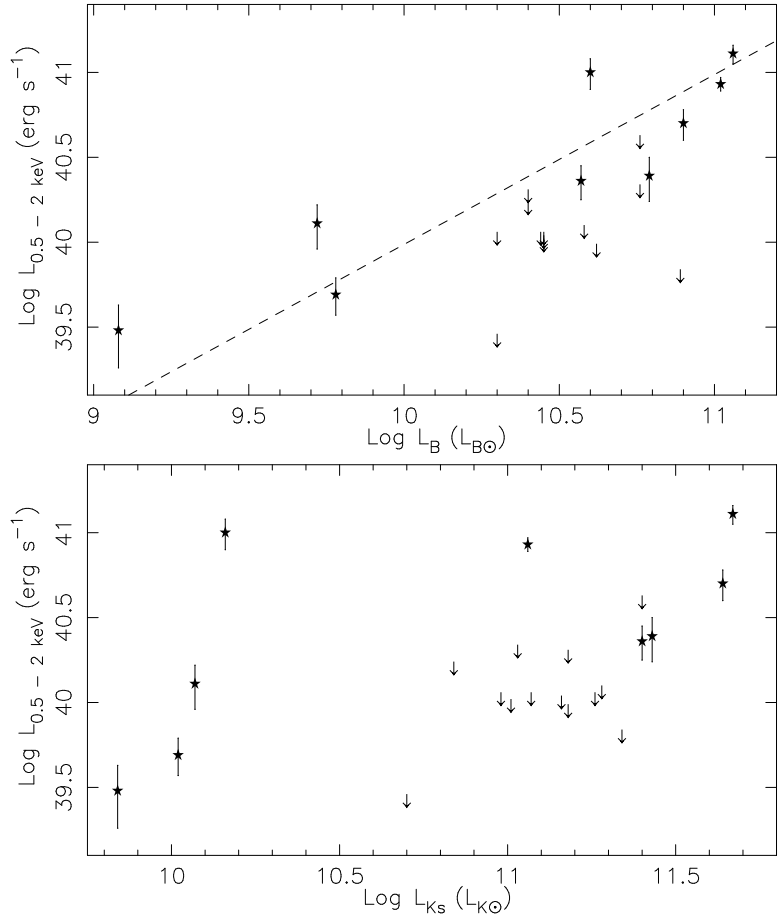}}
\caption{The stellar luminosity $L_{\rm B}$ (upper) and $L_{\rm Ks}$ (lower)
- the X-ray coronal gas luminosity $L_{\rm 0.5 - 2 keV}$ relations of 22
late-type galaxies in our sample. A better correlation is found in the
$L_{\rm B} - L_{\rm X, bol}$ plot than in the $L_{\rm Ks} - L_{\rm X, bol}$
plot, although there are only 9 detections. The dash line represents
$L_{\rm X, bol} \propto L_{\rm B}$.
}
\end{figure}
\clearpage

\begin{figure}
\vspace{-6cm}
 \centerline{\includegraphics[height=1.0\linewidth]{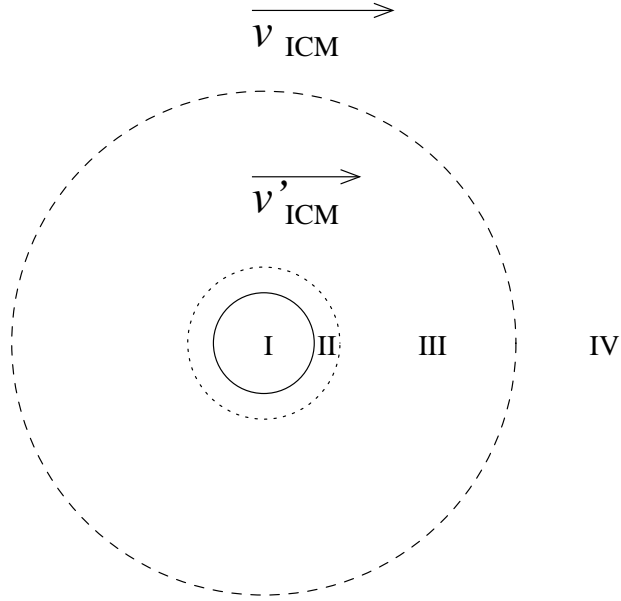}}
\vspace{-4.5cm}
\caption{A cartoon of an embedded corona in the frame of the corona.
Region I represents a cool corona with a 3 kpc radius.
Region II is a boundary layer separating the ICM and the ISM, where the
speed of the gas decreases to zero at the inner boundary of II.
Region III is still within the optical galaxy (15 kpc for this example)
and the ICM wind is reduced by the stellar mass loss there. Region IV is
the free ICM wind. The transition from III to IV is gradual
as the stellar light decreases outwards.
}
\end{figure}

\begin{figure}
\vspace{-3.2cm}
 \centerline{\includegraphics[height=0.6\linewidth,angle=270]{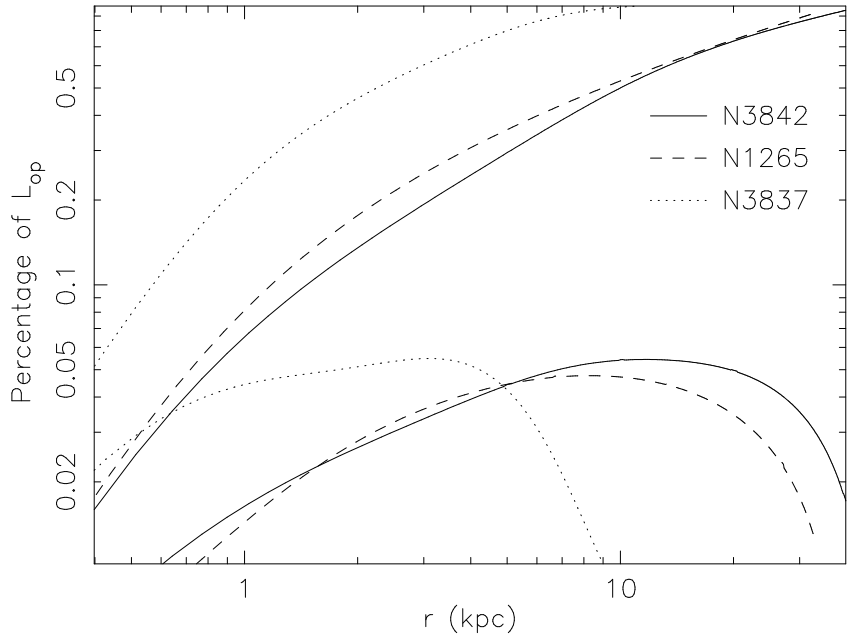}}
\caption{The fractions of stellar light in regions of interest for three
galaxy examples. A sphere with
a radius of r is assumed to be centered at the galactic nucleus. The lower curves
represent the fraction of the stellar light (relative to the total luminosity)
ahead of the sphere and within a cross section of $\pi$r$^{2}$ as a function
of r. For the coronae of these three galaxies,
the fraction is 4\% - 5\%. The upper curves represent the lower ones adding
the fraction of stellar light within the sphere with a radius of r. For coronae
of NGC~3842 (a radius of $\sim$ 4.1 kpc) and NGC~1265 (a radius of $\sim$ 2.3 kpc),
the fractions are 20\% - 25\%, while the fraction is $\sim$ 50\% for the corona
of NGC~3837 (a radius of $\sim$ 2.5 kpc).
} 
\end{figure}
\clearpage

\begin{figure}
\vspace{-2.4cm}
 \centerline{\includegraphics[height=0.6\linewidth,angle=270]{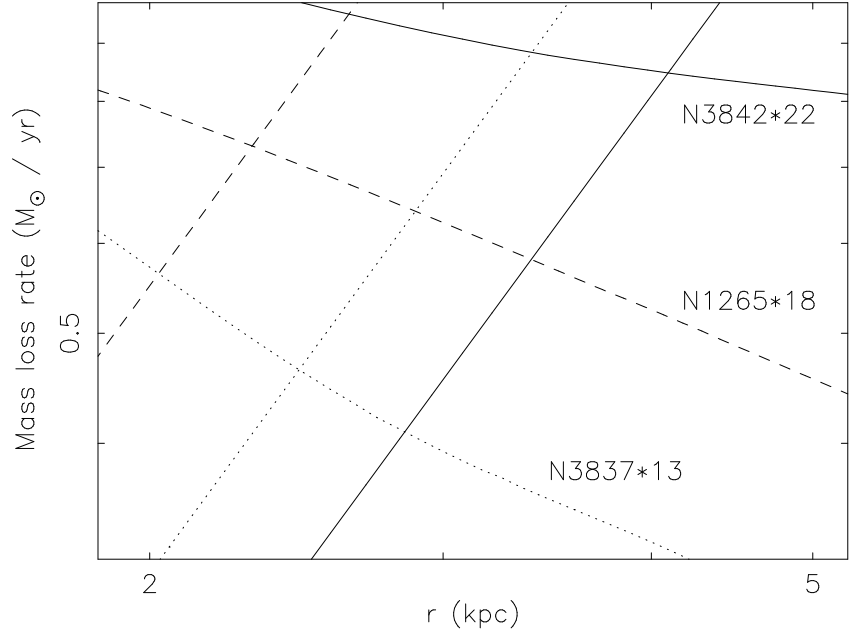}}
\caption{The predicted mass loss rate by stripping from equ. 5 (lines
decreasing towards the center, $\dot{M}_{\rm strip} \propto r^{2}$) vs.
the rescaled stellar mass loss rate in a boundary layer with a width of
0.5 kpc for three examples of coronae (solid lines: NGC~3842; dotted lines:
NGC~3837; dashed lines: NGC~1265; see $\S$7.1). The scaling factors are
chosen to balance the stellar mass loss rate with the stripping rate
at the current boundaries of these coronae. A suppression factor of
13 - 22 on stripping is required. When the size of a corona is reduced
by stripping, the mass loss rate by stripping decreases quickly, while
the stellar mass loss rate in the boundary layer increases in the
regions of interest. Therefore, their balance can always be achieved
as long as stripping has been suppressed significantly from equ. 5
(e.g., over ten times).
}
\end{figure}

\begin{figure}
\vspace{-4.4cm}
 \centerline{\includegraphics[height=0.6\linewidth,angle=270]{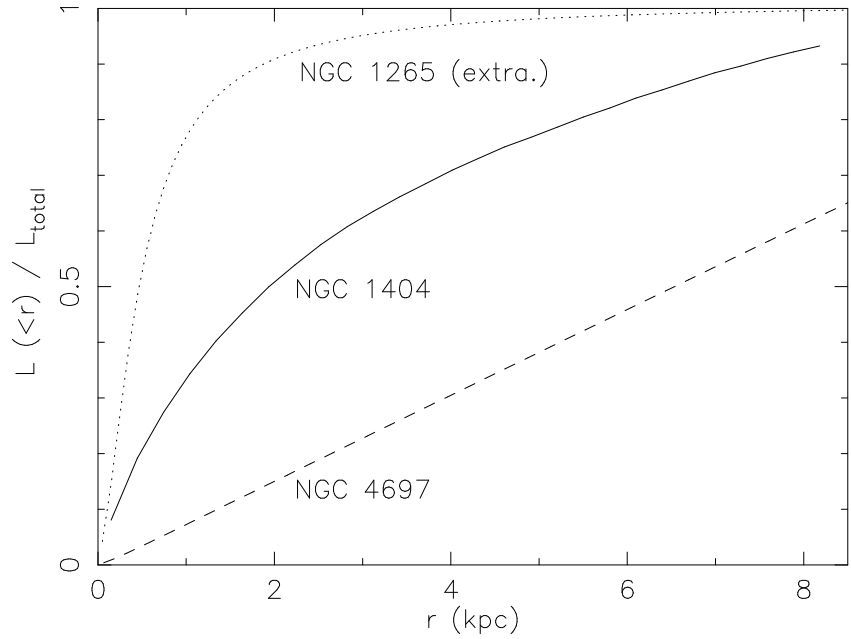}}
\caption{The enclosed X-ray luminosity as a function of radius for two
representative coronae in poor environments. For NGC~4697, we used the
properties derived by Sarazin et al. (2001): r$_{0}$=0.16 kpc, $\beta$=0.335
and a coronal size of 13 kpc. Most X-ray light is on the outskirts. For
NGC~1404 in the 1.5 keV Fornax cluster, we derived r$_{0}$=0.31 kpc and
$\beta$=0.481. The X-ray light of NGC~1404 is much more concentrated than
that of NGC~4697. As a comparison, we also show an example of embedded
coronae with small r$_{0}$ and large $\beta$ by extrapolating NGC~1265's
profile (within $\sim$ 2 kpc, SJJ05) to 10 kpc (r$_{0}$=0.4 kpc and $\beta$=0.73).
}
\end{figure}
\clearpage

\begin{figure}
\vspace{-4cm}
 \centerline{\includegraphics[height=0.95\linewidth,angle=270]{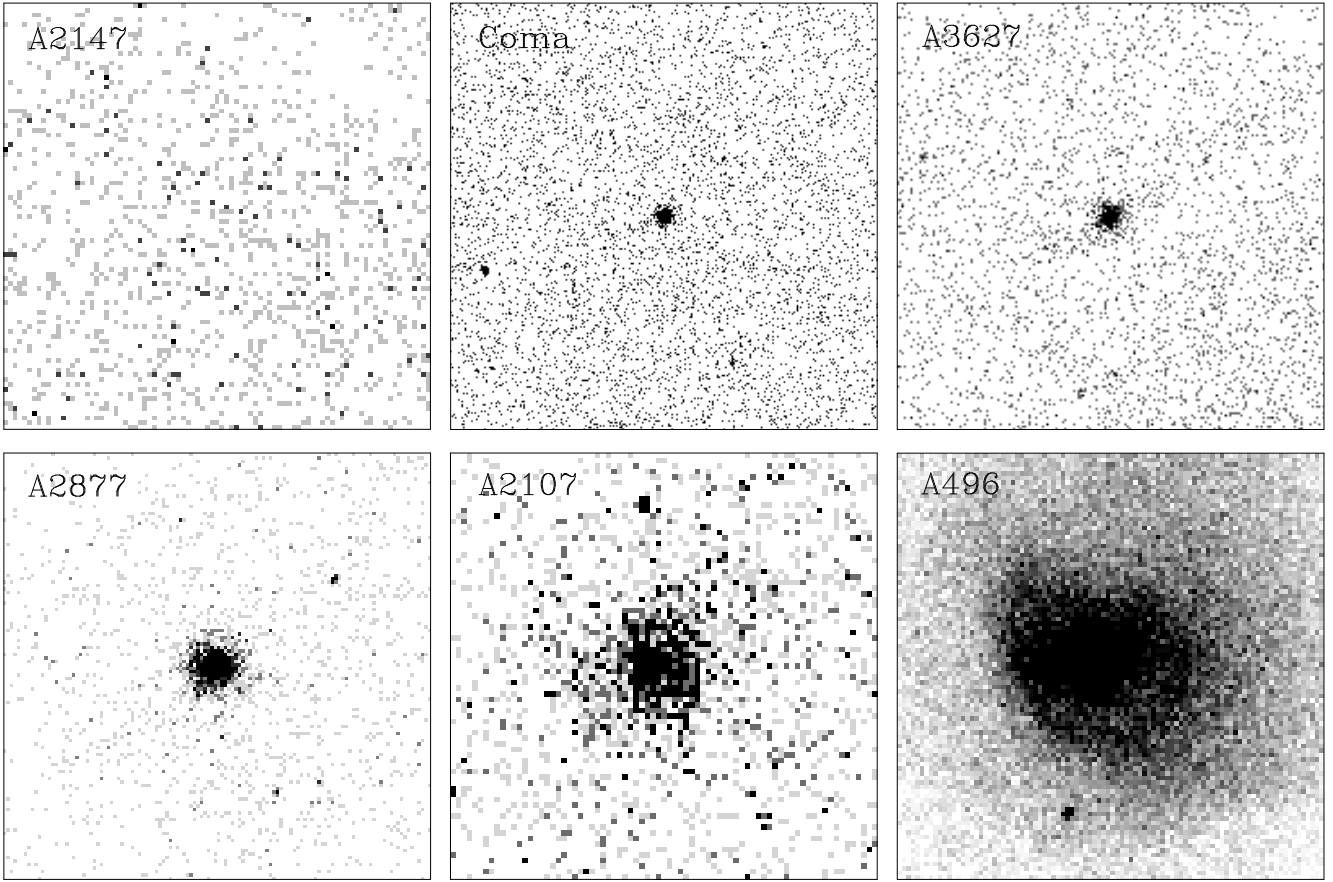}}
\vspace{0.2cm}
\caption{Different appearances of cD galaxies in X-rays. A2147: cD galaxy
UGC~10143 embedded in the 4.7 keV ICM, without a corona
($L_{\rm 1.4GHz}=4.6\times10^{22}$ W Hz$^{-1}$); Coma: an 1 keV corona of NGC~4874
(2 kpc radius) embedded in the 8 - 9 keV ICM
($L_{\rm 1.4GHz}=2.2\times10^{23}$ W Hz$^{-1}$); A3627: an 1 keV corona of
ESO~137-006 (3-4 kpc radius) embedded in the 6 keV ICM
($L_{\rm 1.4GHz}=2.4\times10^{25}$ W Hz$^{-1}$); A2877: an 1.5 keV X-ray
source of IC~1633 (9 kpc radius) embedded in the 3.6 keV ICM
($L_{\rm 1.4GHz}=2.1\times10^{21}$ W Hz$^{-1}$); A2107: a 2.7 keV
X-ray source of UGC~0995 ($\sim$ 18 kpc radius) embedded in the 4 keV ICM
($L_{\rm 1.4GHz}<5.9\times10^{21}$ W Hz$^{-1}$);
A496: a cluster cooling core with complicated internal
gas motion centered on the cD galaxy GIN~189
($L_{\rm 1.4GHz}=3.0\times10^{23}$ W Hz$^{-1}$). Each box is in the same physical
scale, 60 kpc $\times$ 60 kpc, while the nucleus of the cD galaxy is at the center.
}
\end{figure}
\clearpage

\begin{figure}
\vspace{-7cm}
 \centerline{\includegraphics[scale=0.45]{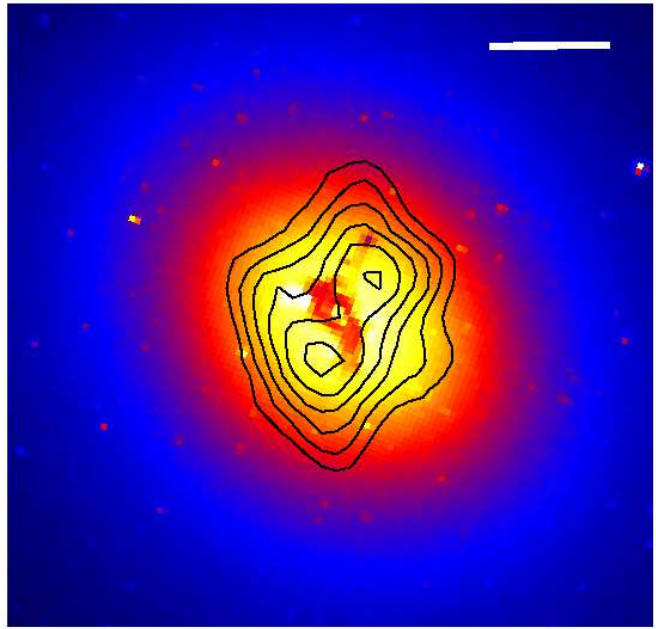}}
\vspace{-0.5cm}
\caption{The \chandra\ contours of the NGC~3311 corona (in linear scale)
superposed on
the \hst\ F555W image of the nuclear region of NGC~3311. The NGC~3311
corona appears disturbed while a dust filament lies between the two X-ray
peaks, implying a nature of multi-phase for NGC~3311's embedded ISM.
The scalebar is 4 arcsec (or 1.03 kpc).
}
\end{figure}

\begin{figure}
\vspace{-7cm}
 \centerline{\includegraphics[scale=0.8]{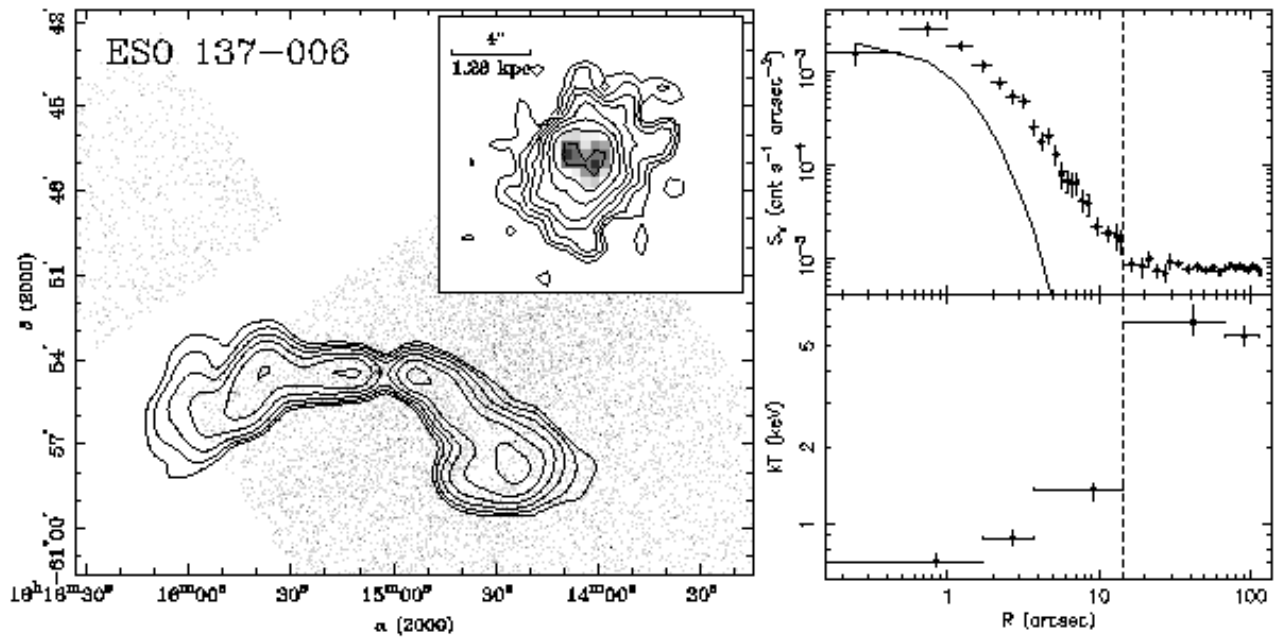}}
\caption{{\bf Left:} The 843 MHz contours of PKS~1610-608 (from SUMMS)
superposed on ESO~137-006's \chandra\ image. PKS~1610-608 is a powerful
radio source associated with ESO~137-006. The zoom-in contours of the
ESO~137-006 corona (in square-root scale) is also shown. The corona is 
double-peaked.
{\bf Right:} 0.5 - 2 keV surface brightness profile (upper) and temperature
profile (lower) around ESO~137-006 (1$''$ = 0.32 kpc). The 1 keV \chandra\
local PSF is also shown (the solid curve). Beyond the central 14$''$, the
ICM emission is uniform with a temperature of $\sim$ 6 keV. The corona
may have been disturbed within the central 0.2 kpc. The coronal emission is
truncated at 4.2$\pm$0.2 kpc if the local PSF is included in the fits to
the 0.5 - 2 keV profile. For the temperature profile, the ICM background
on the inner three annuli has been subtracted.
}
\end{figure}
\clearpage

\begin{figure}
\vspace{-4cm}
 \centerline{\includegraphics[scale=0.8]{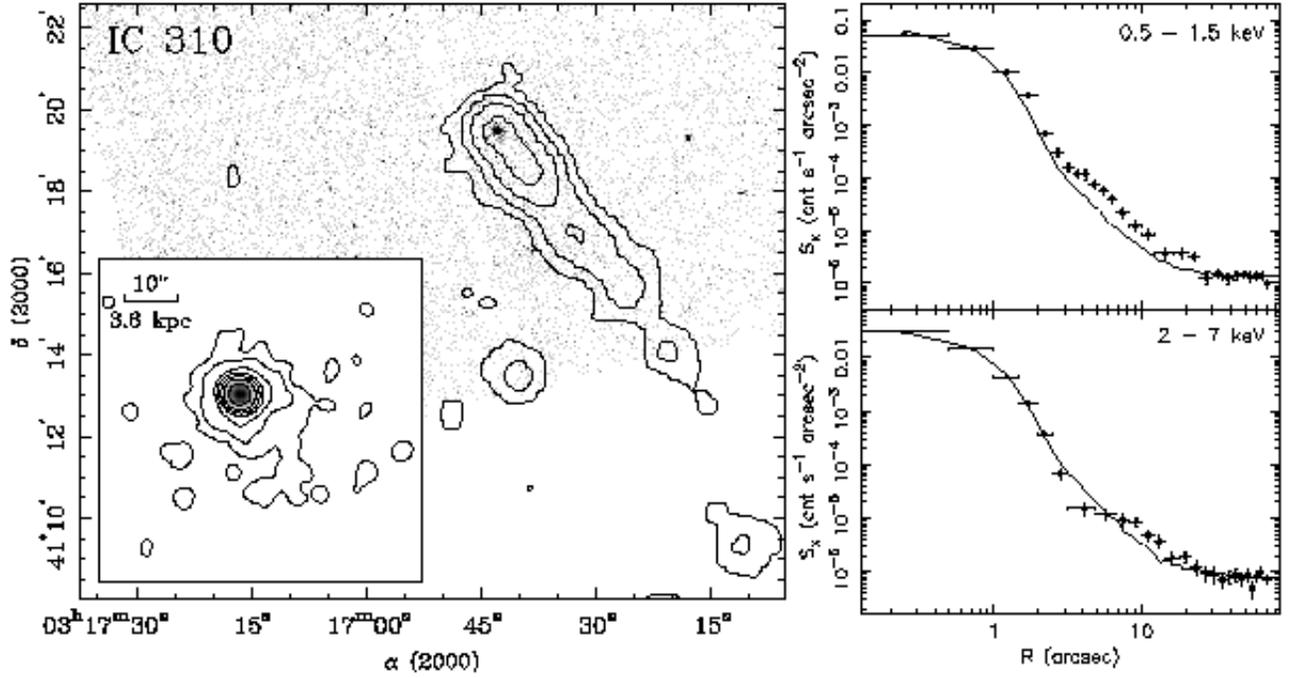}}
\caption{{\bf Left:} The 1.4 GHz contours of IC~310 (from NVSS) superposed on
its \chandra\ image. The zoom-in of the \chandra\ contours (in square-root scale)
reveal a small X-ray tail to the southwest at a significance level of 3.4 $\sigma$.
{\bf Right:} Surface brightness profiles of IC~310 in the soft and hard bands,
with PSFs (plus a constant local background) shown as the solid lines.
While the model (PSF+background) fits the average level of the hard X-ray
profile well, an excess between 2.5$''$ and 12$''$ of the soft X-ray
profile is clearly observed (1$''$ = 0.36 kpc). The follow-up spectroscopic
analysis confirms the existence of a $\sim$ 0.7 keV corona.
}
\end{figure}

\begin{figure}
\vspace{-6cm}
 \centerline{\includegraphics[scale=0.5,angle=270]{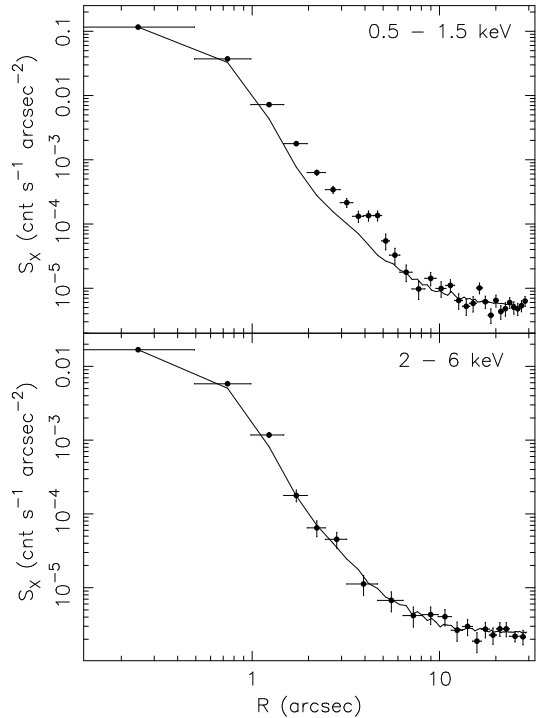}}
\caption{Surface brightness profiles of 3C~264 in the soft and hard bands,
with PSFs (plus a constant local background) shown as the solid lines. While
the model (PSF+background) accurately reconstructs the hard band profile, a soft
excess between 1.5$''$ and 6$''$ is clearly observed (1$''$=0.45 kpc). The
follow-up spectroscopy analysis confirms the existence of a $\sim$ 0.7 keV
corona.
}
\end{figure}
\clearpage

\begin{figure}
\vspace{-2cm}
 \centerline{\includegraphics[scale=0.8]{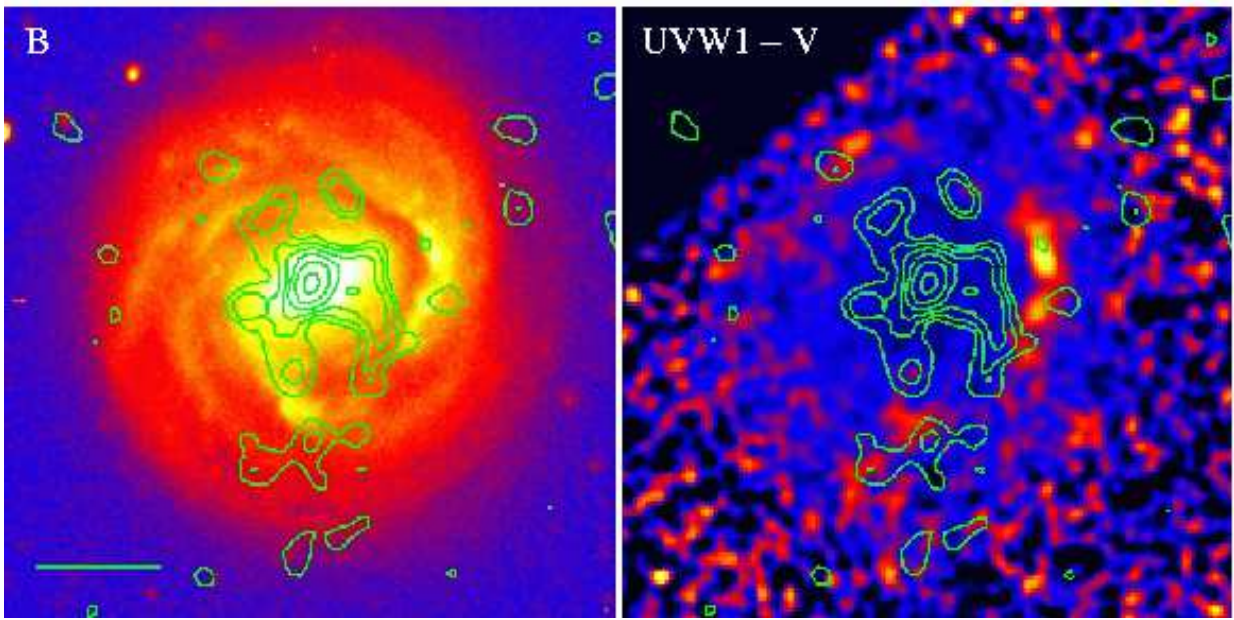}}
\caption{The 0.5 - 1.5 keV \chandra\ contours of NGC~4921 (in Coma) superposed
on the optical B band and the \xmm\ OM UVW1 - V color images.
The scale-bar is 30$''$ (or 14.0 kpc).
NGC~4921 is $\sim 23.4'$ (or 656 kpc) east of Coma's gas core. 
The star formation in NGC~4921 is enhanced in the west arm, while
the HI data implies motion of the galaxy to somewhere west.
}
\end{figure}

\begin{figure}
\vspace{-4cm}
 \centerline{\includegraphics[height=0.45\linewidth,angle=270]{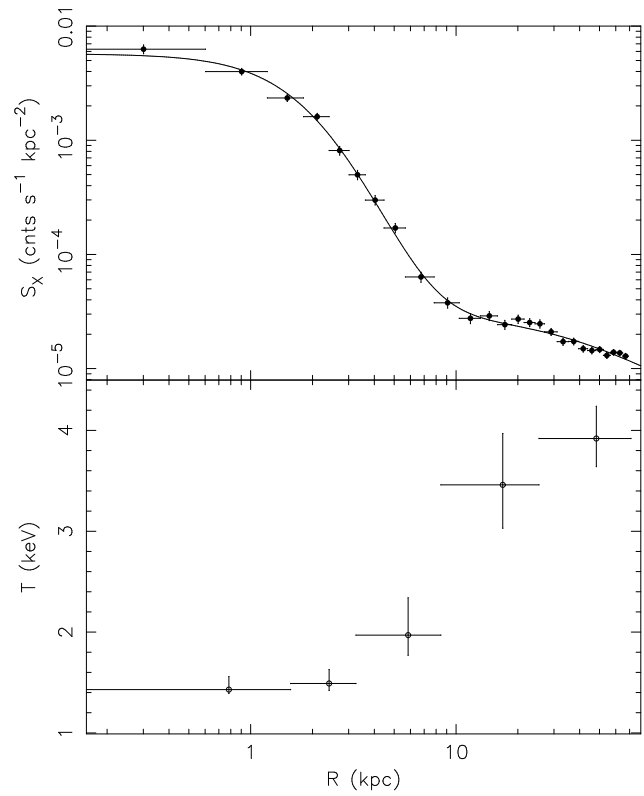}}
\caption{The 0.5 - 3 keV surface brightness profile (upper) and the temperature
profile (lower) of IC~1633 and its surroundings (0.49 kpc = 1$''$).
The solid line is the best-fit of a two-$\beta$ model to the
surface brightness profile. There is a break at $\sim$ 9 kpc. For the
temperature profile, the ICM background on the inner three annuli has been subtracted.
}
\end{figure}
\clearpage

\begin{figure}
\vspace{-2.5cm}
 \centerline{\includegraphics[scale=0.9]{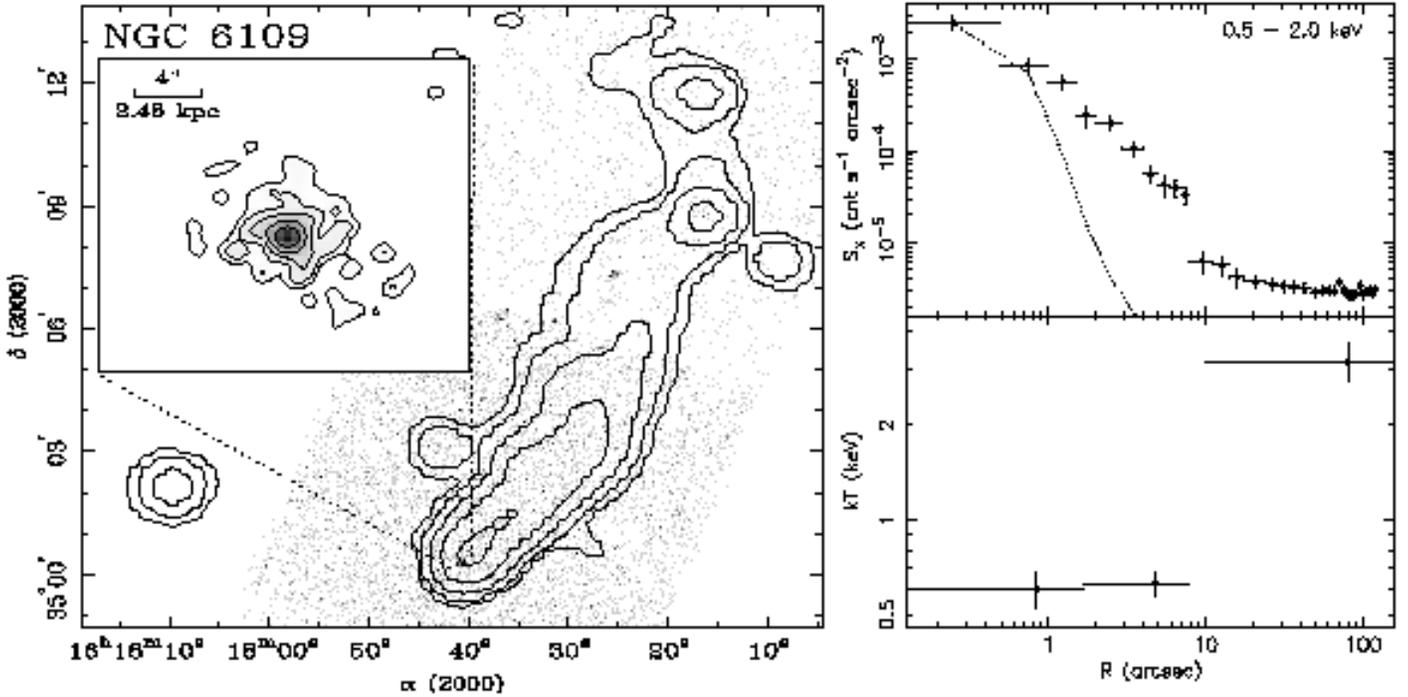}}
\caption{{\bf Left:} The 1.4 GHz contours of B2~1615+35 (from NVSS) superposed
on the \chandra\ image of NGC~6109. The zoom-in of the \chandra\ contours
(in square-root scale) are also shown. {\bf Right:} The soft (0.5 - 2 keV) 
X-ray surface brightness profile of NGC~6109, with the local PSF shown as 
the dotted line. The radius of the corona is 4.6$\pm$0.3 kpc.
}
\end{figure}

\begin{figure}
\vspace{-5cm}
 \centerline{\includegraphics[scale=0.54,angle=270]{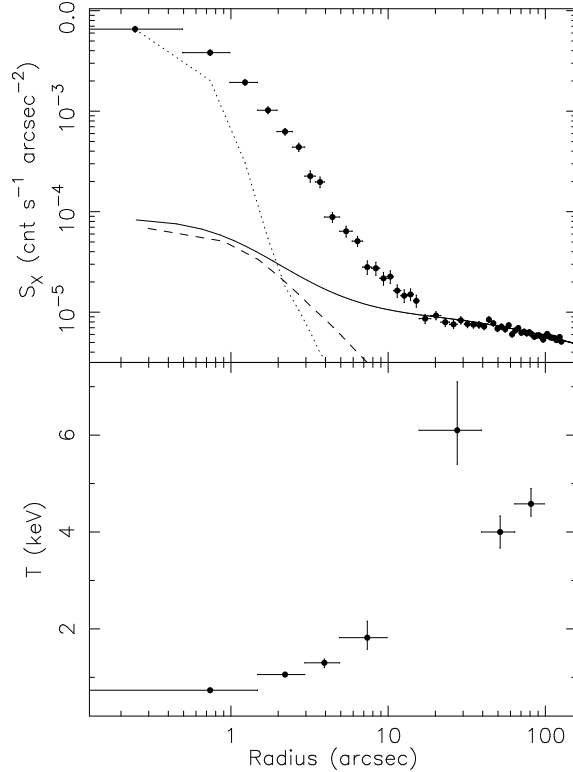}}
\caption{{\bf Upper}: The 0.5 - 2 keV surface brightness profile of NGC~7720,
with the local PSF shown as the dotted line. The dashed line represents the I
band stellar light from the \hst\ data, scaled to matched the predicted total
LMXB light of NGC~7720. The solid line represents the sum of the predicted
LMXB light and the local background (slightly increasing toward the center).
{\bf Lower}: the temperature profile (1$''$ = 0.629 kpc). The ICM background on the
inner four annuli has been subtracted.
}
\end{figure}
\clearpage

\begin{figure}
\vspace{-2.0cm}
 \centerline{\includegraphics[scale=0.55]{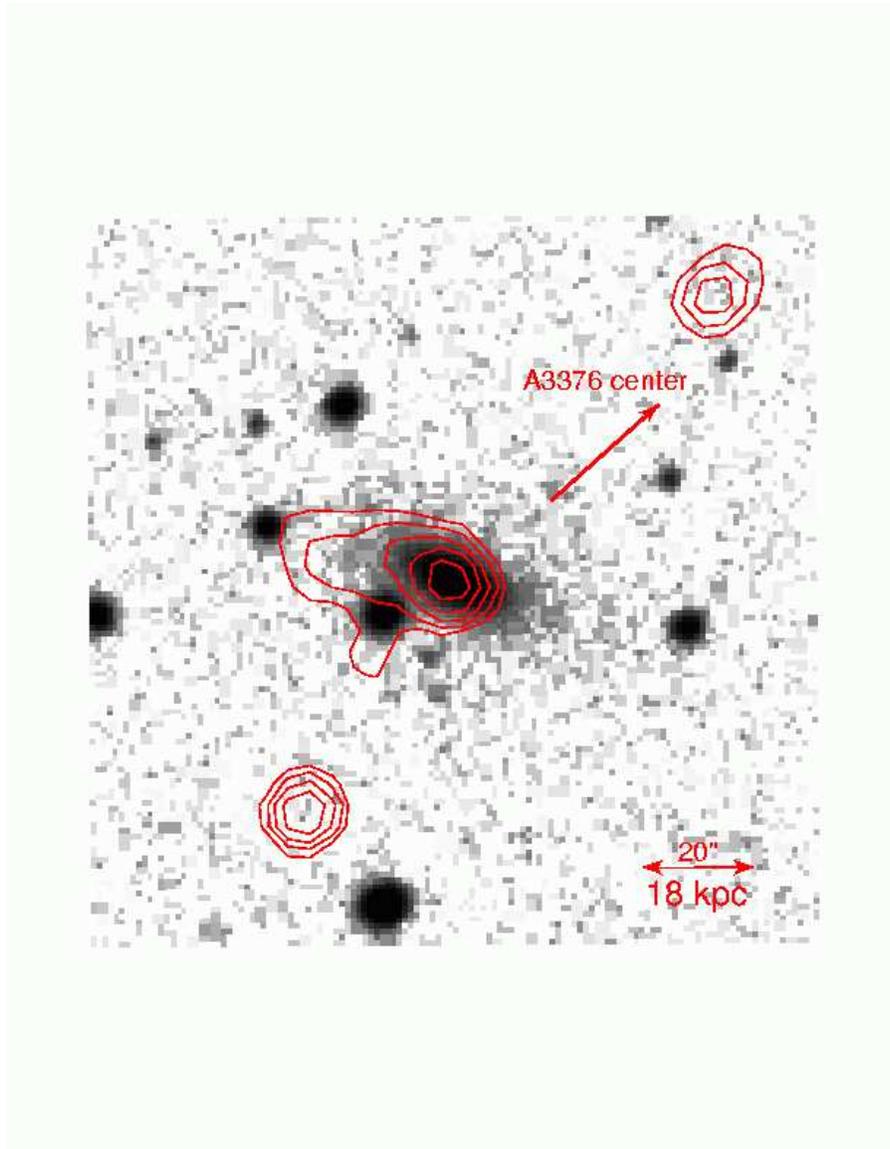}}
\vspace{-0.2cm}
\caption{X-ray contours of PGC~018313 (in square-root scale) superposed on the
DSS II image of the galaxy. An X-ray tail to the east of the galaxy is significant,
extending to $\sim$ 30 kpc from the galaxy center.
}
\end{figure}

\clearpage

Fig. 26 - 28 (void) \\
\\
Please check the complete version at:
http://www.pa.msu.edu/\~{}sunm/coronae\_all\_v1.5\_emuapj.ps.gz

\clearpage

\end{document}